\documentclass[twocolumn,english,superscriptaddress,nofootinbib]{revtex4-1}
\usepackage[T1]{fontenc}
\usepackage[latin9]{inputenc}
\usepackage{color}
\usepackage{textcomp}
\usepackage{mathrsfs}
\usepackage{mathtools}
\usepackage{amsmath}
\usepackage{amssymb}
\usepackage{graphicx}

\makeatletter

\providecommand{\tabularnewline}{\\}

\usepackage{mathrsfs}
\usepackage{amstext}
\usepackage{tabularx}
\providecommand{\tabularnewline}{\\}

\usepackage{natbib}

\usepackage{babel}

\makeatother

\usepackage{babel}
\begin{document}

\title{Shell model based deformation analysis of light Cadmium isotopes}

\date{\today}

\author{T. Schmidt}

\affiliation{Institut f\"{u}r Kernphysik, Universit\"{a}t zu K\"{o}ln, Germany}

\author{K. L. G. Heyde}

\affiliation{Department of Physics and Astronomy, Ghent University, Belgium}

\author{A. Blazhev}

\affiliation{Institut f\"{u}r Kernphysik, Universit\"{a}t zu K\"{o}ln, Germany}

\author{J. Jolie}

\affiliation{Institut f\"{u}r Kernphysik, Universit\"{a}t zu K\"{o}ln, Germany}
\begin{abstract}
Large-scale shell-model calculations for the even-even Cadmium isotopes
$^{98}${\normalsize{}Cd} - $^{108}${\normalsize{}Cd} have been performed
with the ANTOINE code in the $\pi(2p_{1/2};\,1g_{9/2})$ $\nu(2d_{5/2};\,3s_{1/2};\,2d_{3/2};\,1g_{7/2};\,1h_{11/2})$
model space without further truncation. Known experimental energy
levels and $B(E2)$ values could be well reproduced. Taking these
calculations as a starting ground we analyze the deformation parameters
predicted for the {\normalsize{}Cd} isotopes as a function of neutron
number $N$ and spin $J$ using the methods of model independent invariants
introduced by K. Kumar and D. Cline. 
\end{abstract}
\maketitle

\section{Introduction}

In nuclei with a single closed shell (either protons or neutrons)
the energy spectra are characterized by the pairing energy of the
valence particles in the open shell, making seniority an approximate
quantum number. For e.g., the Sn ($Z$ = 50) and Pb ($Z$ = 82) isotopes
and the $N$ = 82 isotones this turns out to be the case. The pair-breaking
energy (energy gap 2$\Delta$) separates the ground state from a rapidly
increasing level density at energies of $E_{x}\sim$1.5 - 2 MeV. Moreover,
the excitation energy of the first excited $2^{+}$ state stays remarkably
constant with changing neutron (in the $Z$ = 50 and $Z$ = 82 isotopes)
or proton number (in the $N$ = 82 isotones). Pairing was incorporated
using the BCS theory of superconductivity as applied to even-even
atomic nuclei \citep{001} and is described in detail by \citep{002,003}.
This pairing fingerprint is well-covered by many present-day large-scale
shell-model calculations \citep{004,005}.

A few valence particle (or holes) away from the closed shell an onset
of quadrupole collectivity appears, which is indicated by the change
in excitation energy of the low-lying $2_{1}^{+}$, $4_{1}^{+}$,
$6_{1}^{+}$,... states as well as by the increase of corresponding
$B(E2)$ values. This is an interesting issue to explore in detail
how nuclei with just two protons outside the closed shell (or missing)
behave, see, e.g., the Cd ($Z$ = 48), Te ($Z$ = 52), the Hg ($Z$
= 80) and the Po ($Z$ = 84) isotopes, as well as isotones with $N$
= 80 and $N$ = 84.

For the Cd nuclei, an extensive set of experimental data has been
obtained over the years, covering essentially the whole $N$ = 50
- 82 neutron major shell, and even going beyond the $N$ = 82 closed
shell, both on low- and high-spin states, $B(E2)$ values, g-factors
(see the detailed set of references: \citep{006,007,008,009,010,011,012_Boelaert1,013_Boelaert2,014,133free,134free,015,016,017,018,019,020,021,022,023,024,025,026,027,028,029,030,031,032,033,034,035,036,037,038,039,040,041,042,043,044,045,046,047,135Jolie01,136Jolie02,137Jolie03,138Jolie04,139Jolie05,140Jolie06,141Jolie07,142Jolie08,143Jolie09,144Jolie10,145Jolie11,146Jolie12,147Jolie13,148Jolie14,152Jolie22_not_listed}),
as well as the systematics for those data (see refs. \citep{048,049,050_NNDC}).
See also reference \citep{051} for a recent review on the structure
of $^{100}$Sn and neighboring nuclei including the light Cd isotopes.

The Cd nuclei have been studied using shell-model calculations for
the lighter mass region \citep{007,008,009,010,011,012_Boelaert1,014,016,052}
and also using the Interacting Boson Model (IBM) \citep{015,016,019,021,097,136Jolie02,149Jolie16,098,099,101,100,150Jolie21,112,152Jolie22_not_listed}.
Besides that, other studies, starting from a general collective Bohr
Hamiltonian, derived from a microscopic starting point using a Skyrme
force, calculations using the Adiabatic Time-Dependent Hartree-Fock-Bogoliubov
(ATDHFB) method (for the nuclei $^{106-116}$Cd) \citep{054}, as
well as using a self-consistent HFB approach, starting from the finite
range Gogny interaction \citep{055} have been carried out.

Our aim, in the present paper, is to show how, starting from an extensive
and large-scale shell-model calculation, it becomes possible to characterize
the onset of quadrupole collectivity with increasing number of valence
neutrons outside of the ${\normalcolor {\normalcolor _{38}^{88}}}$Sr$_{50}$
core nucleus. We concentrate on the calculation of both quadratic
and cubic rotational invariants (constructed starting from the E2
transition and diagonal matrix element) as was originally proposed
by Kumar \citep{056_Kumar} and Cline and Flaum. \citep{057_Cline,058,059,060}.
Subsequently, using those matrix elements as input, we can extract
quantitative information about the changing collective properties
of the low-lying states (band structure if possible), through the
quadrupole parameters ($\beta$, $\gamma$) \citep{061_Bohr&Mottelson},
mainly used to characterize the intrinsic deformation properties of
the Cd nuclei studied in the present paper. 

The detailed spectroscopic results for the Cd nuclei, studied in this
paper, such as energy spectra (covering both the low-spin and high-spin
regions), indications of ``collective bands\textquotedblright{} and
related electromagnetic properties (mainly electric quadrupole and
magnetic dipole), as well as a detailed comparison with the extensive
set of data, will form the content for a forthcoming paper. 

\section{Shell model calculations}

Large-scale shell-model calculations (LSSM) of Cadmium isotopes have
been performed using the complete neutron model space ($N$ = 50 -
82), i.e. filling the $2d_{5/2}$, $3s_{1/2}$, $2d_{3/2}$, $1g_{7/2}$
and $1h_{11/2}$ orbitals with neutrons while ten protons remain distributed
in the $2p_{1/2}$ and $1g_{9/2}$ orbitals ($Z$ = 48) forming the
proton model space ($Z$ = 38 - 50). This way $^{88}$Sr acts as an
inert model space core, i.e. we assume no interaction between the
valence particles in the model space and the inert core. Within this
model space we study and explore a multitude of nuclear structure
properties. In particular, the changing quadrupole collective properties,
indicated through the $E_{x}(2_{1}^{+})$ and $B(E2)$ values along
the yrast band, with increasing number of valence neutrons moving
outside the $N$ = 50 closed shell. The nucleon-nucleon interaction
used is an effective realistic force originating from the Bonn-CD
potential, resulting in a G-matrix called the $v3sb$ effective interaction
(see \citep{012_Boelaert1} page 2, left column for more details).
This interaction has been modified implying a slight adjustment of
the monopoles such as to exhibit the correct propagation of the single-particle
neutron energies moving from $N=51$ ($^{89}$Sr) towards the end
of the shell at $N=81$ ($^{131}$Sn) as well as some changes in the
effective pp, np and nn matrix elements as outlined in ref. \citep{012_Boelaert1}
resulting from a fit of the force to 189 data points (excitation energies)
in the mass region considered here. This interaction is called later
on $v3sbm$.

An important point is the right choice of the proton and neutron effective
charges throughout the full set of Cd nuclei. The procedure used is
to fix the proton effective charge $e_{\pi}$ fitting the theoretical
$B(E2)$ value for the $8_{1}^{+}\rightarrow6_{1}^{+}$ transition
to the known experimental value in $^{98}$Cd \citep{008}. Having
fixed this value, the neutron effective charge $e_{\nu}$ was fixed
by comparing the experimentally known and theoretically calculated
$B(E2)$ values for the $2_{1}^{+}\rightarrow0_{1}^{+}$ transitions
in $^{102,104}$Cd \citep{013_Boelaert2}. The effective charges used
in the current work are $e_{\pi}=1.7e$ and $e_{\nu}=1.1e$, i.e.,
the same effective charges as in the previous shell-model calculations
with this interaction \citep{012_Boelaert1,013_Boelaert2}. A former
shell-model study on light Cd isotopes using v3sb by A. Ekstr\"{o}m
et al.\citep{014} (see also references \citep{159free,160free,161free})
successfully reproduced the experimental values with similar effective
charges.

With these ingredients kept constant in the LSSM study of the Cd nuclei,
it is only the $NN$ interaction acting in the large model space that
produces the nuclear structure properties as a function of increasing
neutron number. The large-scale shell-model calculations were performed
with the code ANTOINE \citep{004}. The calculations presented here
were conducted in the full model space without any additional truncations.
The m-scheme matrix dimension for m = 0 in $^{108}$Cd was about $10^{8}$.
First results have been published in \citep{062}.

\section{Shape invariants\label{sec:Shape-invariants}}

It turns out that rather than comparing a multitude of experimental
data separately with the results from a specific model description
of nuclear structure, almost model-independent methods have been developed
to characterize the nuclear quadrupole properties for all states.
This leads to the possibility to find sets of correlated states based
on the intrinsic properties for those states. The idea is to construct
so-called rotational invariants, which, in the case of studying quadrupole
collective properties, are built from a product of a number of the
E2 operators \citep{056_Kumar,057_Cline} which, starting from the
E2 operator, 
\begin{equation}
P_{2\mu}=\sum_{i=1}^{A}e_{i}r_{i}^{2}Y_{2\mu}\left(\Omega_{i}\right),
\end{equation}
is described as a tensor product of n=2,3,... such operators, coupled
to a rank 0 tensor, defined as 
\begin{equation}
P^{\left(n\right)}=\left[P_{2}\otimes P_{2}\otimes\cdots\otimes P_{2}\right]_{2}\cdot P_{2}.
\end{equation}
When calculating the expectation value of such operators in any given
eigenstate of the nucleus $\left|J,M\right\rangle $ and because of
its zero rank character, this gives rise to a rotational invariant
quantity. As the spectroscopic quadrupole moments for $0^{+}$ states
vanish, invariants are also the only access to study the deformation
of those states. This also implies that the results will be the same,
independent of the reference frame used: be it the laboratory frame
or the frame centered on the principal axis of the nucleus. For a
more general derivation and discussion of these invariants, the reader
is referred to the references \citep{056_Kumar,057_Cline}.

In this section, we give a short review on the building of these invariants
and also lay open some differences in the methods initiated by Kumar
\citep{056_Kumar} and Cline \citep{057_Cline}.

For our analysis, we aim to extract the deformation parameters $\beta$
and $\gamma$ as introduced by Bohr and Mottelson \citep{061_Bohr&Mottelson}.
Therefore, it is sufficient to derive the invariants $P_{s}^{\left(2\right)}$
and $P_{s}^{\left(3\right)}$, though invariants of higher couplings
are possible in principle \citep{057_Cline}, denoted as $P_{s}^{(n)}=\left\langle s,M_{s}\left|P^{(n)}\right|s,M_{s}\right\rangle $,
resulting in the expressions

\begin{equation}
P_{s}^{\left(2\right)}=\frac{1}{2I_{s}+1}\sum_{r}\left(M_{sr}\right)^{2},\label{eq:Kumar P2}
\end{equation}

\setlength{\arraycolsep}{1pt} 
\begin{equation}
P_{s}^{\left(3\right)}=-\frac{\sqrt{5}}{2I_{s}+1}\left(-1\right)^{2I_{s}}\sum_{rt}\left\{ \begin{array}{ccc}
2 & 2 & 2\\
I_{s} & I_{r} & I_{t}
\end{array}\right\} M_{sr}M_{rt}M_{ts},\label{eq:Kumar P3}
\end{equation}
with $r$ and $t$ describing the intermediate states of the coupling
and with $\left\{ \right\} $ denoting a Wigner-$6j$-symbol. Furthermore,
$M_{sr}=\left\langle s\left\Vert P_{2}\right\Vert r\right\rangle $
is a shorthand notation for the reduced $E2$ matrix elements that
are needed as input to calculate the $P_{s}^{\left(2\right)}$ and
$P_{s}^{\left(3\right)}$ invariants. These can be the results extracted
from experimental studies, or, as we are performing LSSM calculations
for the Cd nuclei, the calculated $E2$ reduced matrix elements. In
the above, $s,r,t,...$ are shorthand notations to specify all quantum
numbers necessary to characterize the various nuclear levels.

\setlength{\arraycolsep}{5pt}

Once the invariants are calculated and available, deformation parameters
can be extracted using methods originally proposed by Kumar \citep{056_Kumar}
and Cline and Flaum (\citep{057_Cline,058,059,060}) using slightly
different methods to do so. These methods have been used in a large
number of recent papers (in particular in the region of the Ge, Kr,
Mo, Ru and Pd nuclei \citep{063,064,065,066,067_NPA_766,068}, as
well as in the much heavier Pb mass region (W, Os, Pt, Hg, Po isotopes)
\citep{069,070,071,072,073,074}.

In the present paper, we use both methods to study the sensitivity
of the extracted deformation parameters $\beta$ and $\gamma$, which
we review in a succinct way so as to point out similarities and some
subtle differences. 

Shape invariants have also been studied within the context of the
Interacting Boson Model (IBM) in various mass regions (see e.g. \citep{075,076}
and references therein) in order to extract information on a mean-field
level (nuclear quadrupole deformation parameters,..). Care should
be taken into account when applying the calculation of the quadrupole
invariants within the context of the IBM, as was pointed out by Dobaczewski
et al. \citep{077}.

\subsection{Kumar method\label{subsec:Kumar-method}}

The approach of Kumar makes use of the intrinsic quadrupole moment,
which is: 
\begin{equation}
Q_{s\mu}^{i}=\sqrt{\frac{16\pi}{5}}\int\rho_{s}r^{2}Y_{2\mu}dV.\label{eq:Quadrupolmoment intr. allgem.}
\end{equation}
In this notation $Q_{s\mu}^{i}$ is the $\mu$th component of the
quadrupole moment of an equivalent ellipsoid of the nucleus, with
charge density $\rho_{s}$. Furthermore, because of its reflection
symmetry, for the quadrupole components of the ellipsoid it can be
shown that \citep{056_Kumar,061_Bohr&Mottelson}: 
\begin{eqnarray}
Q_{s2}^{i}=Q_{s,-2,}^{i}\nonumber \\
Q_{s1}^{i}=Q_{s,-1}^{i}=0.\label{eq:Kumar coll. amplitudes quadrup.}
\end{eqnarray}
The non vanishing components are written per definition as \citep{056_Kumar,061_Bohr&Mottelson}:
\begin{eqnarray}
Q_{s0}^{i}=Q_{s}^{i}\cos\gamma_{s},\nonumber \\
Q_{s2}^{i}=Q_{s,-2}^{i}=\frac{Q_{s}^{i}}{\sqrt{2}}\sin\gamma_{s}.\label{eq:Kumar Qi cos sin}
\end{eqnarray}
The invariant $P_{s}^{\left(2\right)}$ of equation (\ref{eq:Kumar P2})
is now rewritten by replacing the reduced matrix elements of the $E2$
operators $P_{2\mu}$ with the intrinsic quadrupole moments given
in (\ref{eq:Quadrupolmoment intr. allgem.}). This way the invariants
for $n=2$ and 3 are expressed by $Q_{s\mu}^{i}$ and $\gamma_{s}$,
and, due to equation (\ref{eq:Kumar Qi cos sin}), they are expressed
by $Q_{s}^{i}$. This way one finally gets the expressions \citep{056_Kumar}:
\begin{equation}
Q_{s}^{i}=\sqrt{\frac{16\pi}{5}}\sqrt{P_{s}^{\left(2\right)}},\label{eq:Quadrupol intr. gleich P2}
\end{equation}
\begin{equation}
\cos3\gamma_{s}=-\sqrt{\frac{7}{2}}\frac{P_{s}^{\left(3\right)}}{\left(P_{s}^{\left(2\right)}\right)^{\frac{3}{2}}}.\label{eq:Kumar cos3Gamma}
\end{equation}
Although the intrinsic quadrupole moment $Q_{s}^{i}$ in (\ref{eq:Quadrupol intr. gleich P2})
already is an expression of the nuclear deformation, it is convenient
to express the nuclear shape by the usual deformation parameters $\beta$
and $\gamma$. Therefore a calculation of $\beta$ defined by the
ratio of the quadrupole moment over the monopole moment is useful:

\begin{equation}
\beta_{s\mu}=\frac{4\pi}{5}\frac{\int\rho_{s}r^{2}Y_{2\mu}dV}{\int\rho_{s}r^{2}dV}=\sqrt{\frac{\pi}{5}}\frac{Q_{s\mu}^{i}}{Z\left\langle s\left|r^{2}\right|s\right\rangle }.\label{eq:beta_quadrupol_monopol}
\end{equation}
The normalization factor $\frac{4\pi}{5}$ has been chosen so that
the deformation parameter matches with $\beta$ of Bohr and Mottelson
\citep{061_Bohr&Mottelson}. Since the proportionality factors in
(\ref{eq:beta_quadrupol_monopol}) are independent of $\mu$, the
tensor $\beta_{s\mu}$ follows the same relations as $Q_{s\mu}$ in
equations (\ref{eq:Kumar coll. amplitudes quadrup.}) and (\ref{eq:Kumar Qi cos sin})
and thus it is sufficient to focus on the magnitude of the deformation
$\beta_{s}$ instead of $\beta_{s\mu}$. The remaining difficulty
now is to calculate the monopole moment $\left\langle r^{2}\right\rangle $.
This can be done by calculating $\left\langle r^{2}\right\rangle $
for an equivalent ellipsoid of equal volume (or $R_{0}^{3}$) with
charge density obtained by uniformly distributing the charge $Ze$
over the ellipsoid and where $P_{s}^{(2)}$ and $P_{s}^{(3)}$ are
equal to those of the nucleus in the given state $\left|s,M_{s}\right\rangle $.
Kumar \citep{056_Kumar} has shown that extracting the value of $\beta$
under the above condition is equivalent to solving the cubic equation:
\begin{equation}
\delta_{s}^{3}\left(g_{s}^{3}-2\cos3\gamma_{s}\right)+3\delta_{s}^{2}-1=0,
\end{equation}
with the relation between $\delta_{s}$ and $\beta_{s}$ expressed
as 
\begin{equation}
\delta_{s}=\beta_{s}/\sqrt{\frac{4\pi}{5}},
\end{equation}
and $g_{s}=\frac{6ZR_{0}^{2}}{5Q_{s}^{i}}$. It is important to mention
that the values $Q_{s}^{i}$ and $cos3\gamma_{s}$ define the deformation
characteristics associated with the given state $s$, the extraction
of deformation parameters $\beta$, $\gamma$ imply a certain model
assumption, which is very general though.

\subsection{Recent analyses methods: Cline-Flaum approach \label{subsec:Recent-analyses-methods:}}

In most of the recent papers it is more common to use a slightly different
notation for the invariants $P_{s}^{\left(2\right)}$ and $P_{s}^{\left(3\right)}$,
which was initially introduced by D. Cline \citep{057_Cline}. Besides
the difference in notation there is a difference in prefactors, which
can result in some confusion, when comparing both notations \footnote{Kumar is deriving the invariants by making a transformation from a
tensor product to a scalar product, when coupling $E2$ matrix elements
($P_{2}\cdot P_{2}=\sqrt{5}\left[P^{(2)}\otimes P^{(2)}\right]_{0}^{(0)}$)
to produce the angular momentum zero coupling. In this section the
invariants are derived by tensor couplings exclusively, resulting
in a general difference in factors of $\sqrt{5}$, when comparing
both methods.}.

Here, one starts from the knowledge that the expectation value of
a tensor rank zero operator, $\left[E2\otimes E2\right]_{0}^{(0)}$,
$[\left[E2\otimes E2\right]^{(2)}\otimes E2]_{0}^{(0)}$, is independent
of the reference frame. Evaluating the above tensor products within
the principal axis frame, the quadrupole operator is described by
two non-vanishing quadrupole operators, only. One makes the choice
of $E2_{0}=Qcos\delta$, $E2_{\pm2}=Qsin\delta\frac{1}{\sqrt{2}}$
, $E2_{\pm1}=0$.

Analogous to (\ref{eq:Kumar P2}) one now derives the result \citep{057_Cline}:
\begin{eqnarray}
\left\langle i\left|\left[E2\otimes E2\right]_{0}^{(0)}\right|i\right\rangle  & = & \frac{1}{\sqrt{5}}\frac{1}{2I_{i}+1}\sum_{t}\left|\left\langle i\left\Vert E2\right\Vert t\right\rangle \right|^{2}\nonumber \\
 & = & \frac{1}{\sqrt{5}}\left\langle Q^{2}\right\rangle .\label{eq:2tensor_coupling}
\end{eqnarray}
The matrix elements involved are now denoted by $\left\langle i\left\Vert E2\right\Vert t\right\rangle $
instead of $M_{sr}$ as before. A comparison of both formulas shows
indeed that $\left\langle Q^{2}\right\rangle =P_{i}^{\left(2\right)}$,
which in turn is equal to a summation of $B\left(E2\right)$-values,
indicating the equivalence of both methods in evaluating the quadratic
invariants. In the same way one obtains for the coupling of three
operators analogous to invariant $P_{s}^{\left(3\right)}$ of equation
(\ref{eq:Kumar P3}) \citep{057_Cline,067_NPA_766}: 
\begin{eqnarray}
 & \left\langle i\left|\left[\left[E2\otimes E2\right]^{(2)}\otimes E2\right]_{0}^{(0)}\right|i\right\rangle \nonumber \\
= & \frac{\left(-1\right)^{2I_{i}}}{2I_{i}+1}\sum_{t,u}\left\langle i\left\Vert E2\right\Vert u\right\rangle \left\langle u\left\Vert E2\right\Vert t\right\rangle \left\langle t\left\Vert E2\right\Vert i\right\rangle \nonumber \\
 & \times\left\{ \begin{array}{ccc}
2 & 2 & 2\\
I_{i} & I_{u} & I_{t}
\end{array}\right\} .
\end{eqnarray}
Carrying out the recoupling when working in the principal axis frame,
the expectation value of the cubic invariant becomes 
\begin{equation}
\left\langle i\left|\left[\left[E2\otimes E2\right]^{(2)}\otimes E2\right]_{0}^{(0)}\right|i\right\rangle =-\sqrt{\frac{2}{35}}\left\langle Q^{3}\cos\left(3\delta\right)\right\rangle .
\end{equation}
Here the deformation parameter expressing triaxiality is denoted by
$\delta$. Keeping in mind that $\left\langle Q^{3}\right\rangle =(P_{i}^{\left(2\right)})^{\frac{3}{2}}$
one obtains exactly the same expressions using the Kumar convention
as when using the notation of the present subsection to calculate
the invariant $P_{i}^{\left(3\right)}$ (or $\bigl\langle i\bigr|\bigl[\left[E2\times E2\right]_{2}^{(2)}\times E2\bigr]\bigl|i\bigr\rangle$
respectively) and, thus, $\gamma_{s}$ equals $\delta$.

In most experimental papers, starting from reduced $E2$ matrix elements
$\left\langle \left\Vert E2\right\Vert \right\rangle $ obtained using
Coulomb excitation, the values of $\left\langle Q^{2}\right\rangle $
and $\left\langle Q^{3}\cos(3\delta)\right\rangle $, can be extracted
for each individual excited state. Having come this far, it is important
to mention that the quadrupole invariants do play a role as extracted
``observables\textquotedblright , being constructed from measured
reduced $E2$ matrix elements, and as such express the nuclear deformation
characteristics in a condensed way through the sums $P_{s}^{(2)}$
, $P_{s}^{(3)}$ and this for a given nuclear state described by the
quantum numbers $s$ (Kumar notation), or $i$ (Cline notation).

They are both useful and significant in presenting the way in which
the nuclear excited states are ``correlated\textquotedblright{} through
a subset of collective degrees of freedom, expressed mostly using
the Bohr-Mottelson $\beta$, $\gamma$ parameters. A problem, at present,
is still the knowledge of too few data because of the difficulties
to extract rather complete sets of reduced $E2$ matrix elements using
Coulomb excitation \citep{014,133free,134free}. Therefore, there
is need for experiments that use higher-energy Coulomb excitation,
which is one of the goals of HIE-ISOLDE project \citep{153Blazhev01_not_listet_HIE-ISOLDE}.

It is interesting though (see Kumar method) to transform the $\left\langle Q^{2}\right\rangle $
and $\left\langle Q^{3}\cos(3\delta)\right\rangle $ invariants into
corresponding $\left\langle \beta^{2}\right\rangle $ and $\left\langle \beta^{3}\cos(3\gamma)\right\rangle $
values. An extensive study has been carried out by Srebrny \textit{et
al. }\citep{067_NPA_766} describing deformed nuclei by a Nilsson
ellipsoidal deformed potential \citep{078,079,080}, equating the
experimental (or theoretical, as derived from our present LSSM calculations)
$P_{s}^{(2)}$ , $P_{s}^{(3)}$ invariants with the corresponding
theoretical values. This results in the relations 
\begin{equation}
Q^{2}=\left(\frac{3}{4\pi}ZeR_{0}^{2}\right)^{2}\left(\beta^{2}+\mathcal{O}(\beta^{3})\right),\label{eq:Q^2-expansion}
\end{equation}
and

\begin{eqnarray}
Q^{3}\cos\left(3\delta\right) & = & \left(\frac{3}{4\pi}ZeR_{0}^{2}\right)^{3}\left(\beta^{3}\cos(3\gamma)+\mathcal{O}(\beta^{4})\right)\label{eq:Q^3Cos3delta-expansion}
\end{eqnarray}
(whereas in Kumar's approach, no such expansion is needed).

It can also be shown that within the context of Hartree-Fock-Bogoliubov
microscopic calculations, only the lowest order contribution results,
for both equation (\ref{eq:Q^2-expansion}) and (\ref{eq:Q^3Cos3delta-expansion}).
In general, small differences in the extracted values of $\beta$,
$\gamma$ may result depending on the precise definition of the collective
variables (see e.g. \citep{066}). The higher order terms of equations
(\ref{eq:Q^2-expansion}) and (\ref{eq:Q^3Cos3delta-expansion}) can
be found in the appendix of \citep{067_NPA_766}.

We note that in the treatments of Kumar and Cline \citep{057_Cline,067_NPA_766,070},
sums of products of $E2$ matrix elements are defined for fluctuations
in the invariants. This is important when addressing experimental
data. Herein, we present detailed maps of all of the $E2$ strength
which, de facto, represent a very high resolution decomposition of
the fluctuational content of the centroids.

\subsection{Deformation analysis: application to the harmonic oscillator}

Before moving into a detailed discussion of the deformation properties
for the light Cd nuclei, spanning the region in between mass number
$A=98$ and $A=108$, we have carried out a schematic analysis of
the Kumar-Cline sum-rule method extracting the invariants when applied
to a harmonic vibrator model. Consequently, we have an exact test
of the sum-rule method when applied for vibrational nuclei in the
evaluation of the value of $\left\langle \beta^{2}\right\rangle $
\citep{003,081}.

Within the harmonic vibrator model with an homogeneous charge distribution,
the resulting collective model electric operator, describing harmonic
vibrational motion \citep{061_Bohr&Mottelson} with multipolarity
$\lambda$ is defined by:

\begin{equation}
\mathscr{M}\left(E\lambda\right)=\frac{3}{4\pi}ZeR^{\lambda}\hat{\alpha}_{\lambda\mu},
\end{equation}
with $\hat{\alpha}_{\lambda\mu}$, the collective coordinates describing
the oscillatory behavior.

This results in the well-known variation of the mean-square charge
radius, which can be derived as \citep{081}:

\begin{eqnarray}
\beta_{N}^{2} & = & \bigl\langle\alpha,N,JM\bigl|\sum_{\mu}\hat{\alpha}_{2\mu}^{*}\hat{\alpha}_{2\mu}\bigr|\alpha,N,JM\bigr\rangle,\nonumber \\
 & = & \frac{\hbar\omega_{2}}{2{\normalcolor {\color{red}{\normalcolor C}}_{2}}}\left(5+2N\right),\label{eq:E=000026G mean square beta}
\end{eqnarray}
with the phonon number $N$, spin $J$ and spin projection $M$.

It can easily be shown in an example calculation, using the restricted
framework of an ideal vibrator, where both states and transition strength
of the phonons are known, that a simple application of sum rules to
(\ref{eq:E=000026G mean square beta}) results in the calculation
of invariants for the ideal vibrator including the correct energy
dependency as in (\ref{eq:E=000026G mean square beta}).

We first consider the ground state $0_{1}^{+}$. Here, the sum in
eq. (\ref{eq:2tensor_coupling}) only contains the $2_{1}^{+}$ state
as intermediate state, with a result similar to $\left\langle Q^{2}\right\rangle =q_{0}^{2}\frac{5\hbar\omega_{2}}{2{\normalcolor {\color{red}{\normalcolor C}}}_{{\normalcolor {\color{red}{\normalcolor 2}}}}}$,
where $q_{0}$ according to equation \eqref{eq:Q^2-expansion} is
defined by $q_{0}=\frac{3}{4\pi}ZeR_{0}^{2}$. This straightforwardly
identifies the invariant analyses with the correct value by $\left\langle \beta^{2}\right\rangle $
for the ground-state vibrational mode. As a proof of concept the example
of the sum rule calculation for the first phonon $\left(N=1\right)$
state $2_{1}^{+}$ will be shown too: 
\begin{equation}
\bigl\langle2_{1}^{+}\bigl|\sum_{\mu}\bigl|\alpha_{2\mu}\bigr|^{2}\bigr|2_{1}^{+}\bigr\rangle=\frac{1}{5}\sum_{J,f}\left(-1\right)^{J}\left|\bigl\langle2_{1}^{+}\bigl\Vert\hat{\alpha}_{2}\bigr\Vert J_{f}\bigr\rangle\right|^{2}.\label{eq:example sum rule}
\end{equation}

Let us remark that the right-hand side can also be written as $\sum_{J,f}B(E2;2_{1}^{+}\rightarrow J_{f})$.
In this case the sum of (\ref{eq:example sum rule}) runs over $J_{f}=4_{1}^{+},\,2_{2}^{+},\,0_{2}^{+}$
and $0_{1}^{+}$ only. Thus, with respect to the normalized transition
strength of the harmonic oscillator phonon levels to the first $\left(2_{1}^{+}\rightarrow0_{1}^{+}\right)$
phonon transition \citep{061_Bohr&Mottelson}: 
\begin{eqnarray}
 & \sum_{J_{N-1}}B\left(E2;N,J_{n}\rightarrow N-1,J_{N-1}\right)\nonumber \\
 & =N\cdot B\left(E2;N=1\rightarrow N=0\right),
\end{eqnarray}
and taking $B\left(E2;N=1\rightarrow N=0\right)\coloneqq1$ the sum
of (\ref{eq:example sum rule}) results into: 
\begin{equation}
\left(1+\frac{2}{5}+\frac{10}{5}+\frac{18}{5}\right)\frac{\hbar\omega_{2}}{2{\color{red}{\normalcolor C}}_{{\color{red}{\normalcolor 2}}}}=7\frac{\hbar\omega_{2}}{2{\color{red}{\normalcolor C}}_{2}}.\label{eq:result example}
\end{equation}
We recall, that the quadrupole moment of a pure harmonic oscillator
is zero and thus in the sum of (\ref{eq:example sum rule}) $\left\langle 2_{1}^{+}\left\Vert \hat{\alpha}_{2}\right\Vert 2_{1}^{+}\right\rangle =0$.
The outcome of (\ref{eq:result example}) is fully consistent with
the expected result from (\ref{eq:E=000026G mean square beta}).

This example demonstrates the direct connection between the invariant
$P_{s}^{\left(2\right)}$, the deformation parameter $\beta_{s}=\sqrt{\left\langle \beta^{2}\right\rangle }$
and the collective coordinates $\alpha_{2\mu}$. Although the mean
square deformation $\beta_{s}$, derived from the invariants in the
lab frame, and $\beta$, originally defined as a collective coordinate,
are of the same physical quality, by construction they are not the
same physical quantity.

\section{Deformation analysis}

\subsection{Deformation correlated to neutron number}

\begin{figure}[th]
\includegraphics[scale=0.116]{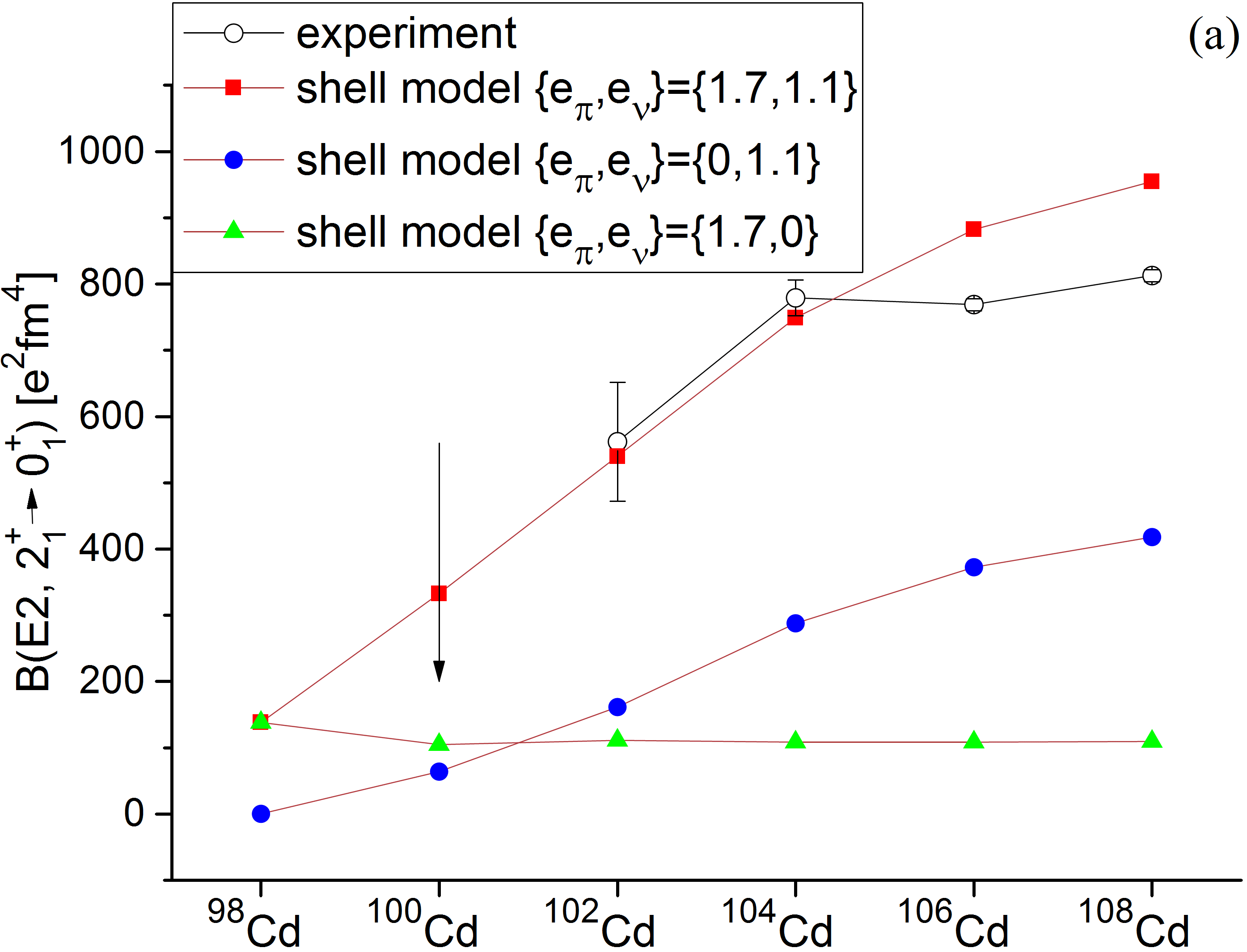}

\includegraphics[scale=0.116]{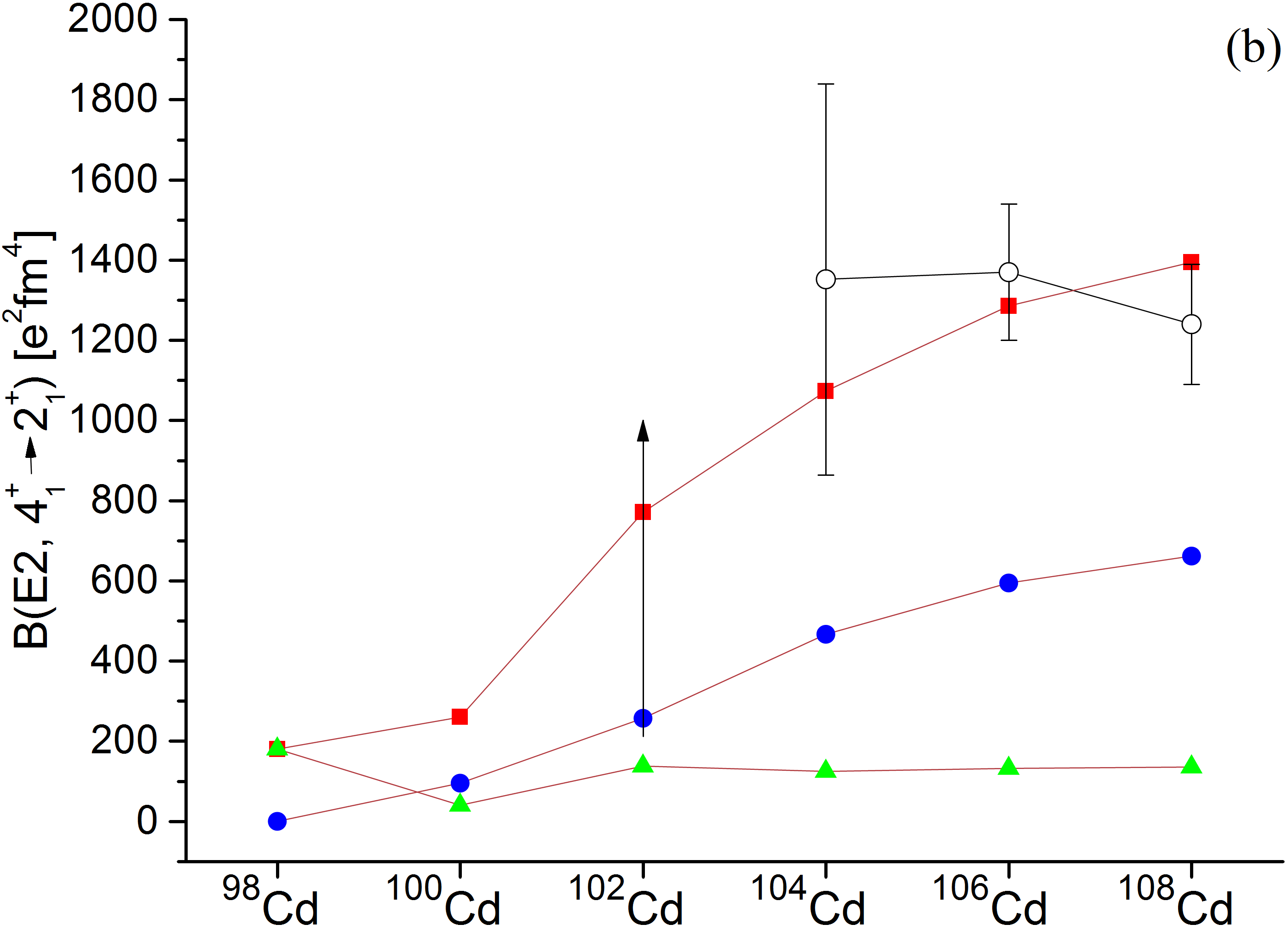}

\caption{(a): The theoretical $B\left(E2;\,2_{1}^{+}\rightarrow0_{1}^{+}\right)$
values with red lines to guide the eye. We also present the separate
contributions of protons (green triangles), neutrons (blue dots) and
overall contribution to transition strength (red squares). The black
circles present experimental results as well as data taken from \citep{014}
($^{100}$Cd), \citep{013_Boelaert2} ($^{102,104}$Cd) and \citep{134free}
($^{106,108}$Cd). (Color online). \protect \\
(b): The theoretical $B\left(E2;\,4_{1}^{+}\rightarrow2_{1}^{+}\right)$values
compared to experimental results taken from \citep{011} ($^{102}$Cd),
\citep{013_Boelaert2} ($^{104}$Cd) and \citep{133free} ($^{106,108}$Cd).
(Color online).}

\label{2 to 0 over N} 
\end{figure}

A common tool for the analysis of collectivity in general is the $E2$
transition strength of the first $2_{1}^{+}$ state to the ground
state. In the harmonic vibrator picture, this transition is also considered
to correspond to the one phonon transition and, because of its collective
character, is expected to increase with the number of valence particles.
In Figure \ref{2 to 0 over N} we present a comparison between the
theoretical and experimental $B(E2)$ values as well as the separate
contributions corresponding with the proton and neutron part, in the
shell-model calculations. The evolution of the theoretical $B(E2)$
values follows a steady increase as a function of neutron number up
to $^{104}$Cd. Here the experimental data indicate a leveling off
whereas the theoretical B(E2) values are still increasing albeit less
steep. The experimental data for the heavier Cd nuclei (beyond $A=108$)
indicate a further increase, coming to a maximal value of 1140 $e^{2}fm^{4}$
at neutron number 70 ($A=118$) before dropping again (see ref. \citep{014}).
On the other hand, these effective charges do not result in too large
$B(E2;\ensuremath{4_{1}^{+}\rightarrow}2\ensuremath{_{1}^{+}})$ values
for the $\ensuremath{^{104-108}}$Cd nuclei as shown in Figure \ref{2 to 0 over N}
(b). A slight reduction of the effective charges from$(e_{\pi},e_{\nu})=(1.7e,1.1e)$
to $(1.6e,1.0e)$ as in \citep{014} would result in a better agreement
with the $B\left(E2;\,2_{1}^{+}\rightarrow0_{1}^{+}\right)$ values
for $^{106,108}$Cd, but the reproduction of the $B(E2;\ensuremath{4_{1}^{+}\rightarrow}2\ensuremath{_{1}^{+}})$
data would deteriorate. Any small variation of the effective charges
would result in small changes of the SM $E2$ strengths used as an
input for the deformation analysis performed in this work, thus preserving
the overall validity of the deformation analysis results.

\begin{figure}[t]
\includegraphics[scale=0.124]{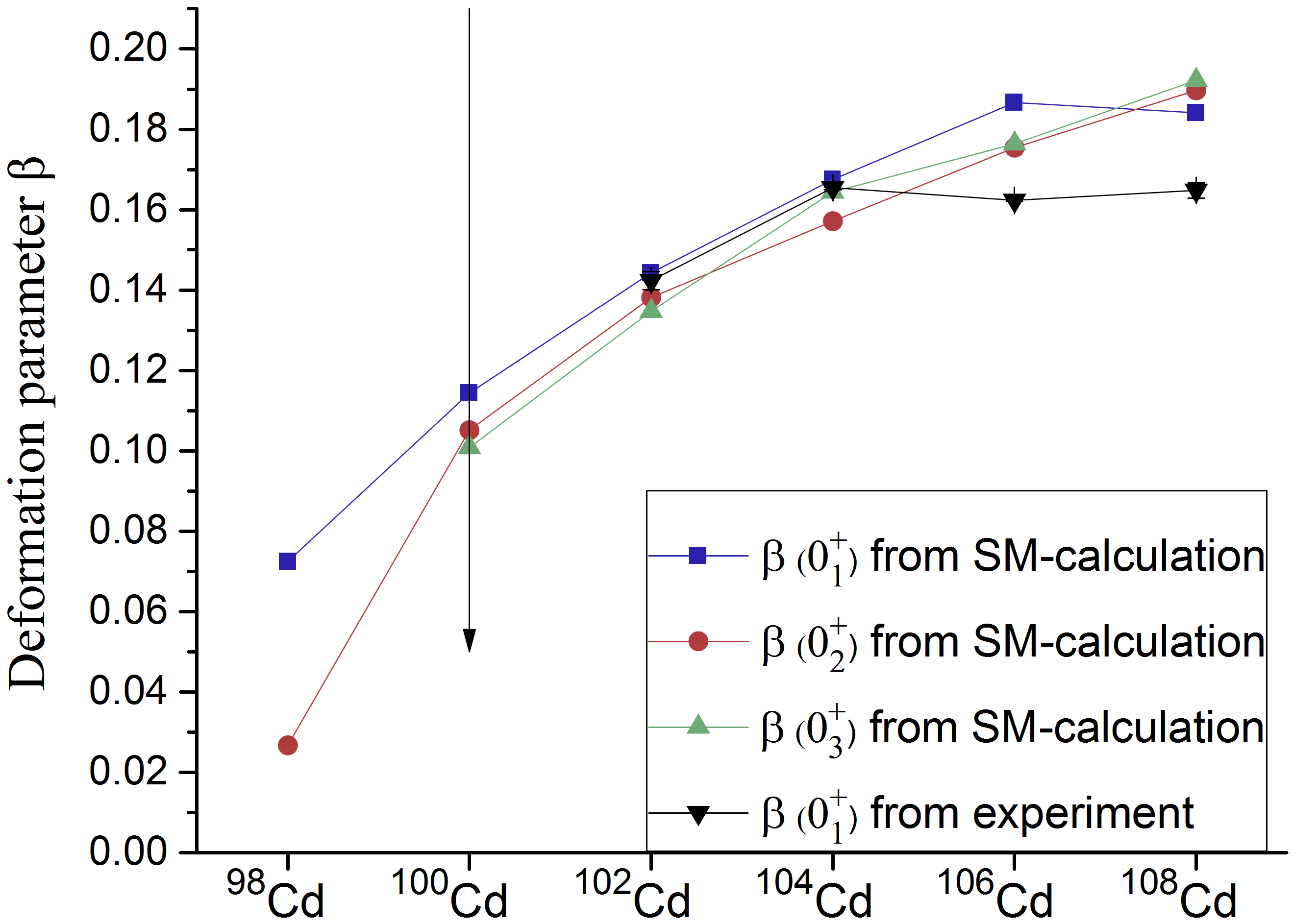}

\caption{The calculated quadrupole deformation for the $0_{1}^{+}$ (blue squares),
$0_{2}^{+}$ (red dots) and $0_{3}^{+}$ (green triangles), compared
with the experimental data (\citep{014} ($^{100}$Cd), \citep{013_Boelaert2}
($^{102,104}$Cd) and \citep{134free} ($^{106,108}$Cd)) for the
ground state only considering the $2_{1}^{+}\rightarrow0_{1}^{+}$
$E2$ transition strength, to extract the value of $\beta$. (Color
online).}

\label{Abb. deformation over N} 
\end{figure}

Considering the $B\left(E2;\,0_{1}^{+}\rightarrow2_{1}^{+}\right)$
transition as the only allowed transition starting from the $0_{1}^{+}$
state, is of course an idealized picture in terms of exciting possible
higher-lying $2^{+}$ states for the Cd nuclei discussed here. On
the other hand it is expected that the $0_{1}^{+}\rightarrow2_{1}^{+}$
transition, on average, covers $\sim97\%$ of the summed $E2$ transition
strength $\sum_{f}B\left(E2;\,0_{1}^{+}\rightarrow2_{f}^{+}\right)$
{[}\citealp{131free,082_Casten}~see page 449{]} for medium and heavy
mass nuclei. Because of the relation between the $E2$ transition
matrix elements and the quadrupole deformation, as discussed in Sections
\ref{subsec:Kumar-method} and \ref{subsec:Recent-analyses-methods:},
a behavior of the deformation describing the intrinsic properties
of the $0_{1}^{+}$ ground state, as a function of neutron number
$N$, similar to Figure \ref{2 to 0 over N} (a) is expected (see
Figure \ref{Abb. deformation over N}). 

\begin{figure*}
\centering

\includegraphics[scale=0.272]{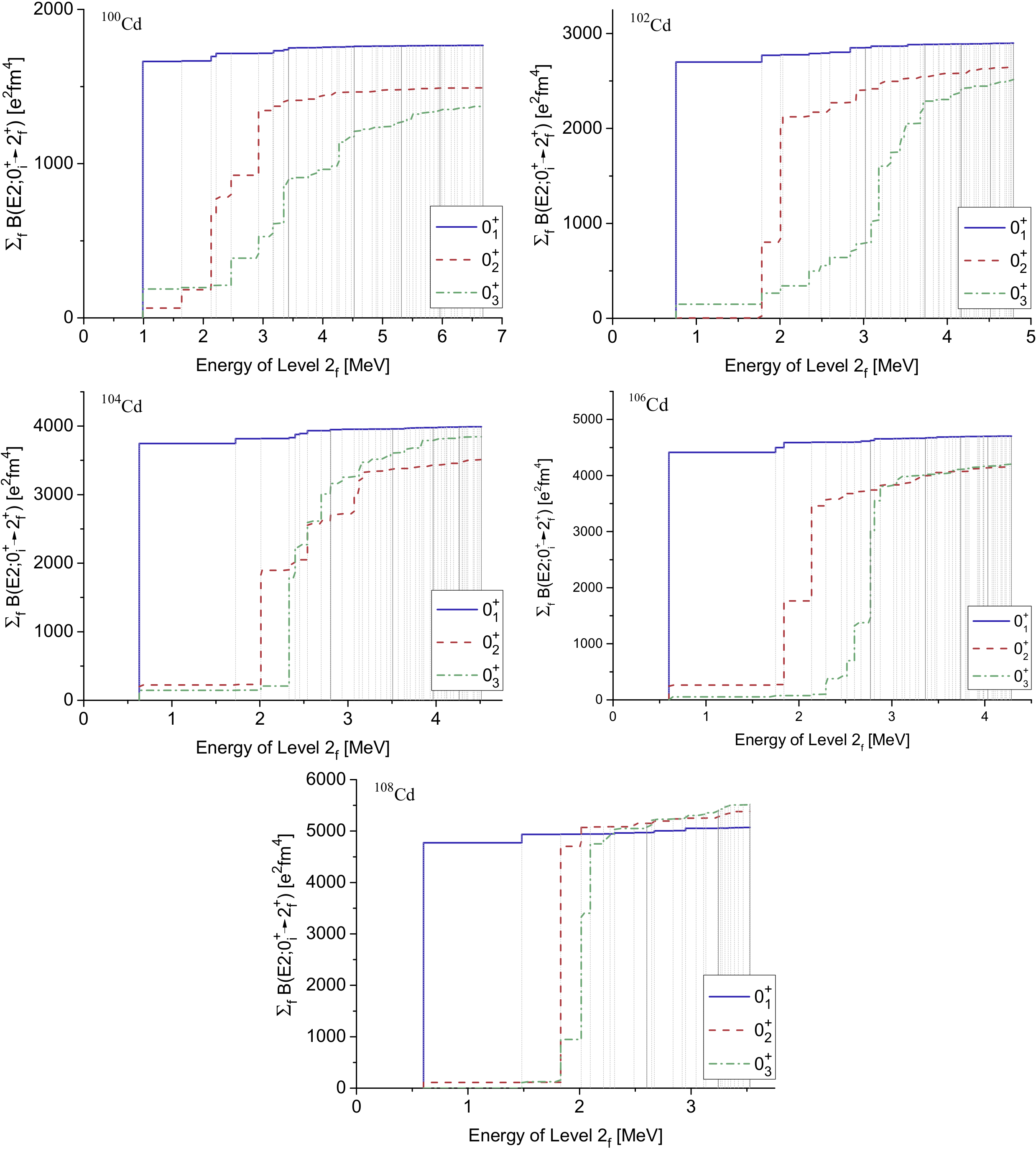}\caption{A graphical illustration of the contributions for each $0_{i}^{+}$
state (with $i=1,\,2,\,3$) to the sums of equation (\ref{eq:Kumar P2})
and (\ref{eq:2tensor_coupling}) respectively, as a function of energy
of the various $2_{f}^{+}$ states (with $f=1,...50$ for $^{100-106}$Cd
and $f=1,...30$ for $^{108}$Cd, respectively ) on the horizontal
axis. Vertical, dashed drop lines indicate the energies of the $2_{f}^{+}$
states in each figure, with a solid drop line for every full set of
ten $2^{+}$ states. (Color online).}

\label{Abb. Convergence sums} 
\end{figure*}

The present experimental status is such, that the $2_{1}^{+}\rightarrow0_{1}^{+}$
and the $2_{2}^{+}\rightarrow0_{1}^{+}$ transitions are the only
transitions from a $2^{+}$ state to the ground state with experimentally
known $B(E2)$ values in $^{106}$Cd and $^{108}$Cd. In $^{100}$Cd,
$^{102}$Cd and $^{104}$Cd only the $B(E2)$ value of the $2_{1}^{+}\rightarrow0_{1}^{+}$
transition is known, whereas for $^{100}$Cd only an upper limit is
available. For $^{98}$Cd, no data about the $2_{1}^{+}\rightarrow0_{1}^{+}$
transition strength is known. On the other hand, the present shell-model
calculations include all $E2$ transitions $0_{1}^{+}\rightarrow2_{f}^{+}$
up to $f=50$ for $^{100-106}$Cd and up to $f=30$ for $^{108}$Cd
\footnote{The $0_{1}^{+}\rightarrow2_{f}^{+}$ transitions of $^{108}$Cd have
only been calculated up to $f=30$ to reduce the computation time,
which is still sufficient to cover the full $E2$ strength, as Figure
\ref{Abb. Convergence sums} exhibits.}, which is a very large number of states, and is expected to cover
the full $E2$ strength as compared to the experimentally known data.

In Figure \ref{Abb. deformation over N} we present a comparison of
the theoretical $\beta$ values corresponding to the $0_{1}^{+}$,
$0_{2}^{+}$ and $0_{3}^{+}$ states, resulting from all calculated
transitions and the experimental $\beta$ value making use of the
only known $2_{1}^{+}\rightarrow0_{1}^{+}$ $E2$ transition \citep{013_Boelaert2,014,134free}
for the present Cd nuclei. (The shell-model based results shown here,
are derived by the method as discussed in section \ref{subsec:Recent-analyses-methods:}).
The fact that the $\beta$ values extracted for the $0_{1}^{+}$ ground
states, making use of many $E2$ transition matrix elements $\left\langle 2_{f}^{+}\left\Vert E2\right\Vert 0_{1}^{+}\right\rangle $,
resulting from the present shell-model calculations, and only one
experimental transition matrix element, are very close for all nuclei
considered, is a result of the dominant contribution of the $0_{1}^{+}\rightarrow2_{1}^{+}$
$E2$ transition strength to the total sum, as argued before. The
shell-model results exhibit the same behavior, as the sums $\sum_{f=2}^{M}B(E2;\,0_{1}^{+}\rightarrow2_{f}^{+})$
over the non-yrast transitions cover only $\sim5\%-6\%$ of the total
sum. Figure \ref{Abb. Convergence sums} presents the sums of the
transition strength for the three lowest $0_{1,2,3}^{+}$ states with
contributions from shell model $2_{f}^{+}$ states as a function of
energy of the $2^{+}$ states for the nuclei $^{100}$Cd up to $^{108}$Cd.
This figure which represents the $E2$ strength function connected
to the first three $0^{+}$ states, at the same time exhibits the
convergence of the sum in $P_{s}^{(2)}$ and $\left\langle Q^{2}\right\rangle $
respectively (equations (\ref{eq:Kumar P2}) and (\ref{eq:2tensor_coupling}))
for the $0_{1,2,3}^{+}$ states, for which transitions from states
of higher energy ($2_{f}^{+};f\gtrsim30$) are only contributing in
negligible amounts. 

The way convergence is reached for the $0_{2,3}^{+}$ levels exhibits
a most interesting behavior with increasing mass. Whereas for the
lighter isotopes ($A=100-104$) the curves are mainly characterized
by a large number of intermediate steps, an increasing concentration
into just a few states shows up in approaching $^{108}$Cd. One observes
a transition from a highly fragmented trend for the $0_{2}^{+},$
and even more so for the $0_{3}^{+}$ state moving through the nuclei
with, to a situation in $^{108}$Cd, with the $0_{2}^{+}$ state exhibiting
a similar character as the ground state. The total strength of each
sum is in line with the extracted value of $\beta$, shown in Figure
\ref{Abb. deformation over N}. What is thus interesting is the fact,
that the strength functions of the $0_{3}^{+}$ state and the lower
$0^{+}$ states are quite different even though the corresponding
$\beta$ values are very close to each other. In the latter case,
one also notices that for the $0_{3}^{+}$ state in $^{108}$Cd, the
full strength is nearly reached within 3 steps, which is different
from the lighter mass Cd nuclei. In those cases, where the full $E2$
strength concentrates in a single state, the deformation parameter
is associated to an intrinsic state. Mean field calculations of Prochniak
et al. \citep{054} point also in this direction.

At this place, we emphasize the importance of studying the strength
functions in such detail, in order to understand the underlying structure
of the states and band members. The $E2$ strength functions, shown
in Figure \ref{Abb. Convergence sums}, are adding important information
to the invariants calculated before using the Kumar-Cline method.
Combining the information contained in Figures \ref{Abb. deformation over N}
and \ref{Abb. Convergence sums} and later also Figure\ref{Abb.: Beta_over_bands},
deep insight into the structure of in particular the ground state
band, and to a lesser extent for the higher bands, is obtained.

The values of $\beta$ are compared in Table \ref{Tabelle: beta kumar vs cline}.
The table displays, that the values of both deformation analysis methods
differ only in the order of three places after the decimal point,
which illustrates the excellent congruency of both approaches. 
\begin{table*}
\centering

\begin{ruledtabular} %
\begin{tabular}{crlrlrl}
\multicolumn{1}{c}{nucleus} & \multicolumn{2}{c}{$\beta$ value for $0_{1}^{+}$} & \multicolumn{2}{c}{$\beta$ value for $0_{2}^{+}$} & \multicolumn{2}{c}{$\beta$ value for $0_{3}^{+}$}\tabularnewline
{}  & sect. \ref{subsec:Kumar-method}  & sect. \ref{subsec:Recent-analyses-methods:}$^{a}$  & sect. \ref{subsec:Kumar-method}  & sect. \ref{subsec:Recent-analyses-methods:}$^{a}$  & sect. \ref{subsec:Kumar-method}  & sect. \ref{subsec:Recent-analyses-methods:}$^{a}$\tabularnewline
\hline 
\noalign{\vskip 1mm}$^{98}$Cd & 0.0723  & 0.0725  & -  & 0.0267  & -  & -\tabularnewline
$^{100}$Cd  & 0.1138  & 0.1144  & 0.1047  & 0.1051  & 0.1004  & 0.1008\tabularnewline
$^{102}$Cd & 0.1435  & 0.1446  & 0.1371  & 0.1381  & 0.1338  & 0.1347\tabularnewline
$^{104}$Cd  & 0.1657  & 0.1675  & 0.1557  & 0.1571  & 0.1628  & 0.1644\tabularnewline
$^{106}$Cd  & 0.1842  & 0.1866  & 0.1734  & 0.1754  & 0.1744  & 0.1764\tabularnewline
$^{108}$Cd  & 0.1818  & 0.1841  & 0.1872  & 0.1890 & 0.1895  & 0.1922\tabularnewline
\end{tabular}\end{ruledtabular}

\caption{Comparison of $\beta$ values derived from the methods of section
\ref{subsec:Kumar-method} and \ref{subsec:Recent-analyses-methods:}
(see Figure \ref{Abb. deformation over N}). $^{a}$We emphasize,
that the results of section \ref{subsec:Recent-analyses-methods:}
are the root-mean square values, i. e. $\sqrt{\left\langle 0_{i}^{+}\left|\beta^{2}\right|0_{i}^{+}\right\rangle }$.}

\label{Tabelle: beta kumar vs cline} 

\medskip{}
\end{table*}

We notice, that no $\beta$ value could be derived for the $0_{2}^{+}$
state in $^{98}$Cd using the method of section \ref{subsec:Kumar-method}
in this LSSM model space. The nucleus $^{98}$Cd has a closed neutron
shell and two proton holes within the $1g_{9/2}$ and $2p_{1/2}$
orbitals, which can couple to form excited states within the used
model space. The $^{98}$Cd $E2$ transition scheme therefore consists
mainly of an yrast band, characterized by seniority $\nu=2$ excitations,
with the two proton holes placed in the $1g_{9/2}$ orbital for each
state of the band. The only other possible particle distribution producing
positive parity states is, when the two proton holes are placed in
the $2p_{1/2}$ orbital, making up for a $0_{2}^{+}$state. Therefore
besides the yrast-band the $0_{2}^{+}\rightarrow2_{1}^{+}$ transition
is the only possible $E2$ transition. In comparison the $0_{1}^{+}\rightarrow2_{1}^{+}$
$B(E2)$ value is approximately seven times stronger than the $0_{2}^{+}\rightarrow2_{1}^{+}$
$B(E2)$ value. This leads to a ''lack'' in $E2$ transition strength
when calculating the $P_{s}^{\left(2\right)}$ invariant for the $s=0_{2}^{+}$
state, relative to the $P_{s}^{\left(3\right)}$ invariant, which
involves, in addition, the quadrupole moment of the $2_{1}^{+}$ intermediate
state, with a three times stronger diagonal $E2$ matrix element as
compared to the $0_{2}^{+}\rightarrow2_{1}^{+}$ transition $E2$
matrix element. This results, according to equation (\ref{eq:Kumar cos3Gamma}),
into a situation with $P_{s}^{\left(3\right)}/\bigl(P_{s}^{\left(2\right)}\bigr)^{\frac{3}{2}}>1$
and, consequently, $\gamma$ cannot be calculated for the $0_{2}^{+}$which,
in turn, is necessary for the derivation of $\beta$.

\begin{figure}[t]
\includegraphics[scale=0.129]{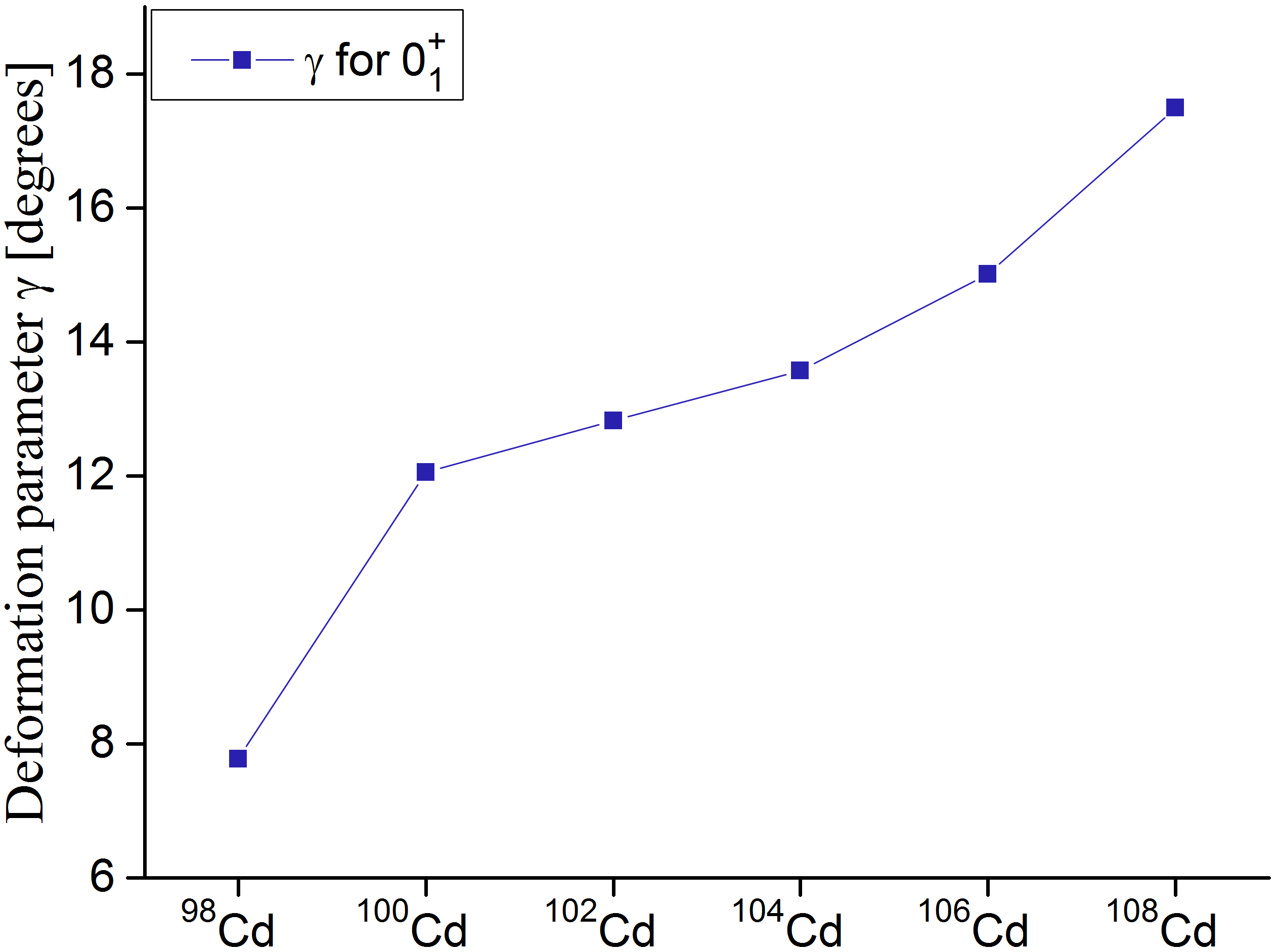}\caption{The deformation parameter $\gamma$ extracted for the different isotopes.
(Color online).}

\label{Abb. gamma deformation over N} 
\end{figure}

Using the concept that most nuclei exhibit some softness and, consequently,
exhibit a tendency for deformation (be it static as for strongly deformed
nuclei, or, in a dynamic way for soft nuclei in transitional regions
and near to closed shells), any excited state is prone to be described
using collective modes of motion (rotation, shape oscillations), implying
that a nuclear shape is not a net ``observable\textquotedblright .
Although the magnitude of the overall nuclear deformation $\beta$
remains a well defined variable, the uncertainty in the nuclear shape
$\gamma$ (which can be quantified by calculating the variance, defined
as $\sigma(\left\langle Q^{3}\right\rangle )\equiv\sqrt{\left\langle Q^{6}\right\rangle -(\left\langle Q^{3}\right\rangle )^{2}}$
(see also references \citep{057_Cline,067_NPA_766,070})) is in general
increasing with the increasing nuclear spin. Thus we do not consider
$\gamma$ values of other states than $0_{1}^{+}$, though they could
in principle be calculated from the shell-model results. In Figure
\ref{Abb. gamma deformation over N} the deformation parameter $\gamma$
is shown for the ground state as a function of increasing mass number
$A(N)$. Together with the information of Figure \ref{Abb. deformation over N},
one notices that $\gamma$ is increasing with $N$, starting at a
slightly prolate deformation in $^{98}$Cd of $\gamma=7.8\text{\textdegree}$.
In $^{108}$Cd the $\gamma$ value reaches a value $17.5\text{\textdegree}$
and thus is approaching the maximal triaxiality value of 30\textdegree ,
a value that separates the regions of prolate and oblate deformation.
Therefore the overall shape of the ground states is to be considered
as prolate with a growing triaxiality as the number of valence neutrons
increases when starting to fill the $N=50-82$ shell. This result
may support the picture of a $\gamma$ soft structure in the light
Cd isotopes, opposite to the traditional, vibrational picture. 

Besides LSSM calculations, only few studies in the context of (beyond)
mean-field studies have been performed. For the Cd nuclei mean-field
calculations have been carried out by Prochniak et.al. \citep{054}
and Rodriguez and Egido \citep{055}. In particular, in reference
\citep{054}, total energy surfaces have been calculated for the $^{106-116}$Cd
nuclei. In the mass span of $A=$106 to $A=$108, a slight prolate
minimum appears at a value of $\beta\sim$ 0.15-0.2, a value quite
close to the results of our shell-model based deformation analysis.
The correspondence becomes even more pronounced when comparing the
spectroscopic quadrupole moments derived from the mean-field results
of ref. \citep{054} for the $2_{1,2,3}^{+}$ states. The resulting
negative values for the $2_{1,3}^{+}$ and a positive value for the
$2_{2}^{+}$ state, as well as the magnitudes are comparable to our
SM results. This shows that both descriptions are very much consistent.

\subsection{Deformation correlated to spin }

\begin{figure*}
\centering

\includegraphics[scale=0.2745]{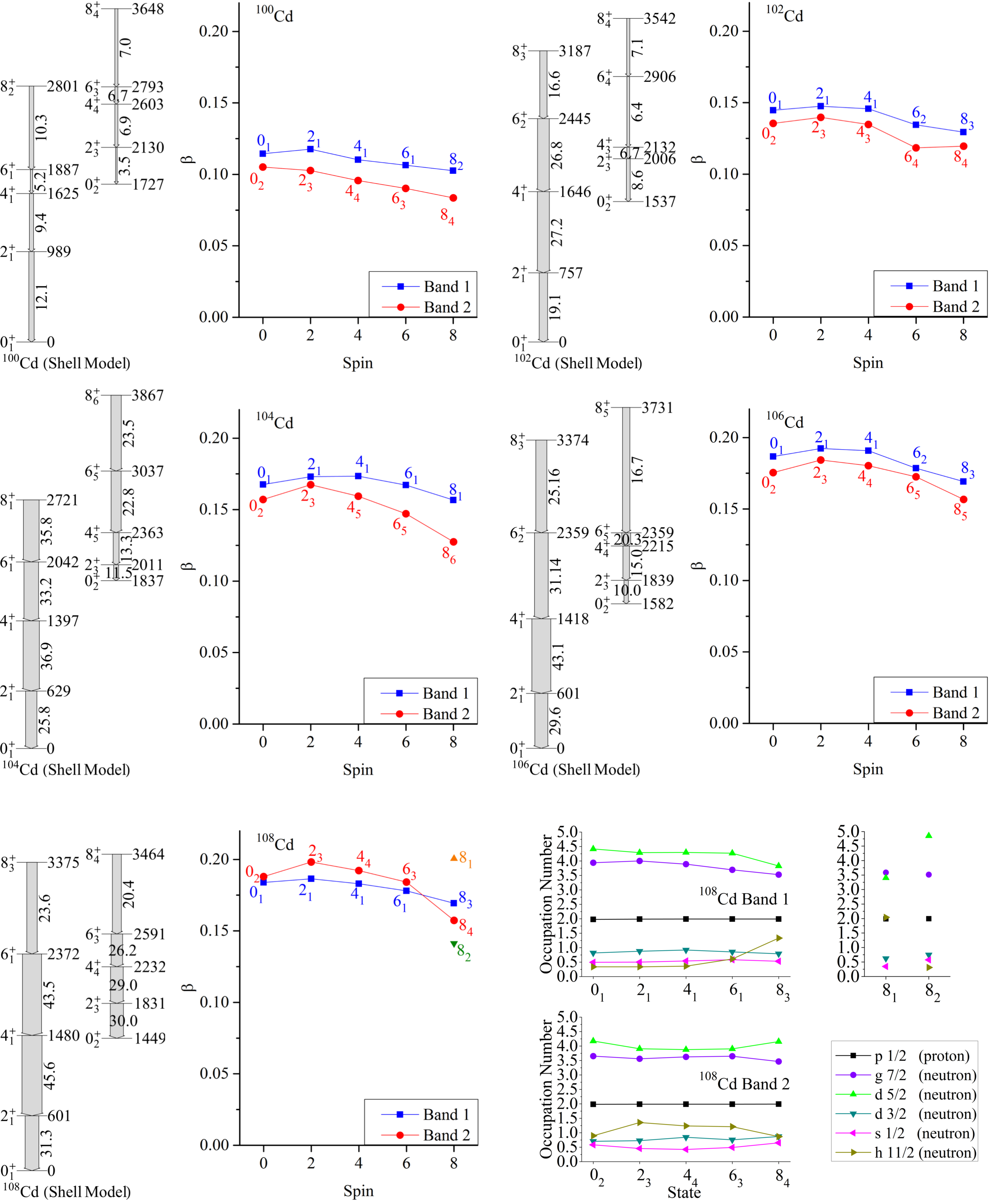}

\caption{Partial level schemes with energies in {[}keV{]} and transition strengths
in {[}W.u.{]} grouped in band structures for $^{100}$Cd -$^{108}$Cd.
On the right-hand side of each figure the $\beta$ deformation as
a function of increasing spin is displayed, whereas blue squares denote
values for members of band 1, connected to the ground state and red
dots denote values for members of band 2 ending up in the $0_{2}^{+}$
state. In the bottom right part of the figure, the specific orbital
occupation of the two bands in $^{108}$Cd is presented as a function
of increasing spin. The deformation and orbital occupation of the
$8_{1}^{+}$ and $8_{2}^{+}$ state in $^{108}$Cd are given as additional
examples. (Color online).}

\label{Abb.: Beta_over_bands} 
\end{figure*}

We have also studied the deformation for various excited states and
the way this indicates the presence of correlations with the nuclear
spin as one moves up in excitation energy. This study has been carried
out by examining the deformation corresponding to the shell-model
wave functions as a function of spin, up to $J=8$, thereby following
the deformation through a set of states defined as a band, as derived
from the shell-model. In Figure \ref{Abb.: Beta_over_bands} such
bands are shown for the nuclei $^{100}$Cd up to $^{108}$Cd , in
all of which two bands could be identified. The way in which we define
a given band is by studying where the strongest $E2$ matrix elements
appear, when moving up in excitation energy through the spin sequences
of $0^{+}$, $2^{+}$, $4^{+}$, $6^{+}$ in steps of $\triangle J=2$
until the $8^{+}$ states and starting at the $0_{1}^{+}$ and $0_{2}^{+}$
states, respectively. In the resulting band structure any possible
deexcitation from any state of band 1 will most probably end in the
ground state, whereas a deexcitation of any state of band 2 will most
probably end up into $0_{2}^{+}$ respectively ($M1$ branches are
excluded in this study). A detailed study of the spectroscopy of these
Cd nuclei, including comparisons with the large set of experimental
data available at present, on issues such as low-lying states, high-spin
bands, detailed spectroscopic information on moments, will form the
content of a forthcoming paper.

In Figure \ref{Abb.: Beta_over_bands} results of the deformation
parameter $\beta$ are given for the members of bands 1 and 2 in $^{100}$Cd
- $^{108}$Cd, as well as for the $8_{1}^{+}$ and $8_{2}^{+}$ states
in $^{108}$Cd. When analyzing the evolution of state deformation
for band 1 of all considered nuclei and comparing the deformation
curves, a number of characteristic similarities in all the nuclei
show up. Starting at the $0_{1}^{+}$ level one notices that the deformation
is slightly increasing when moving into the $2^{+}$ level for each
band. Then a decrease in deformation follows, when going from $2^{+}$
to $4^{+}$, except for $^{104}$Cd where one again observes a slight
increase but weaker than before, even close to stagnation in deformation
as compared to the former step. For higher spins, beyond $4^{+}$
and from $^{102}$Cd on, the decrease in deformation is enhanced until
$J^{\pi}=8^{+}$. Such a trend could be explained qualitatively by
looking at the sources of $E2$ transition strength not only but especially
between the band members. As the number of valence neutron pairs increases,
going from $^{100}$Cd to $^{108}$Cd, the contribution of configurations
allowing seniority changing transitions also increases, which affects
stronger transitions between the low spin states, i.e. $0^{+},2^{+},4^{+}$.
In addition, even small admixtures of neutron stretched $E2$ transitions
($\triangle j=\triangle l=2$), which in this model space are only
of the type $2d_{5/2}\longleftrightarrow3s_{1/2}$ will increase the
total $E2$ strength and consequently the deformation $\beta$. As
the lowest seniority configurations, for which one or two neutrons
are in the $3s_{1/2}$ orbital, and the other valence neutrons occupying
the rest of the neutron orbitals, can only produce low spins (for
example the coupling of $2d_{5/2}\otimes3s_{1/2}$ is limited to $2^{+}$
and $3^{+}$, while the members of the $1g_{7/2}\otimes3s_{1/2}$
multiplet have spins $3^{+}$ and $4^{+}$), this additional $E2$
strength is concentrated between the low spin states up to $4^{+}$.

Generally, band 2 exhibits characteristics similar to the behavior
of band 1, where an increase in deformation is observed in the step
$J=0\rightarrow J=2$, except for $^{100}$Cd. The nuclei $^{104}$Cd,
$^{106}$Cd and $^{108}$Cd are very good examples for this behavior
with $\Delta\beta\approx0.01$ in the step when going from $0_{2}^{+}$
to the $2^{+}$ member of band 2. For the steps from $2^{+}$ until
$8^{+}$, similar to the case with band 1, an overall decrease in
deformation is observed. In $^{102}$Cd for $J=4\rightarrow J=6$
the decrease in deformation escalates, resulting in a less smooth
deformation curve compared to band 1. With these similarities, it
can be stated that the maximum in deformation is located at the $2^{+}$
band member (except for band 1 in $^{104}$Cd where the $4^{+}$ is
at the maximum value, and band 2 in $^{100}$Cd where deformation
is decreasing from $0_{2}^{+}$ onwards).

Besides these similarities in the deformation characteristics of the
examined nuclei, the overall deformation strength of band 2 in $^{108}$Cd
exceeds the deformation of band 1, which exhibits a change in the
nuclear structure, as this behavior is not present in the lighter
nuclei. These changes in deformation characteristics may well be associated
with the particular occupation numbers of the various orbitals in
$^{108}$Cd. 

At the bottom right part of Figure \ref{Abb.: Beta_over_bands} the
related orbital occupation numbers of band 1 and 2 of $^{108}$Cd
are displayed. The orbital occupation numbers shown in Figure \ref{Abb.: Beta_over_bands}
have been calculated using the shell-model wave functions resulting
from the present calculations. A detailed examination of the orbital
filling shown in Figure \ref{Abb.: Beta_over_bands} indicates, that
the increasing occupation of the $1h_{11/2}$ orbital originates mainly
from a depletion of the $2d_{5/2}$ orbital, and this for all spin
values in the two bands. This is especially obvious, when going from
the $6_{1}^{+}$ to the $8_{3}^{+}$ state in band 1. The occupation
number of the $2d_{5/2}$ is lowered by an amount of $\Delta\approx-0.5$
and the occupation of the $1h_{11/2}$ is raised by $\Delta\approx0.7$.
It can nearly be considered as a neutron moving from the $2d_{5/2}$
to the $1h_{11/2}$ orbital. Also minor contributions from other orbitals
can be observed, e.g. resulting from excitations from the $1g_{7/2}$,
$2d_{3/2}$ and $3s_{1/2}$ orbitals to the neutron $1h_{11/2}$ orbital.

When comparing the deformation curve of band 2 in $^{108}$Cd to the
curve showing the occupation of the $1h_{11/2}$ orbital in band 2,
it is obvious that neutrons in the $1h_{11/2}$ orbital are affecting
the deformation. In those states, where the $1h_{11/2}$ orbital occupation
is increasing, the corresponding deformation of band 2 is enhanced.
Although both curves show a similar shape, the relation between the
deformation of band 2 and the occupation of the $1h_{11/2}$ orbital
is not directly proportional.

In band 1 of $^{108}$Cd the deformation strength is also effected
by the filling of the $1h_{11/2}$ orbital. This is not as obvious
from the shape of the deformation curve, but the influence caused
by the $1h_{11/2}$ can be illustrated by comparing the decrease in
deformation over the steps from $4^{+}$ to $8^{+}$ for the various
nuclei presented. In $^{100}$Cd, $^{102}$Cd, $^{104}$Cd and $^{106}$Cd
these drops in deformation amount to $\Delta\beta(4^{+}\rightarrow8^{+})\equiv\beta(4^{+})-\beta(8^{+})=0.008$,
0.016, 0.017 and 0.022 respectively, thus exhibiting an increasing
drop of the deformation. In the case of $^{108}$Cd on the other hand,
this trend is hindered by the drop amount of $\Delta\beta(4^{+}\rightarrow8^{+})=0.014$.

Two further examples which confirm the importance of the $1h_{11/2}$
orbital on the deformation for the high-spin $8_{1}^{+}$ and $8_{2}^{+}$
states, resulting from the shell-model calculations, are highlighted.
The calculated excitation energies are 2755 keV and 3136 keV, respectively.
The $8_{1}^{+}$ state exhibits the largest deformation $\beta=0.20$
found in the examined states in $^{98}$Cd - $^{108}$Cd, whereas
the $8_{2}^{+}$ shows the lowest deformation of the examined states
in $^{108}$Cd with $\beta=0.14$. Comparing the orbital occupation
numbers of these two states, one notices that the major differences
result from the filling of the $2d_{5/2}$ and the $1h_{11/2}$ orbitals.
In the $8_{1}^{+}$state the $2d_{5/2}$ and $1h_{11/2}$ orbitals
are filled with $\approx3.4$ and $\approx2$ neutrons respectively,
whereas in the $8_{2}^{+}$ state the $2d_{5/2}$ orbital contains
$\approx4.8$ and the $1h_{11/2}$ orbital $\approx0.3$ neutrons.
This looks like a neutron $2p-2h$ excitation from the $2d_{5/2}$
into the $1h_{11/2}$ orbital, causing a clear difference in the resulting
deformation when comparing the $\beta$ values for the $8_{1}^{+}$
state with the $8_{2}^{+}$ state (notice the $\nu h_{11/2}$ occupation
number in Figure \ref{Abb.: Beta_over_bands} ). It turns out that
the shell-model $8_{2}^{+}$ state exhibits a proton $1g_{9/2}^{-2}$
character, whereas experimental studies of $^{108}$Cd show that the
$8_{1}^{+}$ state at 3111 keV is of proton character and the $8_{3}^{+}$
state at 3862 keV is of neutron $1h_{11/2}^{2}$ character \citep{130free}.
In the same study it was found, that the $1h_{11/2}$ orbital is playing
a dominant role in the low and high spin structure of $^{108}$Cd
with a shape driving effect.

\section{Summary and Conclusions}

In the present paper, we have studied how to extract the changing
quadrupole collectivity and its associated deformation using input
from large-scale shell-model calculations (LSSM) of the light $^{98-108}$Cd
nuclei. The effective interaction used in the present study succeeds
rather well in describing the overall variations in the excitation
energy of the low-spin states in the Cd nuclei (spanning the $A=98$
to $A=108$ region) \citep{062}, in particular the excitation energy
for the $2_{1}^{+}$ state, as well a the increasing trend in the
$B(E2;2_{1}^{+}\rightarrow0{}_{1}^{+})$ value. This is an indication
of a well-balanced description in which both the monopole and quadrupole
components of the force and the induced polarization for the protons
and neutrons, as obtained here, form a well-balanced system.

We emphasize that large-scale shell-model calculations (LSSM) can
and have been carried out within a symmetry-dictated truncation basis
(whenever the single-particle states spanning the model space are
prone to such truncations). More in particular, both the quasi-SU(3)
\citep{162Heyde,163Heyde} as well as the pseudo-SU(3) \citep{164Heyde,165Heyde}
variants of Elliott\textquoteright s SU(3) model \citep{166Heyde,167Heyde}
provide such options. Applications have been carried out in this spirit
for the $N=$20, 40, 48,.. and even heavier nuclei \citep{005,161free,170Heyde,171Heyde} 

The sdg (excluding the $1h_{11/2}$ unnatural parity orbital) neutron
shell is apt to such an approach. Such calculations may allow to obtain
a deeper insight in the results from a LSSM study about how quadrupole
collectivity is developing in the Cd isotopes, and this as a function
of increasing neutron number.

We have been able, using the calculated $E2$ reduced matrix elements
starting from the shell-model wave functions, defined within the laboratory
framework, to derive both the quadratic and cubic quadrupole invariants.
Using the fact that these invariants contain information about the
nuclear deformation, defined within a frame of the principal axis
of the nucleus, we are able to quantify the changing quadrupole deformation
parameters $\beta$ and $\gamma$ in an almost model-independent way.

The first part (section \ref{sec:Shape-invariants}) presented a rather
detailed comparison of two methods that have been (and are) used to
derive the intrinsic deformation characteristics, extracted from the
quadrupole invariants. Here, the aim was to discuss the slight and
often subtle differences between both approaches that lead to apparent
slightly different expressions in the major papers and how, precisely,
the intrinsic shape parametrization is extracted. This method is called
in most papers \textquotedbl{}model independent\textquotedbl{}, an
issue which we discuss in some detail pointing out how, even though
very general, generic features related to quadrupole deformed and
intrinsic shape can be described and parametrized.

The comparison between the calculated $B(E2;2_{1}^{+}\rightarrow0_{1}^{+})$
reduced transition probabilities and the known data for the Cd nuclei
with $50\le N\le60$, showed that the LSSM reproduces well the collectivity.
For the presented variation of the extracted deformation, characterized
by $\beta$ starting at $^{98}$Cd, one observes an expected, rapid
increase, followed by saturation when reaching the $^{106,108}$Cd
isotopes. Moreover, inspecting the $\beta$ value extracted from the
set of levels that are strongly connected to the $0_{1}^{+}$ and
$0_{2}^{+}$ ``band-head levels'', through a sequence of particularly
strong $E2$ transition matrix elements, one observes two separate
bands characterized by $\beta$ values, that are strongly \textquotedbl{}correlated\textquotedbl{}
as a function of increasing angular momentum up to spin $8^{+}$ states.
We also notice that within each of those bands, there is not a particular
increase in the value of $\beta$ as a function of increasing angular
momentum, which would be expected, even for close to harmonic vibrational
collective quadrupole motion, at least up to mass number $A=102$.
In the $^{106,108}$Cd nuclei, an initial slight increase is observed,
followed by a decrease with increasing spin up to $8^{+}$. In $^{108}$Cd,
a specific upslope in $\beta$ shows up and it turns out that this
effect can be associated with a redistribution of neutrons from the
$2d_{5/2}$ into the $1h_{11/2}$ shell-model orbital (the latter
orbital is characterized by a larger value of $\left\langle \left|r^{2}\right|\right\rangle $
when using harmonic oscillator radial wave functions).

An interesting conclusion from the present LSSM calculations is the
observation that the $0_{2}^{+}$ and $0_{3}^{+}$ as well as associated
band members exhibit similar $\beta$ values as the ones obtained
for the $0_{1}^{+}$ ground state. Different values would be expected
from a purely collective vibrational model approach to describe the
quadrupole collective characteristics of these light Cd nuclei. We
do not exclude that part of this may well be due to the fact that
the model space does not contain proton $n$p - $n$h excitations
across the $Z=50$ closed shell (even though such correlations are
implicitly included by the use of proton as well as neutron effective
charges). Consequently, explicit breaking of the proton $Z=50$ shell
is not incorporated in a direct way (we refer to \citep{049} for
an extensive study of so-called intruder states as well as to a recent
focus issue on shape coexistence \citep{151free_shape_coex_issue}).
This is an issue to be explored in more detail, using more specific
shell-model truncation schemes and model spaces which include $n$p
- $n$h proton excitation to study Cd isotopes.

The present paper has concentrated on the deformation characteristics
for the lighter Cd nuclei. It is to be understood that a more detailed
comparison of the extensive spectroscopic information for the set
of Cd isotopes with mass number ranging from $A=98$ up to $A=108$
will follow. Thereby, both the low-spin energy spectra as well as
the high-spin structure is studied. Moreover, when known, electromagnetic
moments (electric quadrupole and magnetic dipole moments) will be
compared with the present LSSM calculations as carried out at present. 

We note that shortly before the submission of this work, new experimental
results on $B(E2)$ values and g-factors have become available (see
\citep{158Blazhev02_not_listed}) but have not been addressed in our
current paper. The reasons are (i) the fact that the $B(E2)$ values
in table III of \citep{158Blazhev02_not_listed} are inconsistent
with previously published $^{106}$Cd values, and, (ii) these $B(E2)$
values are inconsistent with the Cd systematics for the light Cd isotopes.
\begin{acknowledgments}
Financial support from the Interuniversity Attraction Poles Program
of the Belgian State-Federal Office for Scientific and Cultural Affairs
(IAP Grant P7/12) is acknowledged. We thank N. A. Smirnova for providing
the effective shell-model interaction as well as intensive discussions
in preparing the final version of this paper. We are grateful to J.
L. Wood for extensive discussions during the course of this work,
as well as for careful reading of the manuscript. 
\end{acknowledgments}

\bibliographystyle{apsrev}
\bibliography{Refs}

\begin{thebibliography}{122}
\expandafter\ifx\csname natexlab\endcsname\relax\def\natexlab#1{#1}\fi
\expandafter\ifx\csname bibnamefont\endcsname\relax
  \def\bibnamefont#1{#1}\fi
\expandafter\ifx\csname bibfnamefont\endcsname\relax
  \def\bibfnamefont#1{#1}\fi
\expandafter\ifx\csname citenamefont\endcsname\relax
  \def\citenamefont#1{#1}\fi
\expandafter\ifx\csname url\endcsname\relax
  \def\url#1{\texttt{#1}}\fi
\expandafter\ifx\csname urlprefix\endcsname\relax\def\urlprefix{URL }\fi
\providecommand{\bibinfo}[2]{#2}
\providecommand{\eprint}[2][]{\url{#2}}

\bibitem[{\citenamefont{Bohr et~al.}(1958)\citenamefont{Bohr, Mottelson, and
  Pines}}]{001}
\bibinfo{author}{\bibfnamefont{A.}~\bibnamefont{Bohr}},
  \bibinfo{author}{\bibfnamefont{B.}~\bibnamefont{Mottelson}},
  \bibnamefont{and} \bibinfo{author}{\bibfnamefont{D.}~\bibnamefont{Pines}},
  \bibinfo{journal}{Phys. Rev.} \textbf{\bibinfo{volume}{110}},
  \bibinfo{pages}{936} (\bibinfo{year}{1958}).

\bibitem[{\citenamefont{Ring and Schuck}(1980)}]{002}
\bibinfo{author}{\bibfnamefont{P.}~\bibnamefont{Ring}} \bibnamefont{and}
  \bibinfo{author}{\bibfnamefont{P.}~\bibnamefont{Schuck}},
  \emph{\bibinfo{title}{The Nuclear Many-Body problem}}
  (\bibinfo{publisher}{Springer, Berlin-Heidelberg}, \bibinfo{year}{1980}).

\bibitem[{\citenamefont{Rowe and Wood}(2010)}]{003}
\bibinfo{author}{\bibfnamefont{D.~J.} \bibnamefont{Rowe}} \bibnamefont{and}
  \bibinfo{author}{\bibfnamefont{J.~L.} \bibnamefont{Wood}},
  \emph{\bibinfo{title}{Fundamentals of Nuclear Models; Foundational Models}}
  (\bibinfo{publisher}{World Scientific Publishing}, \bibinfo{year}{2010}).

\bibitem[{\citenamefont{Caurier and Nowacki}(1999)}]{004}
\bibinfo{author}{\bibfnamefont{E.}~\bibnamefont{Caurier}} \bibnamefont{and}
  \bibinfo{author}{\bibfnamefont{F.}~\bibnamefont{Nowacki}},
  \bibinfo{journal}{Act. Phys. Pol. B} \textbf{\bibinfo{volume}{30}},
  \bibinfo{pages}{705} (\bibinfo{year}{1999}).

\bibitem[{\citenamefont{Caurier et~al.}(2005)\citenamefont{Caurier,
  Martinez-Pinedo, Nowacki, Poves, and Zuker}}]{005}
\bibinfo{author}{\bibfnamefont{E.}~\bibnamefont{Caurier}},
  \bibinfo{author}{\bibfnamefont{G.}~\bibnamefont{Martinez-Pinedo}},
  \bibinfo{author}{\bibfnamefont{F.}~\bibnamefont{Nowacki}},
  \bibinfo{author}{\bibfnamefont{A.}~\bibnamefont{Poves}}, \bibnamefont{and}
  \bibinfo{author}{\bibfnamefont{A.~P.} \bibnamefont{Zuker}},
  \bibinfo{journal}{Rev. Mod. Phys.} \textbf{\bibinfo{volume}{77}},
  \bibinfo{pages}{427} (\bibinfo{year}{2005}).

\bibitem[{\citenamefont{G\'{o}rska et~al.}(1994)\citenamefont{G\'{o}rska,
  Schubart, Grawe, Fitzgerald, Fossan, Heese, Maier, Rejmund, Spohr, and
  Rzaca-Urban}}]{006}
\bibinfo{author}{\bibfnamefont{M.}~\bibnamefont{G\'{o}rska}},
  \bibinfo{author}{\bibfnamefont{R.}~\bibnamefont{Schubart}},
  \bibinfo{author}{\bibfnamefont{H.}~\bibnamefont{Grawe}},
  \bibinfo{author}{\bibfnamefont{J.~B.} \bibnamefont{Fitzgerald}},
  \bibinfo{author}{\bibfnamefont{D.~B.} \bibnamefont{Fossan}},
  \bibinfo{author}{\bibfnamefont{J.}~\bibnamefont{Heese}},
  \bibinfo{author}{\bibfnamefont{K.~H.} \bibnamefont{Maier}},
  \bibinfo{author}{\bibfnamefont{M.}~\bibnamefont{Rejmund}},
  \bibinfo{author}{\bibfnamefont{K.}~\bibnamefont{Spohr}}, \bibnamefont{and}
  \bibinfo{author}{\bibfnamefont{T.}~\bibnamefont{Rzaca-Urban}},
  \bibinfo{journal}{Z. Phys. A} \textbf{\bibinfo{volume}{350}},
  \bibinfo{pages}{181} (\bibinfo{year}{1994}).

\bibitem[{\citenamefont{G\'{o}rska et~al.}(1997)\citenamefont{G\'{o}rska,
  Lipoglav\v{s}ek, Grawe, Nyberg, Atac., Axelsson, Bark, Blomqvist,
  Cederk\"{a}ll, Cederwall et~al.}}]{007}
\bibinfo{author}{\bibfnamefont{M.}~\bibnamefont{G\'{o}rska}},
  \bibinfo{author}{\bibfnamefont{M.}~\bibnamefont{Lipoglav\v{s}ek}},
  \bibinfo{author}{\bibfnamefont{H.}~\bibnamefont{Grawe}},
  \bibinfo{author}{\bibfnamefont{J.}~\bibnamefont{Nyberg}},
  \bibinfo{author}{\bibfnamefont{A.}~\bibnamefont{Atac.}},
  \bibinfo{author}{\bibfnamefont{A.}~\bibnamefont{Axelsson}},
  \bibinfo{author}{\bibfnamefont{R.}~\bibnamefont{Bark}},
  \bibinfo{author}{\bibfnamefont{J.}~\bibnamefont{Blomqvist}},
  \bibinfo{author}{\bibfnamefont{J.}~\bibnamefont{Cederk\"{a}ll}},
  \bibinfo{author}{\bibfnamefont{B.}~\bibnamefont{Cederwall}},
  \bibnamefont{et~al.}, \bibinfo{journal}{Phys. Rev. Lett.}
  \textbf{\bibinfo{volume}{79}}, \bibinfo{pages}{2415} (\bibinfo{year}{1997}).

\bibitem[{\citenamefont{Blazhev et~al.}(2004)\citenamefont{Blazhev, G\'{o}rska,
  Grawe, Nyberg, Palacz, Caurier, Dorvaux, Gadea, Nowacki, Andreoiu
  et~al.}}]{008}
\bibinfo{author}{\bibfnamefont{A.}~\bibnamefont{Blazhev}},
  \bibinfo{author}{\bibfnamefont{M.}~\bibnamefont{G\'{o}rska}},
  \bibinfo{author}{\bibfnamefont{H.}~\bibnamefont{Grawe}},
  \bibinfo{author}{\bibfnamefont{J.}~\bibnamefont{Nyberg}},
  \bibinfo{author}{\bibfnamefont{M.}~\bibnamefont{Palacz}},
  \bibinfo{author}{\bibfnamefont{E.}~\bibnamefont{Caurier}},
  \bibinfo{author}{\bibfnamefont{O.}~\bibnamefont{Dorvaux}},
  \bibinfo{author}{\bibfnamefont{A.}~\bibnamefont{Gadea}},
  \bibinfo{author}{\bibfnamefont{F.}~\bibnamefont{Nowacki}},
  \bibinfo{author}{\bibfnamefont{C.}~\bibnamefont{Andreoiu}},
  \bibnamefont{et~al.}, \bibinfo{journal}{Phys. Rev. C}
  \textbf{\bibinfo{volume}{69}}, \bibinfo{pages}{064304}
  (\bibinfo{year}{2004}).

\bibitem[{\citenamefont{Blazhev et~al.}(2010)\citenamefont{Blazhev, Braun,
  Grawe, Boutachkov, S.{\ }Nara{\ }Singh, Brock, Liu, Wadsworth, G\'{o}rska,
  Jolie et~al.}}]{009}
\bibinfo{author}{\bibfnamefont{A.}~\bibnamefont{Blazhev}},
  \bibinfo{author}{\bibfnamefont{N.}~\bibnamefont{Braun}},
  \bibinfo{author}{\bibfnamefont{H.}~\bibnamefont{Grawe}},
  \bibinfo{author}{\bibfnamefont{P.}~\bibnamefont{Boutachkov}},
  \bibinfo{author}{\bibfnamefont{B.}~\bibnamefont{S.{\ }Nara{\ }Singh}},
  \bibinfo{author}{\bibfnamefont{T.}~\bibnamefont{Brock}},
  \bibinfo{author}{\bibfnamefont{Z.}~\bibnamefont{Liu}},
  \bibinfo{author}{\bibfnamefont{R.}~\bibnamefont{Wadsworth}},
  \bibinfo{author}{\bibfnamefont{M.}~\bibnamefont{G\'{o}rska}},
  \bibinfo{author}{\bibfnamefont{J.}~\bibnamefont{Jolie}},
  \bibnamefont{et~al.}, \bibinfo{journal}{J. Phys. Conf. Series G}
  \textbf{\bibinfo{volume}{205}}, \bibinfo{pages}{012035}
  (\bibinfo{year}{2010}).

\bibitem[{\citenamefont{Clark et~al.}(2000)\citenamefont{Clark, Wilson,
  Appelbe, Carpenter, Chiara, Cromaz, Deleplanque, Devlin, Diamond, Fallon
  et~al.}}]{010}
\bibinfo{author}{\bibfnamefont{R.~M.} \bibnamefont{Clark}},
  \bibinfo{author}{\bibfnamefont{J.~N.} \bibnamefont{Wilson}},
  \bibinfo{author}{\bibfnamefont{D.}~\bibnamefont{Appelbe}},
  \bibinfo{author}{\bibfnamefont{M.~P.} \bibnamefont{Carpenter}},
  \bibinfo{author}{\bibfnamefont{C.~J.} \bibnamefont{Chiara}},
  \bibinfo{author}{\bibfnamefont{M.}~\bibnamefont{Cromaz}},
  \bibinfo{author}{\bibfnamefont{M.~A.} \bibnamefont{Deleplanque}},
  \bibinfo{author}{\bibfnamefont{M.}~\bibnamefont{Devlin}},
  \bibinfo{author}{\bibfnamefont{R.~M.} \bibnamefont{Diamond}},
  \bibinfo{author}{\bibfnamefont{P.}~\bibnamefont{Fallon}},
  \bibnamefont{et~al.}, \bibinfo{journal}{Phys. Rev. C}
  \textbf{\bibinfo{volume}{61}}, \bibinfo{pages}{044311}
  (\bibinfo{year}{2000}).

\bibitem[{\citenamefont{Lieb et~al.}(2001)\citenamefont{Lieb, Kast, Jungclaus,
  Johnstone, G.{\ }de{\ }Angelis, Fahlander, M.{\ }de{\ }Poli, Bizzeti, Dewald,
  Peusquens et~al.}}]{011}
\bibinfo{author}{\bibfnamefont{K.~P.} \bibnamefont{Lieb}},
  \bibinfo{author}{\bibfnamefont{D.}~\bibnamefont{Kast}},
  \bibinfo{author}{\bibfnamefont{A.}~\bibnamefont{Jungclaus}},
  \bibinfo{author}{\bibfnamefont{I.~P.} \bibnamefont{Johnstone}},
  \bibinfo{author}{\bibnamefont{G.{\ }de{\ }Angelis}},
  \bibinfo{author}{\bibfnamefont{C.}~\bibnamefont{Fahlander}},
  \bibinfo{author}{\bibnamefont{M.{\ }de{\ }Poli}},
  \bibinfo{author}{\bibfnamefont{P.~G.} \bibnamefont{Bizzeti}},
  \bibinfo{author}{\bibfnamefont{A.}~\bibnamefont{Dewald}},
  \bibinfo{author}{\bibfnamefont{R.}~\bibnamefont{Peusquens}},
  \bibnamefont{et~al.}, \bibinfo{journal}{Phys. Rev. C}
  \textbf{\bibinfo{volume}{63}}, \bibinfo{pages}{054304}
  (\bibinfo{year}{2001}).

\bibitem[{\citenamefont{Boelaert
  et~al.}(2007{\natexlab{a}})\citenamefont{Boelaert, Smirnova, Heyde, and
  Jolie}}]{012_Boelaert1}
\bibinfo{author}{\bibfnamefont{N.}~\bibnamefont{Boelaert}},
  \bibinfo{author}{\bibfnamefont{N.}~\bibnamefont{Smirnova}},
  \bibinfo{author}{\bibfnamefont{K.~L.~G.} \bibnamefont{Heyde}},
  \bibnamefont{and} \bibinfo{author}{\bibfnamefont{J.}~\bibnamefont{Jolie}},
  \bibinfo{journal}{Phys. Rev. C} \textbf{\bibinfo{volume}{75}},
  \bibinfo{pages}{014316} (\bibinfo{year}{2007}{\natexlab{a}}).

\bibitem[{\citenamefont{Boelaert
  et~al.}(2007{\natexlab{b}})\citenamefont{Boelaert, Dewald, Fransen, Jolie,
  Linnemann, Melon, M\"{o}ller, Smirnova, and Heyde}}]{013_Boelaert2}
\bibinfo{author}{\bibfnamefont{N.}~\bibnamefont{Boelaert}},
  \bibinfo{author}{\bibfnamefont{A.}~\bibnamefont{Dewald}},
  \bibinfo{author}{\bibfnamefont{C.}~\bibnamefont{Fransen}},
  \bibinfo{author}{\bibfnamefont{J.}~\bibnamefont{Jolie}},
  \bibinfo{author}{\bibfnamefont{A.}~\bibnamefont{Linnemann}},
  \bibinfo{author}{\bibfnamefont{B.}~\bibnamefont{Melon}},
  \bibinfo{author}{\bibfnamefont{O.}~\bibnamefont{M\"{o}ller}},
  \bibinfo{author}{\bibfnamefont{N.}~\bibnamefont{Smirnova}}, \bibnamefont{and}
  \bibinfo{author}{\bibfnamefont{K.~L.~G.} \bibnamefont{Heyde}},
  \bibinfo{journal}{Phys. Rev. C} \textbf{\bibinfo{volume}{75}},
  \bibinfo{pages}{054311} (\bibinfo{year}{2007}{\natexlab{b}}).

\bibitem[{\citenamefont{Ekstr\"{o}m et~al.}(2009)\citenamefont{Ekstr\"{o}m,
  Cederk\"{a}ll, DiJulio, Fahlander, Hjorth-Jensen, Blazhev, Bruyneel, Butler,
  Davinson, Eberth et~al.}}]{014}
\bibinfo{author}{\bibfnamefont{A.}~\bibnamefont{Ekstr\"{o}m}},
  \bibinfo{author}{\bibfnamefont{J.}~\bibnamefont{Cederk\"{a}ll}},
  \bibinfo{author}{\bibfnamefont{D.~D.} \bibnamefont{DiJulio}},
  \bibinfo{author}{\bibfnamefont{C.}~\bibnamefont{Fahlander}},
  \bibinfo{author}{\bibfnamefont{M.}~\bibnamefont{Hjorth-Jensen}},
  \bibinfo{author}{\bibfnamefont{A.}~\bibnamefont{Blazhev}},
  \bibinfo{author}{\bibfnamefont{B.}~\bibnamefont{Bruyneel}},
  \bibinfo{author}{\bibfnamefont{P.~A.} \bibnamefont{Butler}},
  \bibinfo{author}{\bibfnamefont{T.}~\bibnamefont{Davinson}},
  \bibinfo{author}{\bibfnamefont{J.}~\bibnamefont{Eberth}},
  \bibnamefont{et~al.}, \bibinfo{journal}{Phys. Rev. C}
  \textbf{\bibinfo{volume}{80}}, \bibinfo{pages}{054302}
  (\bibinfo{year}{2009}).

\bibitem[{\citenamefont{Milner et~al.}(1969)\citenamefont{Milner, McGowan,
  Stelson, Robinson, and Sayer}}]{133free}
\bibinfo{author}{\bibfnamefont{W.~T.} \bibnamefont{Milner}},
  \bibinfo{author}{\bibfnamefont{F.~K.} \bibnamefont{McGowan}},
  \bibinfo{author}{\bibfnamefont{P.~H.} \bibnamefont{Stelson}},
  \bibinfo{author}{\bibfnamefont{R.~L.} \bibnamefont{Robinson}},
  \bibnamefont{and} \bibinfo{author}{\bibfnamefont{R.~O.} \bibnamefont{Sayer}},
  \bibinfo{journal}{Nucl. Phys. A} \textbf{\bibinfo{volume}{129}},
  \bibinfo{pages}{687} (\bibinfo{year}{1969}).

\bibitem[{\citenamefont{Esat et~al.}(1976)\citenamefont{Esat, Kean, and
  Spear}}]{134free}
\bibinfo{author}{\bibfnamefont{M.~T.} \bibnamefont{Esat}},
  \bibinfo{author}{\bibfnamefont{D.~C.} \bibnamefont{Kean}}, \bibnamefont{and}
  \bibinfo{author}{\bibfnamefont{R.~H.} \bibnamefont{Spear}},
  \bibinfo{journal}{Nucl. Phys. A} \textbf{\bibinfo{volume}{274}},
  \bibinfo{pages}{237} (\bibinfo{year}{1976}).

\bibitem[{\citenamefont{de{\ }Angelis et~al.}(1999)\citenamefont{de{\ }Angelis,
  Fahlander, Vretenar, Brant, Gadea, Algora, Li, Pan, Farnea, Bazzacco
  et~al.}}]{015}
\bibinfo{author}{\bibfnamefont{G.}~\bibnamefont{de{\ }Angelis}},
  \bibinfo{author}{\bibfnamefont{C.}~\bibnamefont{Fahlander}},
  \bibinfo{author}{\bibfnamefont{D.}~\bibnamefont{Vretenar}},
  \bibinfo{author}{\bibfnamefont{S.}~\bibnamefont{Brant}},
  \bibinfo{author}{\bibfnamefont{A.}~\bibnamefont{Gadea}},
  \bibinfo{author}{\bibfnamefont{A.}~\bibnamefont{Algora}},
  \bibinfo{author}{\bibfnamefont{Y.}~\bibnamefont{Li}},
  \bibinfo{author}{\bibfnamefont{Q.}~\bibnamefont{Pan}},
  \bibinfo{author}{\bibfnamefont{E.}~\bibnamefont{Farnea}},
  \bibinfo{author}{\bibfnamefont{D.}~\bibnamefont{Bazzacco}},
  \bibnamefont{et~al.}, \bibinfo{journal}{Phys. Rev. C}
  \textbf{\bibinfo{volume}{60}}, \bibinfo{pages}{014313}
  (\bibinfo{year}{1999}).

\bibitem[{\citenamefont{M\"{u}ller et~al.}(2001)\citenamefont{M\"{u}ller,
  Jungclaus, Yordanov, Galindo, Hausmann, Kast, Lieb, Brant, Krsti\'{c},
  Vretenar et~al.}}]{016}
\bibinfo{author}{\bibfnamefont{G.~A.} \bibnamefont{M\"{u}ller}},
  \bibinfo{author}{\bibfnamefont{A.}~\bibnamefont{Jungclaus}},
  \bibinfo{author}{\bibfnamefont{O.}~\bibnamefont{Yordanov}},
  \bibinfo{author}{\bibfnamefont{E.}~\bibnamefont{Galindo}},
  \bibinfo{author}{\bibfnamefont{M.}~\bibnamefont{Hausmann}},
  \bibinfo{author}{\bibfnamefont{D.}~\bibnamefont{Kast}},
  \bibinfo{author}{\bibfnamefont{K.~P.} \bibnamefont{Lieb}},
  \bibinfo{author}{\bibfnamefont{S.}~\bibnamefont{Brant}},
  \bibinfo{author}{\bibfnamefont{V.}~\bibnamefont{Krsti\'{c}}},
  \bibinfo{author}{\bibfnamefont{D.}~\bibnamefont{Vretenar}},
  \bibnamefont{et~al.}, \bibinfo{journal}{Phys. Rev. C}
  \textbf{\bibinfo{volume}{64}}, \bibinfo{pages}{014305}
  (\bibinfo{year}{2001}).

\bibitem[{\citenamefont{Ashley et~al.}(2007)\citenamefont{Ashley, Regan,
  Andgren, McCutchan, Zamfir, Amon, Cakirli, Casten, Clark, Gelletly
  et~al.}}]{017}
\bibinfo{author}{\bibfnamefont{S.~F.} \bibnamefont{Ashley}},
  \bibinfo{author}{\bibfnamefont{P.~H.} \bibnamefont{Regan}},
  \bibinfo{author}{\bibfnamefont{K.}~\bibnamefont{Andgren}},
  \bibinfo{author}{\bibfnamefont{E.~A.} \bibnamefont{McCutchan}},
  \bibinfo{author}{\bibfnamefont{N.~V.} \bibnamefont{Zamfir}},
  \bibinfo{author}{\bibfnamefont{L.}~\bibnamefont{Amon}},
  \bibinfo{author}{\bibfnamefont{R.~B.} \bibnamefont{Cakirli}},
  \bibinfo{author}{\bibfnamefont{R.~F.} \bibnamefont{Casten}},
  \bibinfo{author}{\bibfnamefont{R.~M.} \bibnamefont{Clark}},
  \bibinfo{author}{\bibfnamefont{W.}~\bibnamefont{Gelletly}},
  \bibnamefont{et~al.}, \bibinfo{journal}{Phys. Rev. C}
  \textbf{\bibinfo{volume}{76}}, \bibinfo{pages}{064302}
  (\bibinfo{year}{2007}).

\bibitem[{\citenamefont{Datta et~al.}(2005)\citenamefont{Datta, Chattopadhyay,
  Bhattacharya, Ghosh, Goswami, Pal, Sarkar, Jain, Joshi, Bhowmik
  et~al.}}]{018}
\bibinfo{author}{\bibfnamefont{P.}~\bibnamefont{Datta}},
  \bibinfo{author}{\bibfnamefont{S.}~\bibnamefont{Chattopadhyay}},
  \bibinfo{author}{\bibfnamefont{S.}~\bibnamefont{Bhattacharya}},
  \bibinfo{author}{\bibfnamefont{T.~K.} \bibnamefont{Ghosh}},
  \bibinfo{author}{\bibfnamefont{A.}~\bibnamefont{Goswami}},
  \bibinfo{author}{\bibfnamefont{S.}~\bibnamefont{Pal}},
  \bibinfo{author}{\bibfnamefont{M.~S.} \bibnamefont{Sarkar}},
  \bibinfo{author}{\bibfnamefont{H.~C.} \bibnamefont{Jain}},
  \bibinfo{author}{\bibfnamefont{P.~K.} \bibnamefont{Joshi}},
  \bibinfo{author}{\bibfnamefont{R.~K.} \bibnamefont{Bhowmik}},
  \bibnamefont{et~al.}, \bibinfo{journal}{Phys. Rev. C}
  \textbf{\bibinfo{volume}{71}}, \bibinfo{pages}{041305}
  (\bibinfo{year}{2005}).

\bibitem[{\citenamefont{Garrett et~al.}(2012)\citenamefont{Garrett, Bangay,
  Diaz{\ }Varela, Ball, Cross, Demand, Finlay, Garnsworthy, Green, Hackman
  et~al.}}]{019}
\bibinfo{author}{\bibfnamefont{P.~E.} \bibnamefont{Garrett}},
  \bibinfo{author}{\bibfnamefont{J.}~\bibnamefont{Bangay}},
  \bibinfo{author}{\bibfnamefont{A.}~\bibnamefont{Diaz{\ }Varela}},
  \bibinfo{author}{\bibfnamefont{G.~C.} \bibnamefont{Ball}},
  \bibinfo{author}{\bibfnamefont{D.~S.} \bibnamefont{Cross}},
  \bibinfo{author}{\bibfnamefont{G.~A.} \bibnamefont{Demand}},
  \bibinfo{author}{\bibfnamefont{P.}~\bibnamefont{Finlay}},
  \bibinfo{author}{\bibfnamefont{A.~B.} \bibnamefont{Garnsworthy}},
  \bibinfo{author}{\bibfnamefont{K.~L.} \bibnamefont{Green}},
  \bibinfo{author}{\bibfnamefont{G.}~\bibnamefont{Hackman}},
  \bibnamefont{et~al.}, \bibinfo{journal}{Phys. Rev. C}
  \textbf{\bibinfo{volume}{86}}, \bibinfo{pages}{044304}
  (\bibinfo{year}{2012}).

\bibitem[{\citenamefont{Kusnezov et~al.}(1987)\citenamefont{Kusnezov, Bruder,
  Ionescu, Kern, Rast, Heyde, P.{\ }Van{\ }Isacker, Moreau, Waroquier, and
  Meyer}}]{020}
\bibinfo{author}{\bibfnamefont{D.}~\bibnamefont{Kusnezov}},
  \bibinfo{author}{\bibfnamefont{A.}~\bibnamefont{Bruder}},
  \bibinfo{author}{\bibfnamefont{V.}~\bibnamefont{Ionescu}},
  \bibinfo{author}{\bibfnamefont{J.}~\bibnamefont{Kern}},
  \bibinfo{author}{\bibfnamefont{M.}~\bibnamefont{Rast}},
  \bibinfo{author}{\bibfnamefont{K.~L.~G.} \bibnamefont{Heyde}},
  \bibinfo{author}{\bibnamefont{P.{\ }Van{\ }Isacker}},
  \bibinfo{author}{\bibfnamefont{J.}~\bibnamefont{Moreau}},
  \bibinfo{author}{\bibfnamefont{M.}~\bibnamefont{Waroquier}},
  \bibnamefont{and} \bibinfo{author}{\bibfnamefont{R.~A.} \bibnamefont{Meyer}},
  \bibinfo{journal}{Helv. Phys. Acta.} \textbf{\bibinfo{volume}{60}},
  \bibinfo{pages}{456} (\bibinfo{year}{1987}).

\bibitem[{\citenamefont{Garrett et~al.}(2007)\citenamefont{Garrett, Green,
  Lehmann, Jolie, McGrath, Yeh, and Yates}}]{021}
\bibinfo{author}{\bibfnamefont{P.~E.} \bibnamefont{Garrett}},
  \bibinfo{author}{\bibfnamefont{K.~L.} \bibnamefont{Green}},
  \bibinfo{author}{\bibfnamefont{H.}~\bibnamefont{Lehmann}},
  \bibinfo{author}{\bibfnamefont{J.}~\bibnamefont{Jolie}},
  \bibinfo{author}{\bibfnamefont{C.~A.} \bibnamefont{McGrath}},
  \bibinfo{author}{\bibfnamefont{M.}~\bibnamefont{Yeh}}, \bibnamefont{and}
  \bibinfo{author}{\bibfnamefont{S.~W.} \bibnamefont{Yates}},
  \bibinfo{journal}{Phys. Rev. C} \textbf{\bibinfo{volume}{75}},
  \bibinfo{pages}{054310} (\bibinfo{year}{2007}).

\bibitem[{\citenamefont{Green et~al.}(2009)\citenamefont{Green, Garrett,
  Austin, Ball, Bandyopadhyay, Colosimo, Cross, Demand, Grinyer, Hackman
  et~al.}}]{022}
\bibinfo{author}{\bibfnamefont{K.~L.} \bibnamefont{Green}},
  \bibinfo{author}{\bibfnamefont{P.~E.} \bibnamefont{Garrett}},
  \bibinfo{author}{\bibfnamefont{R.~A.~E.} \bibnamefont{Austin}},
  \bibinfo{author}{\bibfnamefont{G.~C.} \bibnamefont{Ball}},
  \bibinfo{author}{\bibfnamefont{D.~S.} \bibnamefont{Bandyopadhyay}},
  \bibinfo{author}{\bibfnamefont{S.}~\bibnamefont{Colosimo}},
  \bibinfo{author}{\bibfnamefont{D.}~\bibnamefont{Cross}},
  \bibinfo{author}{\bibfnamefont{G.~A.} \bibnamefont{Demand}},
  \bibinfo{author}{\bibfnamefont{G.~F.} \bibnamefont{Grinyer}},
  \bibinfo{author}{\bibfnamefont{G.}~\bibnamefont{Hackman}},
  \bibnamefont{et~al.}, \bibinfo{journal}{Phys. Rev. C}
  \textbf{\bibinfo{volume}{80}}, \bibinfo{pages}{032502}
  (\bibinfo{year}{2009}).

\bibitem[{\citenamefont{Kumpulainen et~al.}(1992)\citenamefont{Kumpulainen,
  Julin, Kantele, Passoja, Trzaska, Verho, V\"{a}\"{a}r\"{a}m\"{a}ki, Cutoiu,
  and Ivascu}}]{023}
\bibinfo{author}{\bibfnamefont{J.}~\bibnamefont{Kumpulainen}},
  \bibinfo{author}{\bibfnamefont{R.}~\bibnamefont{Julin}},
  \bibinfo{author}{\bibfnamefont{J.}~\bibnamefont{Kantele}},
  \bibinfo{author}{\bibfnamefont{A.}~\bibnamefont{Passoja}},
  \bibinfo{author}{\bibfnamefont{W.~H.} \bibnamefont{Trzaska}},
  \bibinfo{author}{\bibfnamefont{E.}~\bibnamefont{Verho}},
  \bibinfo{author}{\bibfnamefont{J.}~\bibnamefont{V\"{a}\"{a}r\"{a}m\"{a}ki}},
  \bibinfo{author}{\bibfnamefont{D.}~\bibnamefont{Cutoiu}}, \bibnamefont{and}
  \bibinfo{author}{\bibfnamefont{M.}~\bibnamefont{Ivascu}},
  \bibinfo{journal}{Phys. Rev. C} \textbf{\bibinfo{volume}{45}},
  \bibinfo{pages}{640} (\bibinfo{year}{1992}).

\bibitem[{\citenamefont{Garrett et~al.}(2008)\citenamefont{Garrett, Green, and
  Wood}}]{024}
\bibinfo{author}{\bibfnamefont{P.~E.} \bibnamefont{Garrett}},
  \bibinfo{author}{\bibfnamefont{K.~L.} \bibnamefont{Green}}, \bibnamefont{and}
  \bibinfo{author}{\bibfnamefont{J.~L.} \bibnamefont{Wood}},
  \bibinfo{journal}{Phys. Rev. C} \textbf{\bibinfo{volume}{78}},
  \bibinfo{pages}{044307} (\bibinfo{year}{2008}).

\bibitem[{\citenamefont{Bandyopadhyay et~al.}(2007)\citenamefont{Bandyopadhyay,
  Lesher, Fransen, Boukharouba, Garrett, Green, McEllistrem, and Yates}}]{025}
\bibinfo{author}{\bibfnamefont{D.}~\bibnamefont{Bandyopadhyay}},
  \bibinfo{author}{\bibfnamefont{S.~R.} \bibnamefont{Lesher}},
  \bibinfo{author}{\bibfnamefont{C.}~\bibnamefont{Fransen}},
  \bibinfo{author}{\bibfnamefont{N.}~\bibnamefont{Boukharouba}},
  \bibinfo{author}{\bibfnamefont{P.~E.} \bibnamefont{Garrett}},
  \bibinfo{author}{\bibfnamefont{K.~L.} \bibnamefont{Green}},
  \bibinfo{author}{\bibfnamefont{M.~T.} \bibnamefont{McEllistrem}},
  \bibnamefont{and} \bibinfo{author}{\bibfnamefont{S.~W.} \bibnamefont{Yates}},
  \bibinfo{journal}{Phys. Rev. C} \textbf{\bibinfo{volume}{76}},
  \bibinfo{pages}{054308} (\bibinfo{year}{2007}).

\bibitem[{\citenamefont{Schreckenbach et~al.}(1982)\citenamefont{Schreckenbach,
  Mheemeed, Barreau, von Egidy, Faust, B\"{o}rner, Brissot, Stelts, Heyde,
  Van{\ }Isacker et~al.}}]{026}
\bibinfo{author}{\bibfnamefont{K.}~\bibnamefont{Schreckenbach}},
  \bibinfo{author}{\bibfnamefont{A.}~\bibnamefont{Mheemeed}},
  \bibinfo{author}{\bibfnamefont{G.}~\bibnamefont{Barreau}},
  \bibinfo{author}{\bibfnamefont{T.}~\bibnamefont{von Egidy}},
  \bibinfo{author}{\bibfnamefont{H.~R.} \bibnamefont{Faust}},
  \bibinfo{author}{\bibfnamefont{H.~G.} \bibnamefont{B\"{o}rner}},
  \bibinfo{author}{\bibfnamefont{R.}~\bibnamefont{Brissot}},
  \bibinfo{author}{\bibfnamefont{M.~L.} \bibnamefont{Stelts}},
  \bibinfo{author}{\bibfnamefont{K.~L.~G.} \bibnamefont{Heyde}},
  \bibinfo{author}{\bibfnamefont{P.}~\bibnamefont{Van{\ }Isacker}},
  \bibnamefont{et~al.}, \bibinfo{journal}{Phys. Lett. B}
  \textbf{\bibinfo{volume}{110}}, \bibinfo{pages}{364} (\bibinfo{year}{1982}).

\bibitem[{\citenamefont{Mheemeed et~al.}(1984)\citenamefont{Mheemeed,
  Schreckenbach, Barreau, Faust, B\"{o}rner, Brissot, Hungerford, Schmidt,
  Scheerer, Von{\ }Egidy et~al.}}]{027}
\bibinfo{author}{\bibfnamefont{A.}~\bibnamefont{Mheemeed}},
  \bibinfo{author}{\bibfnamefont{K.}~\bibnamefont{Schreckenbach}},
  \bibinfo{author}{\bibfnamefont{G.}~\bibnamefont{Barreau}},
  \bibinfo{author}{\bibfnamefont{H.~R.} \bibnamefont{Faust}},
  \bibinfo{author}{\bibfnamefont{H.~G.} \bibnamefont{B\"{o}rner}},
  \bibinfo{author}{\bibfnamefont{R.}~\bibnamefont{Brissot}},
  \bibinfo{author}{\bibfnamefont{P.}~\bibnamefont{Hungerford}},
  \bibinfo{author}{\bibfnamefont{H.~H.} \bibnamefont{Schmidt}},
  \bibinfo{author}{\bibfnamefont{H.~J.} \bibnamefont{Scheerer}},
  \bibinfo{author}{\bibfnamefont{T.}~\bibnamefont{Von{\ }Egidy}},
  \bibnamefont{et~al.}, \bibinfo{journal}{Nucl. Phys. A}
  \textbf{\bibinfo{volume}{412}}, \bibinfo{pages}{113} (\bibinfo{year}{1984}).

\bibitem[{\citenamefont{Juutinen et~al.}(1996)\citenamefont{Juutinen, Jones,
  Lampinen, Lhersonneau, M\"{a}kel\"{a}, Piiparinen, Savelius, and
  T\"{o}rm\"{a}nen}}]{028}
\bibinfo{author}{\bibfnamefont{S.}~\bibnamefont{Juutinen}},
  \bibinfo{author}{\bibfnamefont{P.}~\bibnamefont{Jones}},
  \bibinfo{author}{\bibfnamefont{A.}~\bibnamefont{Lampinen}},
  \bibinfo{author}{\bibfnamefont{G.}~\bibnamefont{Lhersonneau}},
  \bibinfo{author}{\bibfnamefont{E.}~\bibnamefont{M\"{a}kel\"{a}}},
  \bibinfo{author}{\bibfnamefont{M.}~\bibnamefont{Piiparinen}},
  \bibinfo{author}{\bibfnamefont{A.}~\bibnamefont{Savelius}}, \bibnamefont{and}
  \bibinfo{author}{\bibfnamefont{S.}~\bibnamefont{T\"{o}rm\"{a}nen}},
  \bibinfo{journal}{Phys. Lett. B} \textbf{\bibinfo{volume}{386}},
  \bibinfo{pages}{80} (\bibinfo{year}{1996}).

\bibitem[{\citenamefont{Mach et~al.}(1989)\citenamefont{Mach, Moszy\'{n}ski,
  Casten, Gill, Brenner, Winger, Krips, Wesselborg, B\"{u}scher, Wohn
  et~al.}}]{029}
\bibinfo{author}{\bibfnamefont{H.}~\bibnamefont{Mach}},
  \bibinfo{author}{\bibfnamefont{M.}~\bibnamefont{Moszy\'{n}ski}},
  \bibinfo{author}{\bibfnamefont{R.~F.} \bibnamefont{Casten}},
  \bibinfo{author}{\bibfnamefont{R.~L.} \bibnamefont{Gill}},
  \bibinfo{author}{\bibfnamefont{D.~S.} \bibnamefont{Brenner}},
  \bibinfo{author}{\bibfnamefont{J.~A.} \bibnamefont{Winger}},
  \bibinfo{author}{\bibfnamefont{W.}~\bibnamefont{Krips}},
  \bibinfo{author}{\bibfnamefont{C.}~\bibnamefont{Wesselborg}},
  \bibinfo{author}{\bibfnamefont{M.}~\bibnamefont{B\"{u}scher}},
  \bibinfo{author}{\bibfnamefont{F.~K.} \bibnamefont{Wohn}},
  \bibnamefont{et~al.}, \bibinfo{journal}{Phys. Rev. Lett.}
  \textbf{\bibinfo{volume}{63}}, \bibinfo{pages}{143} (\bibinfo{year}{1989}).

\bibitem[{\citenamefont{Wang et~al.}(2001)\citenamefont{Wang, Dendooven,
  Huikari, Jokinen, Kolhinen, Lhersonneau, Nieminen, Nummela, Penttil\"{a},
  Per\"{a}j\"{a}rvi et~al.}}]{030}
\bibinfo{author}{\bibfnamefont{Y.}~\bibnamefont{Wang}},
  \bibinfo{author}{\bibfnamefont{P.}~\bibnamefont{Dendooven}},
  \bibinfo{author}{\bibfnamefont{J.}~\bibnamefont{Huikari}},
  \bibinfo{author}{\bibfnamefont{A.}~\bibnamefont{Jokinen}},
  \bibinfo{author}{\bibfnamefont{V.~S.} \bibnamefont{Kolhinen}},
  \bibinfo{author}{\bibfnamefont{G.}~\bibnamefont{Lhersonneau}},
  \bibinfo{author}{\bibfnamefont{A.}~\bibnamefont{Nieminen}},
  \bibinfo{author}{\bibfnamefont{S.}~\bibnamefont{Nummela}},
  \bibinfo{author}{\bibfnamefont{H.}~\bibnamefont{Penttil\"{a}}},
  \bibinfo{author}{\bibfnamefont{K.}~\bibnamefont{Per\"{a}j\"{a}rvi}},
  \bibnamefont{et~al.}, \bibinfo{journal}{Phys. Rev. C}
  \textbf{\bibinfo{volume}{64}}, \bibinfo{pages}{054315}
  (\bibinfo{year}{2001}).

\bibitem[{\citenamefont{Kadi et~al.}(2003)\citenamefont{Kadi, Warr, Garrett,
  Jolie, and Yates}}]{031}
\bibinfo{author}{\bibfnamefont{M.}~\bibnamefont{Kadi}},
  \bibinfo{author}{\bibfnamefont{N.}~\bibnamefont{Warr}},
  \bibinfo{author}{\bibfnamefont{P.~E.} \bibnamefont{Garrett}},
  \bibinfo{author}{\bibfnamefont{J.}~\bibnamefont{Jolie}}, \bibnamefont{and}
  \bibinfo{author}{\bibfnamefont{S.~W.} \bibnamefont{Yates}},
  \bibinfo{journal}{Phys. Rev. C} \textbf{\bibinfo{volume}{68}},
  \bibinfo{pages}{031306(R)} (\bibinfo{year}{2003}).

\bibitem[{\citenamefont{Batchelder et~al.}(2009)\citenamefont{Batchelder, Wood,
  Garrett, Green, Rykaczewski, Bilheux, Bingham, Carter, Fong, Grzywacz
  et~al.}}]{032}
\bibinfo{author}{\bibfnamefont{J.~C.} \bibnamefont{Batchelder}},
  \bibinfo{author}{\bibfnamefont{J.~L.} \bibnamefont{Wood}},
  \bibinfo{author}{\bibfnamefont{P.~E.} \bibnamefont{Garrett}},
  \bibinfo{author}{\bibfnamefont{K.~L.} \bibnamefont{Green}},
  \bibinfo{author}{\bibfnamefont{K.~P.} \bibnamefont{Rykaczewski}},
  \bibinfo{author}{\bibfnamefont{J.~C.} \bibnamefont{Bilheux}},
  \bibinfo{author}{\bibfnamefont{C.~R.} \bibnamefont{Bingham}},
  \bibinfo{author}{\bibfnamefont{H.~K.} \bibnamefont{Carter}},
  \bibinfo{author}{\bibfnamefont{D.}~\bibnamefont{Fong}},
  \bibinfo{author}{\bibfnamefont{R.}~\bibnamefont{Grzywacz}},
  \bibnamefont{et~al.}, \bibinfo{journal}{Phys. Rev. C}
  \textbf{\bibinfo{volume}{80}}, \bibinfo{pages}{054318}
  (\bibinfo{year}{2009}).

\bibitem[{\citenamefont{Aprahamian et~al.}(1984)\citenamefont{Aprahamian,
  Brenner, Casten, Gill, Piotrowski, and Heyde}}]{033}
\bibinfo{author}{\bibfnamefont{A.}~\bibnamefont{Aprahamian}},
  \bibinfo{author}{\bibfnamefont{D.~S.} \bibnamefont{Brenner}},
  \bibinfo{author}{\bibfnamefont{R.~F.} \bibnamefont{Casten}},
  \bibinfo{author}{\bibfnamefont{R.~L.} \bibnamefont{Gill}},
  \bibinfo{author}{\bibfnamefont{A.}~\bibnamefont{Piotrowski}},
  \bibnamefont{and} \bibinfo{author}{\bibfnamefont{K.~L.~G.}
  \bibnamefont{Heyde}}, \bibinfo{journal}{Phys. Lett. B}
  \textbf{\bibinfo{volume}{140}}, \bibinfo{pages}{22} (\bibinfo{year}{1984}).

\bibitem[{\citenamefont{Wang et~al.}(2003)\citenamefont{Wang, Rinta-Antila,
  Dendooven, Huikari, Jokinen, Kolhinen, Lhersonneau, Nieminen, Nummela,
  Penttil\"{a} et~al.}}]{034}
\bibinfo{author}{\bibfnamefont{Y.}~\bibnamefont{Wang}},
  \bibinfo{author}{\bibfnamefont{S.}~\bibnamefont{Rinta-Antila}},
  \bibinfo{author}{\bibfnamefont{P.}~\bibnamefont{Dendooven}},
  \bibinfo{author}{\bibfnamefont{J.}~\bibnamefont{Huikari}},
  \bibinfo{author}{\bibfnamefont{A.}~\bibnamefont{Jokinen}},
  \bibinfo{author}{\bibfnamefont{V.~S.} \bibnamefont{Kolhinen}},
  \bibinfo{author}{\bibfnamefont{G.}~\bibnamefont{Lhersonneau}},
  \bibinfo{author}{\bibfnamefont{A.}~\bibnamefont{Nieminen}},
  \bibinfo{author}{\bibfnamefont{S.}~\bibnamefont{Nummela}},
  \bibinfo{author}{\bibfnamefont{H.}~\bibnamefont{Penttil\"{a}}},
  \bibnamefont{et~al.}, \bibinfo{journal}{Phys. Rev. C}
  \textbf{\bibinfo{volume}{67}}, \bibinfo{pages}{064303}
  (\bibinfo{year}{2003}).

\bibitem[{\citenamefont{Luo et~al.}(2012)\citenamefont{Luo, Rasmussen, Nelson,
  Hamilton, Ramayya, Hwang, Liu, Goodin, Stone, Zhu et~al.}}]{035}
\bibinfo{author}{\bibfnamefont{Y.~X.} \bibnamefont{Luo}},
  \bibinfo{author}{\bibfnamefont{J.~O.} \bibnamefont{Rasmussen}},
  \bibinfo{author}{\bibfnamefont{C.~S.} \bibnamefont{Nelson}},
  \bibinfo{author}{\bibfnamefont{J.~H.} \bibnamefont{Hamilton}},
  \bibinfo{author}{\bibfnamefont{A.~V.} \bibnamefont{Ramayya}},
  \bibinfo{author}{\bibfnamefont{J.~K.} \bibnamefont{Hwang}},
  \bibinfo{author}{\bibfnamefont{S.~H.} \bibnamefont{Liu}},
  \bibinfo{author}{\bibfnamefont{C.}~\bibnamefont{Goodin}},
  \bibinfo{author}{\bibfnamefont{N.~J.} \bibnamefont{Stone}},
  \bibinfo{author}{\bibfnamefont{S.~J.} \bibnamefont{Zhu}},
  \bibnamefont{et~al.}, \bibinfo{journal}{Nucl. Phys. A}
  \textbf{\bibinfo{volume}{874}}, \bibinfo{pages}{32} (\bibinfo{year}{2012}).

\bibitem[{\citenamefont{Batchelder et~al.}(2012)\citenamefont{Batchelder,
  Brewer, Goans, Grzywacz, Griffith, Jost, Korgul, Liu, Paulauskas, Spejewski
  et~al.}}]{036}
\bibinfo{author}{\bibfnamefont{J.~C.} \bibnamefont{Batchelder}},
  \bibinfo{author}{\bibfnamefont{N.~T.} \bibnamefont{Brewer}},
  \bibinfo{author}{\bibfnamefont{R.~E.} \bibnamefont{Goans}},
  \bibinfo{author}{\bibfnamefont{R.}~\bibnamefont{Grzywacz}},
  \bibinfo{author}{\bibfnamefont{B.~O.} \bibnamefont{Griffith}},
  \bibinfo{author}{\bibfnamefont{C.}~\bibnamefont{Jost}},
  \bibinfo{author}{\bibfnamefont{A.}~\bibnamefont{Korgul}},
  \bibinfo{author}{\bibfnamefont{S.~H.} \bibnamefont{Liu}},
  \bibinfo{author}{\bibfnamefont{S.~V.} \bibnamefont{Paulauskas}},
  \bibinfo{author}{\bibfnamefont{E.~H.} \bibnamefont{Spejewski}},
  \bibnamefont{et~al.}, \bibinfo{journal}{Phys. Rev. C}
  \textbf{\bibinfo{volume}{86}}, \bibinfo{pages}{064311}
  (\bibinfo{year}{2012}).

\bibitem[{\citenamefont{Kautzsch et~al.}(1996)\citenamefont{Kautzsch, Walters,
  Fedoseyev, Jading, Jokinen, Kl\"{o}ckl, Kratz, Mishin, Ravn, Van{\ }Duppen
  et~al.}}]{037}
\bibinfo{author}{\bibfnamefont{T.}~\bibnamefont{Kautzsch}},
  \bibinfo{author}{\bibfnamefont{W.~B.} \bibnamefont{Walters}},
  \bibinfo{author}{\bibfnamefont{V.~N.} \bibnamefont{Fedoseyev}},
  \bibinfo{author}{\bibfnamefont{Y.}~\bibnamefont{Jading}},
  \bibinfo{author}{\bibfnamefont{A.}~\bibnamefont{Jokinen}},
  \bibinfo{author}{\bibfnamefont{I.}~\bibnamefont{Kl\"{o}ckl}},
  \bibinfo{author}{\bibfnamefont{K.-L.} \bibnamefont{Kratz}},
  \bibinfo{author}{\bibfnamefont{V.~I.} \bibnamefont{Mishin}},
  \bibinfo{author}{\bibfnamefont{H.~L.} \bibnamefont{Ravn}},
  \bibinfo{author}{\bibfnamefont{P.}~\bibnamefont{Van{\ }Duppen}},
  \bibnamefont{et~al.}, \bibinfo{journal}{Phys. Rev. C}
  \textbf{\bibinfo{volume}{54}}, \bibinfo{pages}{R2811} (\bibinfo{year}{1996}).

\bibitem[{\citenamefont{Ilieva et~al.}(2014)\citenamefont{Ilieva, Th\"{u}rauf,
  Kr\"{o}ll, Kr\"{u}cken, Behrens, Bildstein, Blazhev, B\"{o}nig, Butler,
  Cederk\"{a}ll et~al.}}]{038}
\bibinfo{author}{\bibfnamefont{S.}~\bibnamefont{Ilieva}},
  \bibinfo{author}{\bibfnamefont{M.}~\bibnamefont{Th\"{u}rauf}},
  \bibinfo{author}{\bibfnamefont{T.}~\bibnamefont{Kr\"{o}ll}},
  \bibinfo{author}{\bibfnamefont{R.}~\bibnamefont{Kr\"{u}cken}},
  \bibinfo{author}{\bibfnamefont{T.}~\bibnamefont{Behrens}},
  \bibinfo{author}{\bibfnamefont{V.}~\bibnamefont{Bildstein}},
  \bibinfo{author}{\bibfnamefont{A.}~\bibnamefont{Blazhev}},
  \bibinfo{author}{\bibfnamefont{S.}~\bibnamefont{B\"{o}nig}},
  \bibinfo{author}{\bibfnamefont{P.~A.} \bibnamefont{Butler}},
  \bibinfo{author}{\bibfnamefont{J.}~\bibnamefont{Cederk\"{a}ll}},
  \bibnamefont{et~al.}, \bibinfo{journal}{Phys. Rev. C}
  \textbf{\bibinfo{volume}{89}}, \bibinfo{pages}{014313}
  (\bibinfo{year}{2014}).

\bibitem[{\citenamefont{Batchelder et~al.}(2014)\citenamefont{Batchelder,
  Brewer, Gross, Grywacz, Hamilton, Karmy, Fijalkowska, Liu, Miernik, Padgett
  et~al.}}]{039}
\bibinfo{author}{\bibfnamefont{J.~C.} \bibnamefont{Batchelder}},
  \bibinfo{author}{\bibfnamefont{N.~T.} \bibnamefont{Brewer}},
  \bibinfo{author}{\bibfnamefont{C.~J.} \bibnamefont{Gross}},
  \bibinfo{author}{\bibfnamefont{R.}~\bibnamefont{Grywacz}},
  \bibinfo{author}{\bibfnamefont{J.~H.} \bibnamefont{Hamilton}},
  \bibinfo{author}{\bibfnamefont{M.}~\bibnamefont{Karmy}},
  \bibinfo{author}{\bibfnamefont{A.}~\bibnamefont{Fijalkowska}},
  \bibinfo{author}{\bibfnamefont{S.~H.} \bibnamefont{Liu}},
  \bibinfo{author}{\bibfnamefont{K.}~\bibnamefont{Miernik}},
  \bibinfo{author}{\bibfnamefont{S.~W.} \bibnamefont{Padgett}},
  \bibnamefont{et~al.}, \bibinfo{journal}{Phys. Rev. C}
  \textbf{\bibinfo{volume}{89}}, \bibinfo{pages}{054321}
  (\bibinfo{year}{2014}).

\bibitem[{\citenamefont{Kautzsch et~al.}(2000)\citenamefont{Kautzsch, Walters,
  Hannawald, Kratz, Mishin, Fedoseyev, B\"{o}hmer, Jading, Van{\ }Duppen,
  Pfeiffer et~al.}}]{040}
\bibinfo{author}{\bibfnamefont{T.}~\bibnamefont{Kautzsch}},
  \bibinfo{author}{\bibfnamefont{W.~B.} \bibnamefont{Walters}},
  \bibinfo{author}{\bibfnamefont{M.}~\bibnamefont{Hannawald}},
  \bibinfo{author}{\bibfnamefont{K.-L.} \bibnamefont{Kratz}},
  \bibinfo{author}{\bibfnamefont{V.~I.} \bibnamefont{Mishin}},
  \bibinfo{author}{\bibfnamefont{V.~N.} \bibnamefont{Fedoseyev}},
  \bibinfo{author}{\bibfnamefont{W.}~\bibnamefont{B\"{o}hmer}},
  \bibinfo{author}{\bibfnamefont{Y.}~\bibnamefont{Jading}},
  \bibinfo{author}{\bibfnamefont{P.}~\bibnamefont{Van{\ }Duppen}},
  \bibinfo{author}{\bibfnamefont{B.}~\bibnamefont{Pfeiffer}},
  \bibnamefont{et~al.}, \bibinfo{journal}{Eur. Phys. J. A}
  \textbf{\bibinfo{volume}{9}}, \bibinfo{pages}{201} (\bibinfo{year}{2000}).

\bibitem[{\citenamefont{Hoteling et~al.}(2007)\citenamefont{Hoteling, Walters,
  Tomlin, Mantica, Pereira, Becerril, Fleckenstein, Hecht, Lorusso, Quinn
  et~al.}}]{041}
\bibinfo{author}{\bibfnamefont{N.}~\bibnamefont{Hoteling}},
  \bibinfo{author}{\bibfnamefont{W.~B.} \bibnamefont{Walters}},
  \bibinfo{author}{\bibfnamefont{B.~E.} \bibnamefont{Tomlin}},
  \bibinfo{author}{\bibfnamefont{P.~F.} \bibnamefont{Mantica}},
  \bibinfo{author}{\bibfnamefont{J.}~\bibnamefont{Pereira}},
  \bibinfo{author}{\bibfnamefont{A.}~\bibnamefont{Becerril}},
  \bibinfo{author}{\bibfnamefont{T.}~\bibnamefont{Fleckenstein}},
  \bibinfo{author}{\bibfnamefont{A.~A.} \bibnamefont{Hecht}},
  \bibinfo{author}{\bibfnamefont{G.}~\bibnamefont{Lorusso}},
  \bibinfo{author}{\bibfnamefont{M.}~\bibnamefont{Quinn}},
  \bibnamefont{et~al.}, \bibinfo{journal}{Phys. Rev. C}
  \textbf{\bibinfo{volume}{76}}, \bibinfo{pages}{044324}
  (\bibinfo{year}{2007}).

\bibitem[{\citenamefont{C\'{a}ceres et~al.}(2009)\citenamefont{C\'{a}ceres,
  G\'{o}rska, Jungclaus, Pf\"{u}tzner, Grawe, Nowacki, Sieja, Pietri, Rudolph,
  Podoly\'{a}k et~al.}}]{042}
\bibinfo{author}{\bibfnamefont{L.}~\bibnamefont{C\'{a}ceres}},
  \bibinfo{author}{\bibfnamefont{M.}~\bibnamefont{G\'{o}rska}},
  \bibinfo{author}{\bibfnamefont{A.}~\bibnamefont{Jungclaus}},
  \bibinfo{author}{\bibfnamefont{M.}~\bibnamefont{Pf\"{u}tzner}},
  \bibinfo{author}{\bibfnamefont{H.}~\bibnamefont{Grawe}},
  \bibinfo{author}{\bibfnamefont{F.}~\bibnamefont{Nowacki}},
  \bibinfo{author}{\bibfnamefont{K.}~\bibnamefont{Sieja}},
  \bibinfo{author}{\bibfnamefont{S.}~\bibnamefont{Pietri}},
  \bibinfo{author}{\bibfnamefont{D.}~\bibnamefont{Rudolph}},
  \bibinfo{author}{\bibfnamefont{Z.}~\bibnamefont{Podoly\'{a}k}},
  \bibnamefont{et~al.}, \bibinfo{journal}{Phys. Rev. C}
  \textbf{\bibinfo{volume}{79}}, \bibinfo{pages}{011301}
  (\bibinfo{year}{2009}).

\bibitem[{\citenamefont{Jungclaus et~al.}(2007)\citenamefont{Jungclaus,
  C\'{a}ceres, G\'{o}rska, Pf\"{u}tzner, Pietri, Werner-Malento, Grawe,
  Langanke, Mart\'{i}nez-Pinedo, Nowacki et~al.}}]{043}
\bibinfo{author}{\bibfnamefont{A.}~\bibnamefont{Jungclaus}},
  \bibinfo{author}{\bibfnamefont{L.}~\bibnamefont{C\'{a}ceres}},
  \bibinfo{author}{\bibfnamefont{M.}~\bibnamefont{G\'{o}rska}},
  \bibinfo{author}{\bibfnamefont{M.}~\bibnamefont{Pf\"{u}tzner}},
  \bibinfo{author}{\bibfnamefont{S.}~\bibnamefont{Pietri}},
  \bibinfo{author}{\bibfnamefont{E.}~\bibnamefont{Werner-Malento}},
  \bibinfo{author}{\bibfnamefont{H.}~\bibnamefont{Grawe}},
  \bibinfo{author}{\bibfnamefont{K.}~\bibnamefont{Langanke}},
  \bibinfo{author}{\bibfnamefont{G.}~\bibnamefont{Mart\'{i}nez-Pinedo}},
  \bibinfo{author}{\bibfnamefont{F.}~\bibnamefont{Nowacki}},
  \bibnamefont{et~al.}, \bibinfo{journal}{Phys. Rev. Lett.}
  \textbf{\bibinfo{volume}{99}}, \bibinfo{pages}{132501}
  (\bibinfo{year}{2007}).

\bibitem[{\citenamefont{Wang et~al.}(2014)\citenamefont{Wang, Kaneko, and
  Sun}}]{044}
\bibinfo{author}{\bibfnamefont{H.-K.} \bibnamefont{Wang}},
  \bibinfo{author}{\bibfnamefont{K.}~\bibnamefont{Kaneko}}, \bibnamefont{and}
  \bibinfo{author}{\bibfnamefont{Y.}~\bibnamefont{Sun}},
  \bibinfo{journal}{Phys. Rev. C} \textbf{\bibinfo{volume}{89}},
  \bibinfo{pages}{064311} (\bibinfo{year}{2014}).

\bibitem[{\citenamefont{Breitenfeldt et~al.}(2010)\citenamefont{Breitenfeldt,
  Borgmann, Audi, Baruah, Beck, Blaum, B\"{o}hm, Cakirli, Casten, Delahaye
  et~al.}}]{045}
\bibinfo{author}{\bibfnamefont{M.}~\bibnamefont{Breitenfeldt}},
  \bibinfo{author}{\bibfnamefont{C.}~\bibnamefont{Borgmann}},
  \bibinfo{author}{\bibfnamefont{G.}~\bibnamefont{Audi}},
  \bibinfo{author}{\bibfnamefont{S.}~\bibnamefont{Baruah}},
  \bibinfo{author}{\bibfnamefont{D.}~\bibnamefont{Beck}},
  \bibinfo{author}{\bibfnamefont{K.}~\bibnamefont{Blaum}},
  \bibinfo{author}{\bibfnamefont{C.}~\bibnamefont{B\"{o}hm}},
  \bibinfo{author}{\bibfnamefont{R.~B.} \bibnamefont{Cakirli}},
  \bibinfo{author}{\bibfnamefont{R.~F.} \bibnamefont{Casten}},
  \bibinfo{author}{\bibfnamefont{P.}~\bibnamefont{Delahaye}},
  \bibnamefont{et~al.}, \bibinfo{journal}{Phys. Rev. C}
  \textbf{\bibinfo{volume}{81}}, \bibinfo{pages}{034313}
  (\bibinfo{year}{2010}).

\bibitem[{\citenamefont{Dillmann et~al.}(2003)\citenamefont{Dillmann, Kratz,
  W\"{o}hr, Arndt, Brown, Hoff, Hjorth-Jensen, K\"{o}ster, Ostrowski, Pfeiffer
  et~al.}}]{046}
\bibinfo{author}{\bibfnamefont{I.}~\bibnamefont{Dillmann}},
  \bibinfo{author}{\bibfnamefont{K.-L.} \bibnamefont{Kratz}},
  \bibinfo{author}{\bibfnamefont{A.}~\bibnamefont{W\"{o}hr}},
  \bibinfo{author}{\bibfnamefont{O.}~\bibnamefont{Arndt}},
  \bibinfo{author}{\bibfnamefont{B.~A.} \bibnamefont{Brown}},
  \bibinfo{author}{\bibfnamefont{P.}~\bibnamefont{Hoff}},
  \bibinfo{author}{\bibfnamefont{M.}~\bibnamefont{Hjorth-Jensen}},
  \bibinfo{author}{\bibfnamefont{U.}~\bibnamefont{K\"{o}ster}},
  \bibinfo{author}{\bibfnamefont{A.~N.} \bibnamefont{Ostrowski}},
  \bibinfo{author}{\bibfnamefont{B.}~\bibnamefont{Pfeiffer}},
  \bibnamefont{et~al.}, \bibinfo{journal}{Phys. Rev. Lett.}
  \textbf{\bibinfo{volume}{91}}, \bibinfo{pages}{162503}
  (\bibinfo{year}{2003}).

\bibitem[{\citenamefont{Yordanov et~al.}(2013)\citenamefont{Yordanov,
  Balabanski, Biero\'{n}, Bissell, Blaum, Budin\v{c}evi\'{c}, Fritzsche,
  Fr\"{o}mmgen, Georgiev, Geppert et~al.}}]{047}
\bibinfo{author}{\bibfnamefont{D.~T.} \bibnamefont{Yordanov}},
  \bibinfo{author}{\bibfnamefont{D.~L.} \bibnamefont{Balabanski}},
  \bibinfo{author}{\bibfnamefont{J.}~\bibnamefont{Biero\'{n}}},
  \bibinfo{author}{\bibfnamefont{M.~L.} \bibnamefont{Bissell}},
  \bibinfo{author}{\bibfnamefont{K.}~\bibnamefont{Blaum}},
  \bibinfo{author}{\bibfnamefont{I.}~\bibnamefont{Budin\v{c}evi\'{c}}},
  \bibinfo{author}{\bibfnamefont{S.}~\bibnamefont{Fritzsche}},
  \bibinfo{author}{\bibfnamefont{N.}~\bibnamefont{Fr\"{o}mmgen}},
  \bibinfo{author}{\bibfnamefont{G.}~\bibnamefont{Georgiev}},
  \bibinfo{author}{\bibfnamefont{C.}~\bibnamefont{Geppert}},
  \bibnamefont{et~al.}, \bibinfo{journal}{Phys. Rev. Lett.}
  \textbf{\bibinfo{volume}{110}}, \bibinfo{pages}{192501}
  (\bibinfo{year}{2013}).

\bibitem[{\citenamefont{Casten et~al.}(1992)\citenamefont{Casten, Jolie,
  Borner, Brenner, Zamfir, Chou, and Aprahamian}}]{135Jolie01}
\bibinfo{author}{\bibfnamefont{R.~F.} \bibnamefont{Casten}},
  \bibinfo{author}{\bibfnamefont{J.}~\bibnamefont{Jolie}},
  \bibinfo{author}{\bibfnamefont{H.~G.} \bibnamefont{Borner}},
  \bibinfo{author}{\bibfnamefont{D.~S.} \bibnamefont{Brenner}},
  \bibinfo{author}{\bibfnamefont{N.~V.} \bibnamefont{Zamfir}},
  \bibinfo{author}{\bibfnamefont{W.~T.} \bibnamefont{Chou}}, \bibnamefont{and}
  \bibinfo{author}{\bibfnamefont{A.}~\bibnamefont{Aprahamian}},
  \bibinfo{journal}{Phys. Lett. B} \textbf{\bibinfo{volume}{297}},
  \bibinfo{pages}{19} (\bibinfo{year}{1992}).

\bibitem[{\citenamefont{D\`{e}l\'{e}ze
  et~al.}(1993{\natexlab{a}})\citenamefont{D\`{e}l\'{e}ze, Drissi, Jolie, Kern,
  and Vorlet}}]{136Jolie02}
\bibinfo{author}{\bibfnamefont{M.}~\bibnamefont{D\`{e}l\'{e}ze}},
  \bibinfo{author}{\bibfnamefont{S.}~\bibnamefont{Drissi}},
  \bibinfo{author}{\bibfnamefont{J.}~\bibnamefont{Jolie}},
  \bibinfo{author}{\bibfnamefont{J.}~\bibnamefont{Kern}}, \bibnamefont{and}
  \bibinfo{author}{\bibfnamefont{J.~P.} \bibnamefont{Vorlet}},
  \bibinfo{journal}{Nucl. Phys. A} \textbf{\bibinfo{volume}{554}},
  \bibinfo{pages}{1} (\bibinfo{year}{1993}{\natexlab{a}}).

\bibitem[{\citenamefont{Bertschy et~al.}(1995)\citenamefont{Bertschy, Drissi,
  Garrett, Jolie, Kern, Mannanal, Vorlet, Warr, and Suhonen}}]{137Jolie03}
\bibinfo{author}{\bibfnamefont{M.}~\bibnamefont{Bertschy}},
  \bibinfo{author}{\bibfnamefont{S.}~\bibnamefont{Drissi}},
  \bibinfo{author}{\bibfnamefont{P.~E.} \bibnamefont{Garrett}},
  \bibinfo{author}{\bibfnamefont{J.}~\bibnamefont{Jolie}},
  \bibinfo{author}{\bibfnamefont{J.}~\bibnamefont{Kern}},
  \bibinfo{author}{\bibfnamefont{S.~J.} \bibnamefont{Mannanal}},
  \bibinfo{author}{\bibfnamefont{J.~P.} \bibnamefont{Vorlet}},
  \bibinfo{author}{\bibfnamefont{N.}~\bibnamefont{Warr}}, \bibnamefont{and}
  \bibinfo{author}{\bibfnamefont{J.}~\bibnamefont{Suhonen}},
  \bibinfo{journal}{Phys. Rev. C\textbf{51}, 103 and C\textbf{52}, 1148}
  (\bibinfo{year}{1995}).

\bibitem[{\citenamefont{Lehmann et~al.}(1996)\citenamefont{Lehmann, Garrett,
  Jolie, McGrath, Yeh, and Yates}}]{138Jolie04}
\bibinfo{author}{\bibfnamefont{H.}~\bibnamefont{Lehmann}},
  \bibinfo{author}{\bibfnamefont{P.~E.} \bibnamefont{Garrett}},
  \bibinfo{author}{\bibfnamefont{J.}~\bibnamefont{Jolie}},
  \bibinfo{author}{\bibfnamefont{C.~A.} \bibnamefont{McGrath}},
  \bibinfo{author}{\bibfnamefont{M.}~\bibnamefont{Yeh}}, \bibnamefont{and}
  \bibinfo{author}{\bibfnamefont{S.~W.} \bibnamefont{Yates}},
  \bibinfo{journal}{Phys. Lett. B} \textbf{\bibinfo{volume}{387}},
  \bibinfo{pages}{259} (\bibinfo{year}{1996}).

\bibitem[{\citenamefont{Warr et~al.}(1997)\citenamefont{Warr, Drissi, Garrett,
  Jolie, Kern, Mannanal, Schenker, and Vorlet}}]{139Jolie05}
\bibinfo{author}{\bibfnamefont{N.}~\bibnamefont{Warr}},
  \bibinfo{author}{\bibfnamefont{S.}~\bibnamefont{Drissi}},
  \bibinfo{author}{\bibfnamefont{P.~E.} \bibnamefont{Garrett}},
  \bibinfo{author}{\bibfnamefont{J.}~\bibnamefont{Jolie}},
  \bibinfo{author}{\bibfnamefont{J.}~\bibnamefont{Kern}},
  \bibinfo{author}{\bibfnamefont{S.~J.} \bibnamefont{Mannanal}},
  \bibinfo{author}{\bibfnamefont{J.-L.} \bibnamefont{Schenker}},
  \bibnamefont{and} \bibinfo{author}{\bibfnamefont{J.~P.}
  \bibnamefont{Vorlet}}, \bibinfo{journal}{Nucl. Phys. A}
  \textbf{\bibinfo{volume}{620}}, \bibinfo{pages}{127} (\bibinfo{year}{1997}).

\bibitem[{\citenamefont{Garrett et~al.}(1999)\citenamefont{Garrett, Lehmann,
  Jolie, McGrath, Yeh, and Yates}}]{140Jolie06}
\bibinfo{author}{\bibfnamefont{P.~E.} \bibnamefont{Garrett}},
  \bibinfo{author}{\bibfnamefont{H.}~\bibnamefont{Lehmann}},
  \bibinfo{author}{\bibfnamefont{J.}~\bibnamefont{Jolie}},
  \bibinfo{author}{\bibfnamefont{C.~A.} \bibnamefont{McGrath}},
  \bibinfo{author}{\bibfnamefont{M.}~\bibnamefont{Yeh}}, \bibnamefont{and}
  \bibinfo{author}{\bibfnamefont{S.~W.} \bibnamefont{Yates}},
  \bibinfo{journal}{Phys. Rev. C} \textbf{\bibinfo{volume}{59}},
  \bibinfo{pages}{2455} (\bibinfo{year}{1999}).

\bibitem[{\citenamefont{Lehmann et~al.}(1999)\citenamefont{Lehmann, Nord, E.{\
  }De{\ }Almeida{\ }Pinto, Beck, Besserer, von Brentano, Drissi, Eckert,
  Herzberg, Jager et~al.}}]{141Jolie07}
\bibinfo{author}{\bibfnamefont{H.}~\bibnamefont{Lehmann}},
  \bibinfo{author}{\bibfnamefont{A.}~\bibnamefont{Nord}},
  \bibinfo{author}{\bibfnamefont{A.}~\bibnamefont{E.{\ }De{\ }Almeida{\
  }Pinto}}, \bibinfo{author}{\bibfnamefont{O.}~\bibnamefont{Beck}},
  \bibinfo{author}{\bibfnamefont{J.}~\bibnamefont{Besserer}},
  \bibinfo{author}{\bibfnamefont{P.}~\bibnamefont{von Brentano}},
  \bibinfo{author}{\bibfnamefont{S.}~\bibnamefont{Drissi}},
  \bibinfo{author}{\bibfnamefont{T.}~\bibnamefont{Eckert}},
  \bibinfo{author}{\bibfnamefont{R.-D.} \bibnamefont{Herzberg}},
  \bibinfo{author}{\bibfnamefont{D.}~\bibnamefont{Jager}},
  \bibnamefont{et~al.}, \bibinfo{journal}{Phys. Rev C}
  \textbf{\bibinfo{volume}{60}}, \bibinfo{pages}{024308}
  (\bibinfo{year}{1999}).

\bibitem[{\citenamefont{Corminboeuf
  et~al.}(2000{\natexlab{a}})\citenamefont{Corminboeuf, Brown, Genilloud,
  Hannant, Jolie, Kern, Warr, and Yates}}]{142Jolie08}
\bibinfo{author}{\bibfnamefont{F.}~\bibnamefont{Corminboeuf}},
  \bibinfo{author}{\bibfnamefont{T.~B.} \bibnamefont{Brown}},
  \bibinfo{author}{\bibfnamefont{L.}~\bibnamefont{Genilloud}},
  \bibinfo{author}{\bibfnamefont{C.~D.} \bibnamefont{Hannant}},
  \bibinfo{author}{\bibfnamefont{J.}~\bibnamefont{Jolie}},
  \bibinfo{author}{\bibfnamefont{J.}~\bibnamefont{Kern}},
  \bibinfo{author}{\bibfnamefont{N.}~\bibnamefont{Warr}}, \bibnamefont{and}
  \bibinfo{author}{\bibfnamefont{S.~W.} \bibnamefont{Yates}},
  \bibinfo{journal}{Phys. Rev. Lett.} \textbf{\bibinfo{volume}{84}},
  \bibinfo{pages}{4060} (\bibinfo{year}{2000}{\natexlab{a}}).

\bibitem[{\citenamefont{Corminboeuf
  et~al.}(2000{\natexlab{b}})\citenamefont{Corminboeuf, Brown, Genilloud,
  Hannant, Jolie, Kern, Warr, and Yates}}]{143Jolie09}
\bibinfo{author}{\bibfnamefont{F.}~\bibnamefont{Corminboeuf}},
  \bibinfo{author}{\bibfnamefont{T.~B.} \bibnamefont{Brown}},
  \bibinfo{author}{\bibfnamefont{L.}~\bibnamefont{Genilloud}},
  \bibinfo{author}{\bibfnamefont{C.~D.} \bibnamefont{Hannant}},
  \bibinfo{author}{\bibfnamefont{J.}~\bibnamefont{Jolie}},
  \bibinfo{author}{\bibfnamefont{J.}~\bibnamefont{Kern}},
  \bibinfo{author}{\bibfnamefont{N.}~\bibnamefont{Warr}}, \bibnamefont{and}
  \bibinfo{author}{\bibfnamefont{S.~W.} \bibnamefont{Yates}},
  \bibinfo{journal}{Phys. Rev. C.} \textbf{\bibinfo{volume}{63}},
  \bibinfo{pages}{014305} (\bibinfo{year}{2000}{\natexlab{b}}).

\bibitem[{\citenamefont{Garrett et~al.}(2001)\citenamefont{Garrett, Lehmann,
  Jolie, McGrath, Yeh, Younes, and Yates}}]{144Jolie10}
\bibinfo{author}{\bibfnamefont{P.~E.} \bibnamefont{Garrett}},
  \bibinfo{author}{\bibfnamefont{H.}~\bibnamefont{Lehmann}},
  \bibinfo{author}{\bibfnamefont{J.}~\bibnamefont{Jolie}},
  \bibinfo{author}{\bibfnamefont{C.~A.} \bibnamefont{McGrath}},
  \bibinfo{author}{\bibfnamefont{M.}~\bibnamefont{Yeh}},
  \bibinfo{author}{\bibfnamefont{W.}~\bibnamefont{Younes}}, \bibnamefont{and}
  \bibinfo{author}{\bibfnamefont{S.~W.} \bibnamefont{Yates}},
  \bibinfo{journal}{Phys. Rev. C} \textbf{\bibinfo{volume}{64}},
  \bibinfo{pages}{024316} (\bibinfo{year}{2001}).

\bibitem[{\citenamefont{Gade et~al.}(2002{\natexlab{a}})\citenamefont{Gade,
  Jolie, and von{\ }Brentano}}]{145Jolie11}
\bibinfo{author}{\bibfnamefont{A.}~\bibnamefont{Gade}},
  \bibinfo{author}{\bibfnamefont{J.}~\bibnamefont{Jolie}}, \bibnamefont{and}
  \bibinfo{author}{\bibfnamefont{P.}~\bibnamefont{von{\ }Brentano}},
  \bibinfo{journal}{Phys. Rev. C} \textbf{\bibinfo{volume}{65}},
  \bibinfo{pages}{041305(R)} (\bibinfo{year}{2002}{\natexlab{a}}).

\bibitem[{\citenamefont{Gade et~al.}(2002{\natexlab{b}})\citenamefont{Gade,
  Fitzler, Fransen, Jolie, Kasemann, Klein, Linnemann, Werner, and von{\
  }Brentano}}]{146Jolie12}
\bibinfo{author}{\bibfnamefont{A.}~\bibnamefont{Gade}},
  \bibinfo{author}{\bibfnamefont{A.}~\bibnamefont{Fitzler}},
  \bibinfo{author}{\bibfnamefont{C.}~\bibnamefont{Fransen}},
  \bibinfo{author}{\bibfnamefont{J.}~\bibnamefont{Jolie}},
  \bibinfo{author}{\bibfnamefont{S.}~\bibnamefont{Kasemann}},
  \bibinfo{author}{\bibfnamefont{H.}~\bibnamefont{Klein}},
  \bibinfo{author}{\bibfnamefont{A.}~\bibnamefont{Linnemann}},
  \bibinfo{author}{\bibfnamefont{V.}~\bibnamefont{Werner}}, \bibnamefont{and}
  \bibinfo{author}{\bibfnamefont{P.}~\bibnamefont{von{\ }Brentano}},
  \bibinfo{journal}{Phys. Rev. C} \textbf{\bibinfo{volume}{66}},
  \bibinfo{pages}{034311} (\bibinfo{year}{2002}{\natexlab{b}}).

\bibitem[{\citenamefont{Kohstall et~al.}(2005)\citenamefont{Kohstall, Belic,
  von{\ }Brentano, Fransen, Gade, Herzberg, Jolie, Kneissl, Linnemann, Nord
  et~al.}}]{147Jolie13}
\bibinfo{author}{\bibfnamefont{C.}~\bibnamefont{Kohstall}},
  \bibinfo{author}{\bibfnamefont{D.}~\bibnamefont{Belic}},
  \bibinfo{author}{\bibfnamefont{P.}~\bibnamefont{von{\ }Brentano}},
  \bibinfo{author}{\bibfnamefont{C.}~\bibnamefont{Fransen}},
  \bibinfo{author}{\bibfnamefont{A.}~\bibnamefont{Gade}},
  \bibinfo{author}{\bibfnamefont{R.-D.} \bibnamefont{Herzberg}},
  \bibinfo{author}{\bibfnamefont{J.}~\bibnamefont{Jolie}},
  \bibinfo{author}{\bibfnamefont{U.}~\bibnamefont{Kneissl}},
  \bibinfo{author}{\bibfnamefont{A.}~\bibnamefont{Linnemann}},
  \bibinfo{author}{\bibfnamefont{A.}~\bibnamefont{Nord}}, \bibnamefont{et~al.},
  \bibinfo{journal}{Phys. Rev. C} \textbf{\bibinfo{volume}{72}},
  \bibinfo{pages}{034302} (\bibinfo{year}{2005}).

\bibitem[{\citenamefont{Linnemann et~al.}(2007)\citenamefont{Linnemann,
  Fransen, Jolie, Kneissl, Knoch, Kohstall, M\"{u}cher, Pitz, Scheck, Scholl
  et~al.}}]{148Jolie14}
\bibinfo{author}{\bibfnamefont{A.}~\bibnamefont{Linnemann}},
  \bibinfo{author}{\bibfnamefont{C.}~\bibnamefont{Fransen}},
  \bibinfo{author}{\bibfnamefont{J.}~\bibnamefont{Jolie}},
  \bibinfo{author}{\bibfnamefont{U.}~\bibnamefont{Kneissl}},
  \bibinfo{author}{\bibfnamefont{P.}~\bibnamefont{Knoch}},
  \bibinfo{author}{\bibfnamefont{C.}~\bibnamefont{Kohstall}},
  \bibinfo{author}{\bibfnamefont{D.}~\bibnamefont{M\"{u}cher}},
  \bibinfo{author}{\bibfnamefont{H.~H.} \bibnamefont{Pitz}},
  \bibinfo{author}{\bibfnamefont{M.}~\bibnamefont{Scheck}},
  \bibinfo{author}{\bibfnamefont{C.}~\bibnamefont{Scholl}},
  \bibnamefont{et~al.}, \bibinfo{journal}{Phys. Rev. C}
  \textbf{\bibinfo{volume}{75}}, \bibinfo{pages}{024310}
  (\bibinfo{year}{2007}).

\bibitem[{\citenamefont{D\`{e}l\'{e}ze
  et~al.}(1993{\natexlab{b}})\citenamefont{D\`{e}l\'{e}ze, Drissi, Kern,
  Tercier, and Vorlet}}]{152Jolie22_not_listed}
\bibinfo{author}{\bibfnamefont{M.}~\bibnamefont{D\`{e}l\'{e}ze}},
  \bibinfo{author}{\bibfnamefont{S.}~\bibnamefont{Drissi}},
  \bibinfo{author}{\bibfnamefont{J.}~\bibnamefont{Kern}},
  \bibinfo{author}{\bibfnamefont{P.~A.} \bibnamefont{Tercier}},
  \bibnamefont{and} \bibinfo{author}{\bibfnamefont{J.~P.}
  \bibnamefont{Vorlet}}, \bibinfo{journal}{Nucl. Phys. A}
  \textbf{\bibinfo{volume}{551}}, \bibinfo{pages}{269}
  (\bibinfo{year}{1993}{\natexlab{b}}).

\bibitem[{\citenamefont{Garrett and Wood}(2010)}]{048}
\bibinfo{author}{\bibfnamefont{P.~E.} \bibnamefont{Garrett}} \bibnamefont{and}
  \bibinfo{author}{\bibfnamefont{J.~L.} \bibnamefont{Wood}},
  \bibinfo{journal}{J. Phys. G} \textbf{\bibinfo{volume}{37}},
  \bibinfo{pages}{064028, and corrigendum 06970128} (\bibinfo{year}{2010}).

\bibitem[{\citenamefont{Heyde and Wood}(2011)}]{049}
\bibinfo{author}{\bibfnamefont{K.~L.~G.} \bibnamefont{Heyde}} \bibnamefont{and}
  \bibinfo{author}{\bibfnamefont{J.~L.} \bibnamefont{Wood}},
  \bibinfo{journal}{Rev. Mod. Phys.} \textbf{\bibinfo{volume}{83}},
  \bibinfo{pages}{1467} (\bibinfo{year}{2011}).

\bibitem[{\citenamefont{National{\ }Nuclear{\ }Data{\ }Center{\ }(NNDC)}(May
  2016)}]{050_NNDC}
\bibinfo{author}{\bibnamefont{National{\ }Nuclear{\ }Data{\ }Center{\
  }(NNDC)}}, \bibinfo{journal}{Brookhaven National Laboratory,
  http://www.nndc.bnl.gov/ensdf/}  (\bibinfo{year}{May 2016}).

\bibitem[{\citenamefont{Faestermann et~al.}(2013)\citenamefont{Faestermann,
  G\'{o}rska, and Grawe}}]{051}
\bibinfo{author}{\bibfnamefont{T.}~\bibnamefont{Faestermann}},
  \bibinfo{author}{\bibfnamefont{M.}~\bibnamefont{G\'{o}rska}},
  \bibnamefont{and} \bibinfo{author}{\bibfnamefont{H.}~\bibnamefont{Grawe}},
  \bibinfo{journal}{Prog. Part. Nucl. Phys.} \textbf{\bibinfo{volume}{69}},
  \bibinfo{pages}{85} (\bibinfo{year}{2013}).

\bibitem[{\citenamefont{Brown and Rykaczewski}(1994)}]{052}
\bibinfo{author}{\bibfnamefont{B.~A.} \bibnamefont{Brown}} \bibnamefont{and}
  \bibinfo{author}{\bibfnamefont{K.}~\bibnamefont{Rykaczewski}},
  \bibinfo{journal}{Phys. Rev. C} \textbf{\bibinfo{volume}{50}},
  \bibinfo{pages}{R2270} (\bibinfo{year}{1994}).

\bibitem[{\citenamefont{Heyde et~al.}(1992)\citenamefont{Heyde, C.{\ }De{\
  }Coster, Jolie, and Wood}}]{097}
\bibinfo{author}{\bibfnamefont{K.~L.~G.} \bibnamefont{Heyde}},
  \bibinfo{author}{\bibnamefont{C.{\ }De{\ }Coster}},
  \bibinfo{author}{\bibfnamefont{J.}~\bibnamefont{Jolie}}, \bibnamefont{and}
  \bibinfo{author}{\bibfnamefont{J.~L.} \bibnamefont{Wood}},
  \bibinfo{journal}{Phys. Rev. C} \textbf{\bibinfo{volume}{46}},
  \bibinfo{pages}{541} (\bibinfo{year}{1992}).

\bibitem[{\citenamefont{Jolie and Lehmann}(1995)}]{149Jolie16}
\bibinfo{author}{\bibfnamefont{J.}~\bibnamefont{Jolie}} \bibnamefont{and}
  \bibinfo{author}{\bibfnamefont{H.}~\bibnamefont{Lehmann}},
  \bibinfo{journal}{Phys. Lett. B} \textbf{\bibinfo{volume}{342}},
  \bibinfo{pages}{1} (\bibinfo{year}{1995}).

\bibitem[{\citenamefont{Heyde et~al.}(1995)\citenamefont{Heyde, Jolie, Lehmann,
  C.{\ }De{\ }Coster, and Wood}}]{098}
\bibinfo{author}{\bibfnamefont{K.~L.~G.} \bibnamefont{Heyde}},
  \bibinfo{author}{\bibfnamefont{J.}~\bibnamefont{Jolie}},
  \bibinfo{author}{\bibfnamefont{H.}~\bibnamefont{Lehmann}},
  \bibinfo{author}{\bibnamefont{C.{\ }De{\ }Coster}}, \bibnamefont{and}
  \bibinfo{author}{\bibfnamefont{J.~L.} \bibnamefont{Wood}},
  \bibinfo{journal}{Nucl. Phys. A} \textbf{\bibinfo{volume}{586}},
  \bibinfo{pages}{1} (\bibinfo{year}{1995}).

\bibitem[{\citenamefont{Lehmann and Jolie}(1995)}]{099}
\bibinfo{author}{\bibfnamefont{H.}~\bibnamefont{Lehmann}} \bibnamefont{and}
  \bibinfo{author}{\bibfnamefont{J.}~\bibnamefont{Jolie}},
  \bibinfo{journal}{Nucl. Phys. A} \textbf{\bibinfo{volume}{588}},
  \bibinfo{pages}{623} (\bibinfo{year}{1995}).

\bibitem[{\citenamefont{C.{\ }De{\ }Coster et~al.}(1996)\citenamefont{C.{\
  }De{\ }Coster, Heyde, Decroix, Van{\ }Isacker, Jolie, Lehmann, and
  Wood}}]{101}
\bibinfo{author}{\bibnamefont{C.{\ }De{\ }Coster}},
  \bibinfo{author}{\bibfnamefont{K.~L.~G.} \bibnamefont{Heyde}},
  \bibinfo{author}{\bibfnamefont{B.}~\bibnamefont{Decroix}},
  \bibinfo{author}{\bibfnamefont{P.}~\bibnamefont{Van{\ }Isacker}},
  \bibinfo{author}{\bibfnamefont{J.}~\bibnamefont{Jolie}},
  \bibinfo{author}{\bibfnamefont{H.}~\bibnamefont{Lehmann}}, \bibnamefont{and}
  \bibinfo{author}{\bibfnamefont{J.~L.} \bibnamefont{Wood}},
  \bibinfo{journal}{Nucl. Phys. A} \textbf{\bibinfo{volume}{600}},
  \bibinfo{pages}{251} (\bibinfo{year}{1996}).

\bibitem[{\citenamefont{Lehmann et~al.}(1997)\citenamefont{Lehmann, Jolie, C.{\
  }De{\ }Coster, Decroix, Heyde, and Wood}}]{100}
\bibinfo{author}{\bibfnamefont{H.}~\bibnamefont{Lehmann}},
  \bibinfo{author}{\bibfnamefont{J.}~\bibnamefont{Jolie}},
  \bibinfo{author}{\bibnamefont{C.{\ }De{\ }Coster}},
  \bibinfo{author}{\bibfnamefont{B.}~\bibnamefont{Decroix}},
  \bibinfo{author}{\bibfnamefont{K.~L.~G.} \bibnamefont{Heyde}},
  \bibnamefont{and} \bibinfo{author}{\bibfnamefont{J.~L.} \bibnamefont{Wood}},
  \bibinfo{journal}{Nucl. Phys. A} \textbf{\bibinfo{volume}{621}},
  \bibinfo{pages}{767} (\bibinfo{year}{1997}).

\bibitem[{\citenamefont{C.{\ }De{\ }Coster et~al.}(1997)\citenamefont{C.{\
  }De{\ }Coster, Decroix, Heyde, Wood, Jolie, and Lehmann}}]{150Jolie21}
\bibinfo{author}{\bibnamefont{C.{\ }De{\ }Coster}},
  \bibinfo{author}{\bibfnamefont{B.}~\bibnamefont{Decroix}},
  \bibinfo{author}{\bibfnamefont{K.~L.~G.} \bibnamefont{Heyde}},
  \bibinfo{author}{\bibfnamefont{J.~L.} \bibnamefont{Wood}},
  \bibinfo{author}{\bibfnamefont{J.}~\bibnamefont{Jolie}}, \bibnamefont{and}
  \bibinfo{author}{\bibfnamefont{H.}~\bibnamefont{Lehmann}},
  \bibinfo{journal}{Nucl. Phys. A} \textbf{\bibinfo{volume}{621}},
  \bibinfo{pages}{802} (\bibinfo{year}{1997}).

\bibitem[{\citenamefont{C.{\ }De{\ }Coster et~al.}(1999)\citenamefont{C.{\
  }De{\ }Coster, Decroix, Heyde, Jolie, Lehmann, and Wood}}]{112}
\bibinfo{author}{\bibnamefont{C.{\ }De{\ }Coster}},
  \bibinfo{author}{\bibfnamefont{B.}~\bibnamefont{Decroix}},
  \bibinfo{author}{\bibfnamefont{K.~L.~G.} \bibnamefont{Heyde}},
  \bibinfo{author}{\bibfnamefont{J.}~\bibnamefont{Jolie}},
  \bibinfo{author}{\bibfnamefont{H.}~\bibnamefont{Lehmann}}, \bibnamefont{and}
  \bibinfo{author}{\bibfnamefont{J.~L.} \bibnamefont{Wood}},
  \bibinfo{journal}{Nucl. Phys. A} \textbf{\bibinfo{volume}{651}},
  \bibinfo{pages}{31} (\bibinfo{year}{1999}).

\bibitem[{\citenamefont{Prochniak et~al.}(2012)\citenamefont{Prochniak,
  Quentin, and Imadalou}}]{054}
\bibinfo{author}{\bibfnamefont{L.}~\bibnamefont{Prochniak}},
  \bibinfo{author}{\bibfnamefont{P.}~\bibnamefont{Quentin}}, \bibnamefont{and}
  \bibinfo{author}{\bibfnamefont{M.}~\bibnamefont{Imadalou}},
  \bibinfo{journal}{Mod. Phys. E} \textbf{\bibinfo{volume}{21}},
  \bibinfo{pages}{1250036} (\bibinfo{year}{2012}).

\bibitem[{\citenamefont{Rodriguez et~al.}(2008)\citenamefont{Rodriguez, Egido,
  and Jungclaus}}]{055}
\bibinfo{author}{\bibfnamefont{T.~R.} \bibnamefont{Rodriguez}},
  \bibinfo{author}{\bibfnamefont{J.~L.} \bibnamefont{Egido}}, \bibnamefont{and}
  \bibinfo{author}{\bibfnamefont{A.}~\bibnamefont{Jungclaus}},
  \bibinfo{journal}{Phys. Lett. B} \textbf{\bibinfo{volume}{668}},
  \bibinfo{pages}{410} (\bibinfo{year}{2008}).

\bibitem[{\citenamefont{Kumar}(1972)}]{056_Kumar}
\bibinfo{author}{\bibfnamefont{K.}~\bibnamefont{Kumar}},
  \bibinfo{journal}{Phys. Rev. Letters} \textbf{\bibinfo{volume}{28}},
  \bibinfo{pages}{249} (\bibinfo{year}{1972}).

\bibitem[{\citenamefont{Cline}(1986)}]{057_Cline}
\bibinfo{author}{\bibfnamefont{D.}~\bibnamefont{Cline}},
  \bibinfo{journal}{Annu. Rev. Nucl. Part. Sci.} \textbf{\bibinfo{volume}{36}},
  \bibinfo{pages}{683} (\bibinfo{year}{1986}).

\bibitem[{\citenamefont{Cline}(1971)}]{058}
\bibinfo{author}{\bibfnamefont{D.}~\bibnamefont{Cline}}, in
  \emph{\bibinfo{booktitle}{Proc. of the Orsay Colloquim on Intermediate
  Nuclei}}, edited by \bibinfo{editor}{\bibfnamefont{R.}~\bibnamefont{Foucher}}
  (\bibinfo{publisher}{Orsay Inst. Phys. Nucl., 4}, \bibinfo{year}{1971}).

\bibitem[{\citenamefont{Cline and Flaum}(1972)}]{059}
\bibinfo{author}{\bibfnamefont{D.}~\bibnamefont{Cline}} \bibnamefont{and}
  \bibinfo{author}{\bibfnamefont{C.}~\bibnamefont{Flaum}}, in
  \emph{\bibinfo{booktitle}{Proc. of Int. Conf. on Nuclear Structure using
  Electron Scattering}}, edited by
  \bibinfo{editor}{\bibfnamefont{K.}~\bibnamefont{Shoda}} \bibnamefont{and}
  \bibinfo{editor}{\bibnamefont{H.Ui}} (\bibinfo{publisher}{Sendai 1972,
  Sendai, Tohoku Univ. 61}, \bibinfo{year}{1972}).

\bibitem[{\citenamefont{Cline}(1983)}]{060}
\bibinfo{author}{\bibfnamefont{D.}~\bibnamefont{Cline}}, in
  \emph{\bibinfo{booktitle}{Proc. of the Int. Conf. on Interacting Bose-Fermi
  systems in Nuclei}}, edited by
  \bibinfo{editor}{\bibfnamefont{F.}~\bibnamefont{Iachello}}
  (\bibinfo{publisher}{Plenum Press, New-York, 241}, \bibinfo{year}{1983}).

\bibitem[{\citenamefont{Bohr and Mottelson}(1998)}]{061_Bohr&Mottelson}
\bibinfo{author}{\bibfnamefont{A.}~\bibnamefont{Bohr}} \bibnamefont{and}
  \bibinfo{author}{\bibfnamefont{B.}~\bibnamefont{Mottelson}},
  \emph{\bibinfo{title}{Nuclear Structure Vol.II: Nuclear Deformation}}
  (\bibinfo{publisher}{World Scientic Publishing}, \bibinfo{year}{1998}).

\bibitem[{\citenamefont{Holt et~al.}(2000)\citenamefont{Holt, Engeland,
  Hjorth-Jensen, and Osnes}}]{159free}
\bibinfo{author}{\bibfnamefont{A.}~\bibnamefont{Holt}},
  \bibinfo{author}{\bibfnamefont{T.}~\bibnamefont{Engeland}},
  \bibinfo{author}{\bibfnamefont{M.}~\bibnamefont{Hjorth-Jensen}},
  \bibnamefont{and} \bibinfo{author}{\bibfnamefont{E.}~\bibnamefont{Osnes}},
  \bibinfo{journal}{Phys. Rev. C} \textbf{\bibinfo{volume}{61}},
  \bibinfo{pages}{064318} (\bibinfo{year}{2000}).

\bibitem[{\citenamefont{Brown et~al.}(1976)\citenamefont{Brown, Lesser, and
  Fossan}}]{160free}
\bibinfo{author}{\bibfnamefont{B.~A.} \bibnamefont{Brown}},
  \bibinfo{author}{\bibfnamefont{P.~M.~S.} \bibnamefont{Lesser}},
  \bibnamefont{and} \bibinfo{author}{\bibfnamefont{D.~B.}
  \bibnamefont{Fossan}}, \bibinfo{journal}{Phys. Rev. C}
  \textbf{\bibinfo{volume}{13}}, \bibinfo{pages}{1900} (\bibinfo{year}{1976}).

\bibitem[{\citenamefont{Sieja et~al.}(2009)\citenamefont{Sieja, Nowacki,
  Langanke, and Mart\'{i}nez-Pinedo}}]{161free}
\bibinfo{author}{\bibfnamefont{K.}~\bibnamefont{Sieja}},
  \bibinfo{author}{\bibfnamefont{F.}~\bibnamefont{Nowacki}},
  \bibinfo{author}{\bibfnamefont{K.}~\bibnamefont{Langanke}}, \bibnamefont{and}
  \bibinfo{author}{\bibfnamefont{G.}~\bibnamefont{Mart\'{i}nez-Pinedo}},
  \bibinfo{journal}{Phys. Rev. C} \textbf{\bibinfo{volume}{79}},
  \bibinfo{pages}{064310} (\bibinfo{year}{2009}).

\bibitem[{\citenamefont{Blazhev et~al.}(2015)\citenamefont{Blazhev, Heyde,
  Jolie, and Schmidt}}]{062}
\bibinfo{author}{\bibfnamefont{A.}~\bibnamefont{Blazhev}},
  \bibinfo{author}{\bibfnamefont{K.~L.~G.} \bibnamefont{Heyde}},
  \bibinfo{author}{\bibfnamefont{J.}~\bibnamefont{Jolie}}, \bibnamefont{and}
  \bibinfo{author}{\bibfnamefont{T.}~\bibnamefont{Schmidt}},
  \bibinfo{journal}{Proc. of the CGS15 conference, EPJ Web of Conferences 93,
  01016}  (\bibinfo{year}{2015}).

\bibitem[{\citenamefont{Ayangeakaa et~al.}(2014)\citenamefont{Ayangeakaa,
  Janssens, Wu, Allmond, Wood, Zhu, Albers, Almaraz-Calderon, Bucher, Carpenter
  et~al.}}]{063}
\bibinfo{author}{\bibfnamefont{A.~D.} \bibnamefont{Ayangeakaa}},
  \bibinfo{author}{\bibfnamefont{R.~V.~F.} \bibnamefont{Janssens}},
  \bibinfo{author}{\bibfnamefont{C.~Y.} \bibnamefont{Wu}},
  \bibinfo{author}{\bibfnamefont{J.~M.} \bibnamefont{Allmond}},
  \bibinfo{author}{\bibfnamefont{J.~L.} \bibnamefont{Wood}},
  \bibinfo{author}{\bibfnamefont{S.}~\bibnamefont{Zhu}},
  \bibinfo{author}{\bibfnamefont{M.}~\bibnamefont{Albers}},
  \bibinfo{author}{\bibfnamefont{S.}~\bibnamefont{Almaraz-Calderon}},
  \bibinfo{author}{\bibfnamefont{B.}~\bibnamefont{Bucher}},
  \bibinfo{author}{\bibfnamefont{M.~P.} \bibnamefont{Carpenter}},
  \bibnamefont{et~al.}, \bibinfo{journal}{Phys. Lett. B}
  \textbf{\bibinfo{volume}{754}}, \bibinfo{pages}{254} (\bibinfo{year}{2014}).

\bibitem[{\citenamefont{G\"{o}rgen and Korten}(2016)}]{064}
\bibinfo{author}{\bibfnamefont{A.}~\bibnamefont{G\"{o}rgen}} \bibnamefont{and}
  \bibinfo{author}{\bibfnamefont{W.}~\bibnamefont{Korten}},
  \bibinfo{journal}{J.Phys.G} \textbf{\bibinfo{volume}{43}},
  \bibinfo{pages}{024002} (\bibinfo{year}{2016}).

\bibitem[{\citenamefont{Cl\'{e}ment et~al.}(2007)\citenamefont{Cl\'{e}ment,
  G\"{o}rgen, Korten, Bouchez, Chatillon, Delaroche, Girod, Goutte,
  H\"{u}rstel, Le{\ }Coz et~al.}}]{065}
\bibinfo{author}{\bibfnamefont{E.}~\bibnamefont{Cl\'{e}ment}},
  \bibinfo{author}{\bibfnamefont{A.}~\bibnamefont{G\"{o}rgen}},
  \bibinfo{author}{\bibfnamefont{W.}~\bibnamefont{Korten}},
  \bibinfo{author}{\bibfnamefont{E.}~\bibnamefont{Bouchez}},
  \bibinfo{author}{\bibfnamefont{A.}~\bibnamefont{Chatillon}},
  \bibinfo{author}{\bibfnamefont{J.-P.} \bibnamefont{Delaroche}},
  \bibinfo{author}{\bibfnamefont{M.}~\bibnamefont{Girod}},
  \bibinfo{author}{\bibfnamefont{H.}~\bibnamefont{Goutte}},
  \bibinfo{author}{\bibfnamefont{A.}~\bibnamefont{H\"{u}rstel}},
  \bibinfo{author}{\bibfnamefont{Y.}~\bibnamefont{Le{\ }Coz}},
  \bibnamefont{et~al.}, \bibinfo{journal}{Phys. Rev. C}
  \textbf{\bibinfo{volume}{75}}, \bibinfo{pages}{054313}
  (\bibinfo{year}{2007}).

\bibitem[{\citenamefont{Wrzosek-Lipska
  et~al.}(2012)\citenamefont{Wrzosek-Lipska, Pr\'{o}chniak, Zieli\'{n}ska,
  Srebrny, Hady\'{n}ska-Kl\c{e}k, Iwanicki, Kisieli\'{n}ski, Kowalczyk,
  Napiorkowski, Pi\c{e}tak et~al.}}]{066}
\bibinfo{author}{\bibfnamefont{K.}~\bibnamefont{Wrzosek-Lipska}},
  \bibinfo{author}{\bibfnamefont{L.}~\bibnamefont{Pr\'{o}chniak}},
  \bibinfo{author}{\bibfnamefont{M.}~\bibnamefont{Zieli\'{n}ska}},
  \bibinfo{author}{\bibfnamefont{J.}~\bibnamefont{Srebrny}},
  \bibinfo{author}{\bibfnamefont{K.}~\bibnamefont{Hady\'{n}ska-Kl\c{e}k}},
  \bibinfo{author}{\bibfnamefont{J.}~\bibnamefont{Iwanicki}},
  \bibinfo{author}{\bibfnamefont{M.}~\bibnamefont{Kisieli\'{n}ski}},
  \bibinfo{author}{\bibfnamefont{M.}~\bibnamefont{Kowalczyk}},
  \bibinfo{author}{\bibfnamefont{P.~J.} \bibnamefont{Napiorkowski}},
  \bibinfo{author}{\bibfnamefont{D.}~\bibnamefont{Pi\c{e}tak}},
  \bibnamefont{et~al.}, \bibinfo{journal}{Phys. Rev. C}
  \textbf{\bibinfo{volume}{86}}, \bibinfo{pages}{064305}
  (\bibinfo{year}{2012}).

\bibitem[{\citenamefont{Srebrny et~al.}(2006)\citenamefont{Srebrny, Czosnyka,
  Droste, Rohozi\'{n}ski, Pr\'{o}chniak, K.Zaj\c{a}c, Pomorski, Cline, Wu,
  B\"{a}cklin et~al.}}]{067_NPA_766}
\bibinfo{author}{\bibfnamefont{J.}~\bibnamefont{Srebrny}},
  \bibinfo{author}{\bibfnamefont{T.}~\bibnamefont{Czosnyka}},
  \bibinfo{author}{\bibfnamefont{C.}~\bibnamefont{Droste}},
  \bibinfo{author}{\bibfnamefont{S.}~\bibnamefont{Rohozi\'{n}ski}},
  \bibinfo{author}{\bibfnamefont{L.}~\bibnamefont{Pr\'{o}chniak}},
  \bibinfo{author}{\bibnamefont{K.Zaj\c{a}c}},
  \bibinfo{author}{\bibfnamefont{K.}~\bibnamefont{Pomorski}},
  \bibinfo{author}{\bibfnamefont{D.}~\bibnamefont{Cline}},
  \bibinfo{author}{\bibfnamefont{C.~Y.} \bibnamefont{Wu}},
  \bibinfo{author}{\bibfnamefont{A.}~\bibnamefont{B\"{a}cklin}},
  \bibnamefont{et~al.}, \bibinfo{journal}{Nucl. Phys. A}
  \textbf{\bibinfo{volume}{766}}, \bibinfo{pages}{2551} (\bibinfo{year}{2006}).

\bibitem[{\citenamefont{Srebrny and Cline}(2011)}]{068}
\bibinfo{author}{\bibfnamefont{J.}~\bibnamefont{Srebrny}} \bibnamefont{and}
  \bibinfo{author}{\bibfnamefont{D.}~\bibnamefont{Cline}},
  \bibinfo{journal}{Int. J. Mod. Phys. E} \textbf{\bibinfo{volume}{20}},
  \bibinfo{pages}{422} (\bibinfo{year}{2011}).

\bibitem[{\citenamefont{Wu et~al.}(1991)\citenamefont{Wu, Cline, Vogt, Kernan,
  Czosnyka, Helmer, Ibbotson, Kavka, Kotlinski, and Diamond}}]{069}
\bibinfo{author}{\bibfnamefont{C.~Y.} \bibnamefont{Wu}},
  \bibinfo{author}{\bibfnamefont{D.}~\bibnamefont{Cline}},
  \bibinfo{author}{\bibfnamefont{E.~G.} \bibnamefont{Vogt}},
  \bibinfo{author}{\bibfnamefont{W.~J.} \bibnamefont{Kernan}},
  \bibinfo{author}{\bibfnamefont{T.}~\bibnamefont{Czosnyka}},
  \bibinfo{author}{\bibfnamefont{K.~G.} \bibnamefont{Helmer}},
  \bibinfo{author}{\bibfnamefont{R.~W.} \bibnamefont{Ibbotson}},
  \bibinfo{author}{\bibfnamefont{A.~E.} \bibnamefont{Kavka}},
  \bibinfo{author}{\bibfnamefont{B.}~\bibnamefont{Kotlinski}},
  \bibnamefont{and} \bibinfo{author}{\bibfnamefont{R.~M.}
  \bibnamefont{Diamond}}, \bibinfo{journal}{Nucl. Phys. A}
  \textbf{\bibinfo{volume}{533}}, \bibinfo{pages}{359} (\bibinfo{year}{1991}).

\bibitem[{\citenamefont{Wu et~al.}(1996)\citenamefont{Wu, Cline, Czosnyka,
  Backlin, Baktash, Diamond, Dracoulis, Hasselgren, Kluge, Kotlinski
  et~al.}}]{070}
\bibinfo{author}{\bibfnamefont{C.~Y.} \bibnamefont{Wu}},
  \bibinfo{author}{\bibfnamefont{D.}~\bibnamefont{Cline}},
  \bibinfo{author}{\bibfnamefont{T.}~\bibnamefont{Czosnyka}},
  \bibinfo{author}{\bibfnamefont{A.}~\bibnamefont{Backlin}},
  \bibinfo{author}{\bibfnamefont{C.}~\bibnamefont{Baktash}},
  \bibinfo{author}{\bibfnamefont{R.~M.} \bibnamefont{Diamond}},
  \bibinfo{author}{\bibfnamefont{G.~D.} \bibnamefont{Dracoulis}},
  \bibinfo{author}{\bibfnamefont{L.}~\bibnamefont{Hasselgren}},
  \bibinfo{author}{\bibfnamefont{H.}~\bibnamefont{Kluge}},
  \bibinfo{author}{\bibfnamefont{B.}~\bibnamefont{Kotlinski}},
  \bibnamefont{et~al.}, \bibinfo{journal}{Nucl. Phys. A}
  \textbf{\bibinfo{volume}{607}}, \bibinfo{pages}{178} (\bibinfo{year}{1996}).

\bibitem[{\citenamefont{Bree et~al.}(2014)\citenamefont{Bree, Wrzosek-Lipska,
  Petts, Andreyev, Bastin, Bender, Blazhev, Bruyneel, Butler, Butterworth
  et~al.}}]{071}
\bibinfo{author}{\bibfnamefont{N.}~\bibnamefont{Bree}},
  \bibinfo{author}{\bibfnamefont{K.}~\bibnamefont{Wrzosek-Lipska}},
  \bibinfo{author}{\bibfnamefont{A.}~\bibnamefont{Petts}},
  \bibinfo{author}{\bibfnamefont{A.}~\bibnamefont{Andreyev}},
  \bibinfo{author}{\bibfnamefont{B.}~\bibnamefont{Bastin}},
  \bibinfo{author}{\bibfnamefont{M.}~\bibnamefont{Bender}},
  \bibinfo{author}{\bibfnamefont{A.}~\bibnamefont{Blazhev}},
  \bibinfo{author}{\bibfnamefont{B.}~\bibnamefont{Bruyneel}},
  \bibinfo{author}{\bibfnamefont{P.~A.} \bibnamefont{Butler}},
  \bibinfo{author}{\bibfnamefont{J.}~\bibnamefont{Butterworth}},
  \bibnamefont{et~al.}, \bibinfo{journal}{Phys. Rev. Lett.}
  \textbf{\bibinfo{volume}{112}}, \bibinfo{pages}{162701}
  (\bibinfo{year}{2014}).

\bibitem[{\citenamefont{Kesteloot et~al.}(2015)\citenamefont{Kesteloot, Bastin,
  Gaffney, Wrzosek-Lipska, Auranen, Bauer, Bender, Bildstein, Blazhev,
  B\"{o}nig et~al.}}]{072}
\bibinfo{author}{\bibfnamefont{N.}~\bibnamefont{Kesteloot}},
  \bibinfo{author}{\bibfnamefont{B.}~\bibnamefont{Bastin}},
  \bibinfo{author}{\bibfnamefont{L.~P.} \bibnamefont{Gaffney}},
  \bibinfo{author}{\bibfnamefont{K.}~\bibnamefont{Wrzosek-Lipska}},
  \bibinfo{author}{\bibfnamefont{K.}~\bibnamefont{Auranen}},
  \bibinfo{author}{\bibfnamefont{C.}~\bibnamefont{Bauer}},
  \bibinfo{author}{\bibfnamefont{M.}~\bibnamefont{Bender}},
  \bibinfo{author}{\bibfnamefont{V.}~\bibnamefont{Bildstein}},
  \bibinfo{author}{\bibfnamefont{A.}~\bibnamefont{Blazhev}},
  \bibinfo{author}{\bibfnamefont{S.}~\bibnamefont{B\"{o}nig}},
  \bibnamefont{et~al.}, \bibinfo{journal}{Phys. Rev. C}
  \textbf{\bibinfo{volume}{92}}, \bibinfo{pages}{054301}
  (\bibinfo{year}{2015}).

\bibitem[{\citenamefont{Garc\'{i}a-Ramos and Heyde}(2014)}]{073}
\bibinfo{author}{\bibfnamefont{J.~E.} \bibnamefont{Garc\'{i}a-Ramos}}
  \bibnamefont{and} \bibinfo{author}{\bibfnamefont{K.~L.~G.}
  \bibnamefont{Heyde}}, \bibinfo{journal}{Phys. Rev. C}
  \textbf{\bibinfo{volume}{89}}, \bibinfo{pages}{014306}
  (\bibinfo{year}{2014}).

\bibitem[{\citenamefont{Garc\'{i}a-Ramos and Heyde}(2015)}]{074}
\bibinfo{author}{\bibfnamefont{J.~E.} \bibnamefont{Garc\'{i}a-Ramos}}
  \bibnamefont{and} \bibinfo{author}{\bibfnamefont{K.~L.~G.}
  \bibnamefont{Heyde}}, \bibinfo{journal}{Phys. Rev. C}
  \textbf{\bibinfo{volume}{92}}, \bibinfo{pages}{034309}
  (\bibinfo{year}{2015}).

\bibitem[{\citenamefont{Jolos et~al.}(1997)\citenamefont{Jolos, von{\
  }Brentano, Pietralla, and Schneider}}]{075}
\bibinfo{author}{\bibfnamefont{R.~V.} \bibnamefont{Jolos}},
  \bibinfo{author}{\bibfnamefont{P.}~\bibnamefont{von{\ }Brentano}},
  \bibinfo{author}{\bibfnamefont{N.}~\bibnamefont{Pietralla}},
  \bibnamefont{and}
  \bibinfo{author}{\bibfnamefont{I.}~\bibnamefont{Schneider}},
  \bibinfo{journal}{Nucl. Phys. A} \textbf{\bibinfo{volume}{618}},
  \bibinfo{pages}{126} (\bibinfo{year}{1997}).

\bibitem[{\citenamefont{Werner et~al.}(2008)\citenamefont{Werner, Williams,
  Casperson, Casten, Scholl, and von Brentano}}]{076}
\bibinfo{author}{\bibfnamefont{V.}~\bibnamefont{Werner}},
  \bibinfo{author}{\bibfnamefont{E.}~\bibnamefont{Williams}},
  \bibinfo{author}{\bibfnamefont{R.~J.} \bibnamefont{Casperson}},
  \bibinfo{author}{\bibfnamefont{R.~F.} \bibnamefont{Casten}},
  \bibinfo{author}{\bibfnamefont{C.}~\bibnamefont{Scholl}}, \bibnamefont{and}
  \bibinfo{author}{\bibfnamefont{P.}~\bibnamefont{von Brentano}},
  \bibinfo{journal}{Phys. Rev. C} \textbf{\bibinfo{volume}{78}},
  \bibinfo{pages}{051303(R)} (\bibinfo{year}{2008}).

\bibitem[{\citenamefont{Dobaczewski et~al.}(1987)\citenamefont{Dobaczewski,
  Rohozinski, and Srebrny}}]{077}
\bibinfo{author}{\bibfnamefont{J.}~\bibnamefont{Dobaczewski}},
  \bibinfo{author}{\bibfnamefont{S.~G.} \bibnamefont{Rohozinski}},
  \bibnamefont{and} \bibinfo{author}{\bibfnamefont{J.}~\bibnamefont{Srebrny}},
  \bibinfo{journal}{Nucl. Phys. A} \textbf{\bibinfo{volume}{462}},
  \bibinfo{pages}{72} (\bibinfo{year}{1987}).

\bibitem[{\citenamefont{Borge}(2016)}]{153Blazhev01_not_listet_HIE-ISOLDE}
\bibinfo{author}{\bibfnamefont{M.}~\bibnamefont{Borge}},
  \bibinfo{journal}{Nucl. Instrum. Methods Phys. Res., Sect. B}
  \textbf{\bibinfo{volume}{376}}, \bibinfo{pages}{408} (\bibinfo{year}{2016}).

\bibitem[{\citenamefont{Nilsson}(1955)}]{078}
\bibinfo{author}{\bibfnamefont{S.~G.} \bibnamefont{Nilsson}},
  \bibinfo{journal}{Kon. Dan. Vidensk Selsk. Mat. Fys. Medd.}
  \textbf{\bibinfo{volume}{29}}, \bibinfo{pages}{n16} (\bibinfo{year}{1955}).

\bibitem[{\citenamefont{Nilsson and Ragnarsson}(1995)}]{079}
\bibinfo{author}{\bibfnamefont{S.~G.} \bibnamefont{Nilsson}} \bibnamefont{and}
  \bibinfo{author}{\bibfnamefont{I.}~\bibnamefont{Ragnarsson}},
  \emph{\bibinfo{title}{Nuclear Shells and Shapes}}
  (\bibinfo{publisher}{Cambridge Univ. Press, New-York}, \bibinfo{year}{1995}).

\bibitem[{\citenamefont{Pr\'{o}chniak and Rohozinski}(2009)}]{080}
\bibinfo{author}{\bibfnamefont{L.}~\bibnamefont{Pr\'{o}chniak}}
  \bibnamefont{and} \bibinfo{author}{\bibfnamefont{S.~G.}
  \bibnamefont{Rohozinski}}, \bibinfo{journal}{J. Phys. G}
  \textbf{\bibinfo{volume}{36}}, \bibinfo{pages}{123101}
  (\bibinfo{year}{2009}).

\bibitem[{\citenamefont{Eisenberg and Greiner}(1987)}]{081}
\bibinfo{author}{\bibfnamefont{J.}~\bibnamefont{Eisenberg}} \bibnamefont{and}
  \bibinfo{author}{\bibfnamefont{W.}~\bibnamefont{Greiner}},
  \emph{\bibinfo{title}{Nuclear Theory Vol. 1 - Nuclear Models}}
  (\bibinfo{publisher}{North-Holland Physics Publishing},
  \bibinfo{year}{1987}).

\bibitem[{\citenamefont{Pietralla et~al.}(1994)\citenamefont{Pietralla, von{\
  }Brentano, Casten, Otsuka, and Zamfir}}]{131free}
\bibinfo{author}{\bibfnamefont{N.}~\bibnamefont{Pietralla}},
  \bibinfo{author}{\bibfnamefont{P.}~\bibnamefont{von{\ }Brentano}},
  \bibinfo{author}{\bibfnamefont{R.~F.} \bibnamefont{Casten}},
  \bibinfo{author}{\bibfnamefont{T.}~\bibnamefont{Otsuka}}, \bibnamefont{and}
  \bibinfo{author}{\bibfnamefont{N.~V.} \bibnamefont{Zamfir}},
  \bibinfo{journal}{Phys. Rev. Lett.} \textbf{\bibinfo{volume}{73}},
  \bibinfo{pages}{2962} (\bibinfo{year}{1994}).

\bibitem[{\citenamefont{Casten}(2005)}]{082_Casten}
\bibinfo{author}{\bibfnamefont{R.~F.} \bibnamefont{Casten}},
  \emph{\bibinfo{title}{Nuclear Structure from a Simple Perspective}}
  (\bibinfo{publisher}{Oxford University Press}, \bibinfo{year}{2005}).

\bibitem[{\citenamefont{Thorslund et~al.}(1993)\citenamefont{Thorslund,
  Fahlander, Nyberg, Juutinen, Julin, Piiparinen, Wyss, Lampinen, L\"{o}nnroth,
  M\"{u}ller et~al.}}]{130free}
\bibinfo{author}{\bibfnamefont{I.}~\bibnamefont{Thorslund}},
  \bibinfo{author}{\bibfnamefont{C.}~\bibnamefont{Fahlander}},
  \bibinfo{author}{\bibfnamefont{J.}~\bibnamefont{Nyberg}},
  \bibinfo{author}{\bibfnamefont{S.}~\bibnamefont{Juutinen}},
  \bibinfo{author}{\bibfnamefont{R.}~\bibnamefont{Julin}},
  \bibinfo{author}{\bibfnamefont{M.}~\bibnamefont{Piiparinen}},
  \bibinfo{author}{\bibfnamefont{R.}~\bibnamefont{Wyss}},
  \bibinfo{author}{\bibfnamefont{A.}~\bibnamefont{Lampinen}},
  \bibinfo{author}{\bibfnamefont{T.}~\bibnamefont{L\"{o}nnroth}},
  \bibinfo{author}{\bibfnamefont{D.}~\bibnamefont{M\"{u}ller}},
  \bibnamefont{et~al.}, \bibinfo{journal}{Nucl. Phys. A}
  \textbf{\bibinfo{volume}{564}}, \bibinfo{pages}{285} (\bibinfo{year}{1993}).

\bibitem[{\citenamefont{Zuker et~al.}(1995)\citenamefont{Zuker, Retamosa,
  Poves, and Caurier}}]{162Heyde}
\bibinfo{author}{\bibfnamefont{A.~P.} \bibnamefont{Zuker}},
  \bibinfo{author}{\bibfnamefont{J.}~\bibnamefont{Retamosa}},
  \bibinfo{author}{\bibfnamefont{A.}~\bibnamefont{Poves}}, \bibnamefont{and}
  \bibinfo{author}{\bibfnamefont{E.}~\bibnamefont{Caurier}},
  \bibinfo{journal}{Phys. Rev. C} \textbf{\bibinfo{volume}{52}},
  \bibinfo{pages}{R1741(R)} (\bibinfo{year}{1995}).

\bibitem[{\citenamefont{Mart\'{i}nez-Pinedo
  et~al.}(1997)\citenamefont{Mart\'{i}nez-Pinedo, Zuker, Poves, and
  Caurier}}]{163Heyde}
\bibinfo{author}{\bibfnamefont{G.}~\bibnamefont{Mart\'{i}nez-Pinedo}},
  \bibinfo{author}{\bibfnamefont{A.~P.} \bibnamefont{Zuker}},
  \bibinfo{author}{\bibfnamefont{A.}~\bibnamefont{Poves}}, \bibnamefont{and}
  \bibinfo{author}{\bibfnamefont{E.}~\bibnamefont{Caurier}},
  \bibinfo{journal}{Phys. Rev. C} \textbf{\bibinfo{volume}{55}},
  \bibinfo{pages}{187} (\bibinfo{year}{1997}).

\bibitem[{\citenamefont{Arima et~al.}(1969)\citenamefont{Arima, Harvey, and
  Shimizu}}]{164Heyde}
\bibinfo{author}{\bibfnamefont{A.}~\bibnamefont{Arima}},
  \bibinfo{author}{\bibfnamefont{M.}~\bibnamefont{Harvey}}, \bibnamefont{and}
  \bibinfo{author}{\bibfnamefont{K.}~\bibnamefont{Shimizu}},
  \bibinfo{journal}{Phys. Lett. B} \textbf{\bibinfo{volume}{30}},
  \bibinfo{pages}{517} (\bibinfo{year}{1969}).

\bibitem[{\citenamefont{Hecht and Adler}(1969)}]{165Heyde}
\bibinfo{author}{\bibfnamefont{K.}~\bibnamefont{Hecht}} \bibnamefont{and}
  \bibinfo{author}{\bibfnamefont{A.}~\bibnamefont{Adler}},
  \bibinfo{journal}{Nucl. Phys. A} \textbf{\bibinfo{volume}{137}},
  \bibinfo{pages}{129} (\bibinfo{year}{1969}).

\bibitem[{\citenamefont{Elliott}(1958{\natexlab{a}})}]{166Heyde}
\bibinfo{author}{\bibfnamefont{J.~P.} \bibnamefont{Elliott}},
  \bibinfo{journal}{Proc. Roy. Soc. A} \textbf{\bibinfo{volume}{245}},
  \bibinfo{pages}{128} (\bibinfo{year}{1958}{\natexlab{a}}).

\bibitem[{\citenamefont{Elliott}(1958{\natexlab{b}})}]{167Heyde}
\bibinfo{author}{\bibfnamefont{J.~P.} \bibnamefont{Elliott}},
  \bibinfo{journal}{Proc. Roy. Soc. A} \textbf{\bibinfo{volume}{254}},
  \bibinfo{pages}{562} (\bibinfo{year}{1958}{\natexlab{b}}).

\bibitem[{\citenamefont{Duflo and Zuker}(1999)}]{170Heyde}
\bibinfo{author}{\bibfnamefont{J.}~\bibnamefont{Duflo}} \bibnamefont{and}
  \bibinfo{author}{\bibfnamefont{A.~P.} \bibnamefont{Zuker}},
  \bibinfo{journal}{Phys. Rev. C} \textbf{\bibinfo{volume}{59}},
  \bibinfo{pages}{R2347(R)} (\bibinfo{year}{1999}).

\bibitem[{\citenamefont{Nowacki}(2007)}]{171Heyde}
\bibinfo{author}{\bibfnamefont{F.}~\bibnamefont{Nowacki}},
  \bibinfo{journal}{Acta Phys. Pol. B} \textbf{\bibinfo{volume}{38}},
  \bibinfo{pages}{1369} (\bibinfo{year}{2007}).

\bibitem[{151(2016)}]{151free_shape_coex_issue}
\bibinfo{journal}{J. Phys. G} \textbf{\bibinfo{volume}{43}},
  \bibinfo{pages}{focus section 020401} (\bibinfo{year}{2016}).

\bibitem[{\citenamefont{Benczer-Koller
  et~al.}(2016)\citenamefont{Benczer-Koller, Kumbartzki, Speidel, Torres,
  Robinson, Sharon, Allmond, Fallon, Abramovic, Bernstein
  et~al.}}]{158Blazhev02_not_listed}
\bibinfo{author}{\bibfnamefont{N.}~\bibnamefont{Benczer-Koller}},
  \bibinfo{author}{\bibfnamefont{G.~J.} \bibnamefont{Kumbartzki}},
  \bibinfo{author}{\bibfnamefont{K.-H.} \bibnamefont{Speidel}},
  \bibinfo{author}{\bibfnamefont{D.~A.} \bibnamefont{Torres}},
  \bibinfo{author}{\bibfnamefont{S.~J.~Q.} \bibnamefont{Robinson}},
  \bibinfo{author}{\bibfnamefont{Y.~Y.} \bibnamefont{Sharon}},
  \bibinfo{author}{\bibfnamefont{J.~M.} \bibnamefont{Allmond}},
  \bibinfo{author}{\bibfnamefont{P.}~\bibnamefont{Fallon}},
  \bibinfo{author}{\bibfnamefont{I.}~\bibnamefont{Abramovic}},
  \bibinfo{author}{\bibfnamefont{L.~A.} \bibnamefont{Bernstein}},
  \bibnamefont{et~al.}, \bibinfo{journal}{Phys. Rev. C}
  \textbf{\bibinfo{volume}{94}}, \bibinfo{pages}{034303}
  (\bibinfo{year}{2016}).

\end{thebibliography}

\end{document}